\documentclass[11pt,fleqn]{article}

\usepackage{bm}
\usepackage{amsmath}
\usepackage{graphicx}
\usepackage{cite}
\usepackage[margin=1.25in]{geometry}
\usepackage{booktabs}
\usepackage{caption}
\usepackage{subcaption}
\usepackage{nicefrac}
\usepackage{placeins}
\usepackage{afterpage}
\usepackage{color}
\usepackage[bottom]{footmisc}

\definecolor{olivedrab}{rgb}{0.42,0.56,0.14}
\definecolor{oxfordblue}{rgb}{0.0, 0.13, 0.28}

\newcommand{\tensr}[1]{\bm{\mathsf{#1}}}
\newcommand{\F}{\tensr{F}}
\newcommand{\PP}{\tensr{P}}

\newcommand{\Fi}{\tensr{F}^{-1}}
\newcommand{\K}{\kappa}
\newcommand{\KMv}[1]{\kappa\subscript{#1}^{Mv}}
\newcommand{\Kp}{\kappa'}
\newcommand{\Sig}{\sigma}

\newcommand{\m}{\mathbf{m}}
\newcommand{\mc}{\mathbf{m^c}}

\newcommand{\Kt}{\tilde{\kappa}}
\newcommand{\Kpt}{\tilde{\kappa}'}

\newcommand{\subscript}[1]{_{#1}}
\newcommand{\f}[1]{f\subscript{#1}}
\newcommand{\ft}[1]{\tilde{f}\subscript{#1}}
\newcommand{\Ks}[1]{\kappa\subscript{#1}}
\newcommand{\Kts}[1]{\tilde{\kappa}\subscript{#1}}
\newcommand{\Ss}[1]{\sigma\subscript{#1}}

\newcommand{\ux}{u_x}
\newcommand{\uy}{u_y}
\newcommand{\uz}{u_z}
\newcommand{\uxx}{u_{x}^{2}}
\newcommand{\uyy}{u_{y}^{2}}
\newcommand{\uzz}{u_{z}^{2}}

\newcommand{\Kps}[1]{\kappa^{\prime}\subscript{#1}}
\newcommand{\Ktps}[1]{\tilde{\kappa}^{\prime}\subscript{#1}}
\title{\vspace{-2.0cm}Fokker-Planck Central Moment Lattice Boltzmann Method for Effective Simulations of Fluid Dynamics}
\author{{William Schupbach}\\Department of Mechanical Engineering\\ University of Colorado Denver \and Kannan Premnath\footnotemark\\Department of Mechanical Engineering\\ University of Colorado Denver}

\begin{document}

\maketitle

\begin{abstract}
We present a new formulation of the central moment lattice Boltzmann (LB) method based on a minimal continuous Fokker-Planck (FP) kinetic model, originally proposed for stochastic diffusive-drift processes (e.g., Brownian dynamics), by adapting it as a collision model for the continuous Boltzmann equation (CBE) for fluid dynamics. The FP collision model has several desirable properties, including its ability to preserve the quadratic nonlinearity of the CBE, unlike that based on the common Bhatnagar-Gross-Krook model. Rather than using an equivalent Langevin equation as a proxy, we construct our approach by directly matching the changes in different discrete central moments independently supported by the lattice under collision to those given by the CBE under the FP-guided collision model. This can be interpreted as a new path for the collision process in terms of the relaxation of the various central moments to “equilibria”, which we term as the Markovian central moment attractors that depend on the products of the adjacent lower order moments and a diffusion coefficient tensor, thereby involving of a chain of attractors; effectively, the latter are nonlinear functions of not only the hydrodynamic variables, but also the non-conserved moments; the relaxation rates are based on scaling the drift coefficient by the order of the moment involved. The construction of the method in terms of the relevant central moments rather than via the drift and diffusion of the distribution functions directly in the velocity space facilitates its numerical implementation and analysis. We show its consistency to the Navier-Stokes equations via a Chapman-Enskog analysis and elucidate the choice of the diffusion coefficient based on the second order moments in accurately representing flows at relatively low viscosities or high Reynolds numbers.  We will demonstrate the accuracy and robustness of our new central moment FP-LB formulation, termed as the FPC-LBM, using the D3Q27 lattice for simulations of a variety of flows, including wall-bounded turbulent flows. We show that the FPC-LBM is more stable than other existing LB schemes based on central moments, while avoiding numerical hyperviscosity effects in flow simulations at relatively very low physical fluid viscosities through a refinement to a model founded on kinetic theory.
\end{abstract}

\let\thefootnote\relax\footnote{*Corresponding author (Email: Kannan.Premnath@ucdenver.edu)}

\newpage
\section{Introduction}

The lattice Boltzmann Method (LBM) is a numerical scheme generally used for simulating fluid flows and various associated physical phenomena~\cite{benzi1992lattice}, and has been shown to deal with complex boundaries and interfacial dynamics, as well as multiscale flows such as turbulence and particulate suspension flows, quite well~\cite{chen1998lattice,aidun2010lattice,lallemand2021lattice}. The key idea in the LBM is that simplified models of the continuous Boltzmann equation~\cite{he1997theory}, are employed to track distributions of fluid particles which are restricted to collide and stream along specific velocity directions designated by lattice links. More generally, the collision process is modeled as to relax the distributions towards an equilibrium state designated by attractors, that are typically defined by the Maxwell distribution. Those particle distributions are then shifted, or streamed, back along the same lattice links. In turn, the hydrodynamic variables are then recovered by taking the relevant statistical moments of those distributions.

Moreover, modeling the collision process in the LBM plays an important role in tuning the physics of the flow characteristics as well as in the numerical stability of the scheme itself. To that end, many different collision models have been developed with the overall goal of improving its ability to represent multi-physics effects or to increase the limits of its applicability in flow simulations that are prone to numerical instabilities. In many practical applications, fluid flows occur with relatively large Reynolds numbers or low viscosities, and thus numerical simulation of those flows requires a robust numerical method capable of dealing with the associated numerical artifacts that arise when using a less robust model.

The first and simplest collision model, known as the single relaxation time (SRT) collision model~\cite{qian1992lattice}, paved the way for treating the collision step as a relaxation process, but it was also found to have sever limitations in terms of numerical stability. More specifically, the SRT model breaks down when applied to problems that involve large Reynolds numbers or very small fluid viscosities. The primary issue being that the distributions are relaxed at the same rate, which causes some of them to incur large numerical errors. As a response to these issues, the multiple relaxation model was developed, which performs the collision step in moment space instead of velocity space, and allows different moments of the distributions to relax at different rates~\cite{d1992generalized,d2002multiple}. The MRT model was found to significantly improve the numerical stability when compared to the SRT collision model. Next, a natural extension of the MRT collision model was developed, which performs the relaxation in a new space known as central moment space that can be described as moment space with a moving reference frame. This new development is commonly known as the central moment collision model or as the cascaded collision model, and the attractors are obtained by matching the discrete central moment equilibria supported by a lattice with the corresponding central moments of the continuous Maxwell distribution~\cite{PhysRevE.73.066705}. This new model again showed great improvements in numerical stability when compared to the MRT collision model.

When simulating flows with relatively low fluid viscosities, some of such LBM collision models involve disparities between the relaxation rates of the second and higher order moments, and as such, some amendments are required to avoid numerical hyper-viscosities. One ad hoc solution in this regard for the central moment family of LB schemes involves constructing the attractors for the higher order central moments by exploiting the factorization property of the continuous Maxwellian and using the products of the post-collision central moments of lower orders to drive the relaxation process, without relying on the Maxwellian by itself~\cite{geier2009factorized}. This scheme is known as the factorized LBM. Furthermore, another solution that deals with the hyper-viscosity artifact, which is much more sophisticated with additional attendant transformations involves relaxing cumulants in favor of central moments under collision~\cite{geier2015cumulant}. In this work, we propose and investigate a new and alternative formulation of the central moment method LB based on the relaxation of central moments towards attractors defined by taking the central moments of the Fokker-Planck model for the Boltzmann collision integral and eliminates these artifacts while achieving highly stable flow simulations through adopting a different perspective and refinements from kinetic theory.

The Fokker-Planck (FP) equation is an important formulation in statistical mechanics originally constructed for describing the evolution of the distribution of particles, whose motion stems from the effects of stochastic forces~\cite{chandrasekhar1943stochastic,wang1945theory,landau1980course,van1992stochastic,gardiner2009stochastic,hornerkamp2002statistical,ebeling2005statistical}, the prototypical example being the Brownian dynamics arising from the stochastic diffusive and drift processes\cite{gillespie2013simple,romanczuk2012active,libchaber2019biology,kubo1986brownian,frey2005brownian}. Thus, the continuous-time FP equation represents the changes in the distribution function due to drift and diffusion in the phase space, and is equivalent to the Langevin equation which is a classical example of the stochastic differential equation containing contributions from deterministic and random terms~\cite{coffey2012langevin}. It is also referred to as the Kolmogorov forward equation in the mathematical literature and can be derived from the master equation, a fundamental approach for describing probabilistically-driven phenomena involving Markovian stochastic processes -- a differential form of the so-called Chapman-Kolmogorov equation, via the Kramers-Moyal expansion. As such, the FP equation has been generalized further and applied to solve a large class of problems~\cite{risken1996fokker,frank2005nonlinear,klimontovich2012statistical}, which includes its recent use in constructing reduced order models for complex systems represented as stochastic cascade processes from empirical data~\cite{peinke2019fokker,friedrich2011approaching}.

\begin{figure}[!ht]
\centering
\includegraphics[trim = 0 0 0 0, clip, width =150mm]{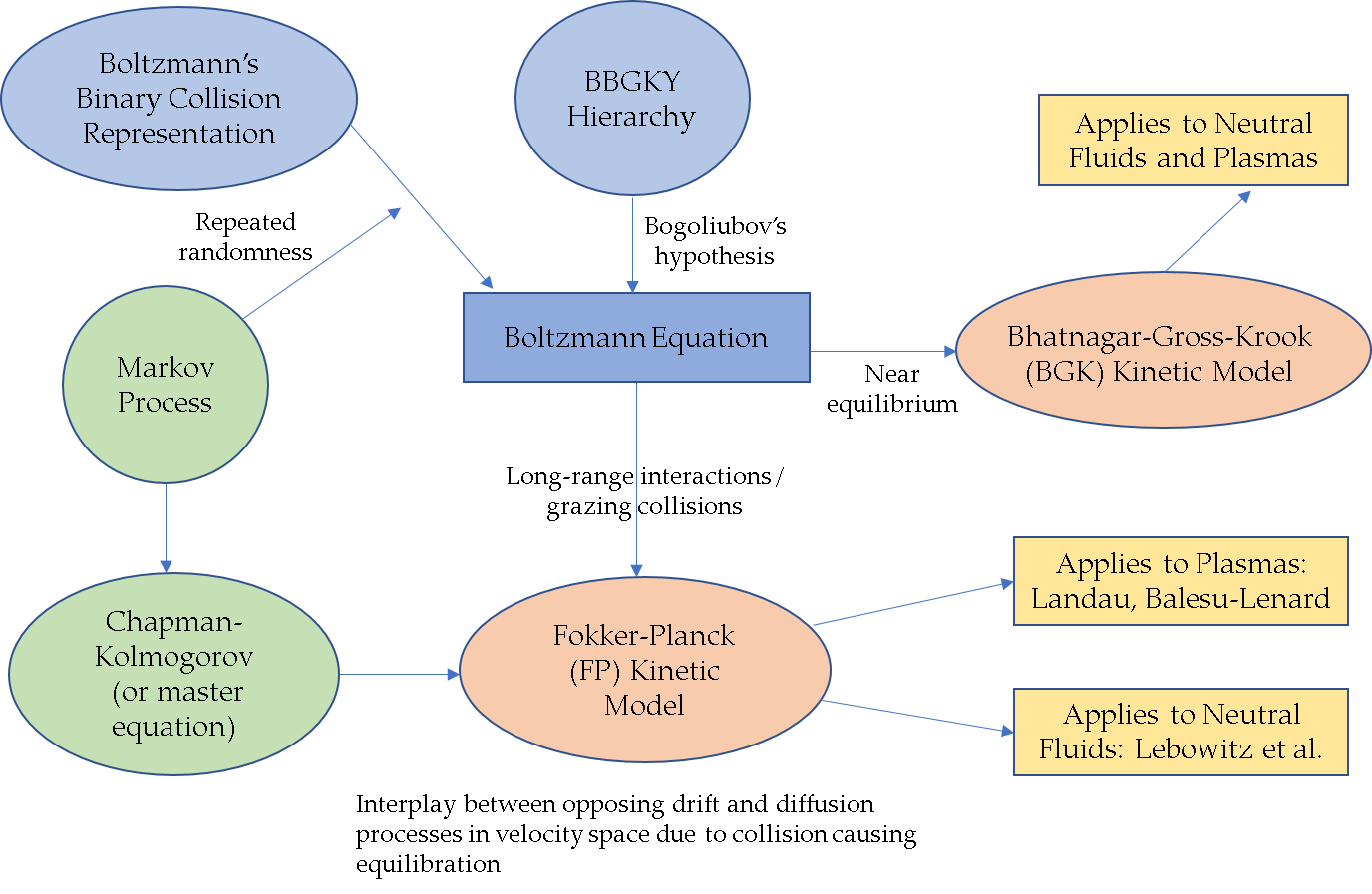}
\caption{Routes for the derivation of the Boltzmann equation and the modeling of its collision term via BGK or FP approach under appropriate approximations, and their applications to representing the dynamics in fluids and plasmas (inspired from~\cite{liboff2003kinetic}).}
\label{fig:routesBE}
\end{figure}

In the context of fluid dynamics, based on stochastic molecular models pioneered by Kirkwood and collaborators~\cite{kirkwood1946statistical,kirkwood1949statistical,zwanzig1953statistical,zwanzig1954statistical}, which are essentially Langevin equations and hence correspond to appropriate FP equations, the equations of fluid motion and their transport coefficients can be derived systematically~\cite{heinz2004molecular}. On the other hand, of specific interest to this work, Lebowitz \emph{et al}~\cite{lebowitz1960nonequilibrium} proposed the FP formulation, viz., the drift-and-diffusion driven changes of a distribution function in the phase space, as a model for the collision term in the continuous Boltzmann equation, where their drift and diffusion coefficients are related to the transport coefficients of the fluid. Moreover, Cercignani~\cite{cercignani1988Boltzmann} demonstrated that such a FP model for collision represents a consistent and rigorous approximation (or a diffusion limit) of the Boltzmann's collision integral term, especially for grazing collisions (see also~\cite{kampen1961power,desloge1963fokker, lewis1969boltzmann,gombosi1994gaskinetic,liboff2003kinetic}). Also, it has been shown by Pawula~\cite{pawula1967approximation} that Boltzmann's collision term leads to a differential operator under a finite truncation via the Kramers-Moyal expansion, involving only the first two terms corresponding to the FP equation. Figure~\ref{fig:routesBE} shows schematically the different routes of how the celebrated Boltzmann equation itself can be derived under the repeated randomness or molecular chaos (stosszahlansatz) assumption~\cite{boltzmannlectures} or via a reduction from the so-called BBGKY hierarchy~\cite{grad1958principles} and the modeling of its collision term via either the popular Bhatnagar-Gross-Krook (BGK) approximation~\cite{bhatnagar1954model} or the FP model under appropriate simplifying considerations; both these models reduces the complexity of using the Boltzmann equation in representing the dynamics of fluids (or plasmas). Clearly, Markov process considerations underlie the derivation of the Boltzmann equation or the development of the FP model, and is thus a crucial aspect in modeling the physical processes during collision.
In addition to the fact that the FP model directly arises as a simplification of the Boltzmann collision term, it has other advantages. The resulting FP collision model preserves the quadratic nonlinearity of the Boltzmann's collision term, unlike the BGK model, which allows only a relaxation to the Maxwell distribution function (see~\cite{cercignani1988Boltzmann}). Moreover, by exploiting the flexibility available in tuning the drift and diffusion coefficients appropriately and with necessary modifications, variety of complex fluid flows with attendant multiphysics effects can be modeled, such as flows involving heat transfer with adjustable Prandtl number, microscale gas flows, polyatomic gas flows, mixtures of gases, etc.

A stochastic numerical scheme for practical implementation, based on using an equivalent Langevin equation in place of
the FP collision model in the Boltzmann equation was constructed by Jenny \emph{et al} in~\cite{jenny2010solution}, which has been extended further and applied to various fluid dynamical applications more recently (see e.g.,~\cite{gorji2011fokker,gorji2013fokker,singh2015fokker,singh2016gaseous,mathiaud2016fokker,mathiaud2017fokker,sadr2017continuous,jiang2018particle,jun2019comparative}). Generally, such algorithms serve as a computationally more efficient alternatives to the direct simulation Monte Carlo method for the solution of the Boltzmann equation for simulation of microscale gas flows at moderate Knudsen numbers. In the LB framework, Moroni \emph{et al}~\cite{moroni2006solving} presented a scheme for the solution of the FP equation whose discretization is effected by the Hermite projections in the velocity space. Such velocity space discretizations, which retain terms up to the third order in the fluid velocities for the changes in the distribution function under collision, was then extended for hydrodynamics in~\cite{moroni2006use} but was found to have stability limitations, which will be addressed through a different approach in this work. More recently, the FP equation has been applied via its proxy, viz., an equivalent Langevin equation and then solved along with an LBM for simulating polymeric fluid flows (see e.g.,~\cite{ammar2010lattice,singh2013lattice,lewis2017lattice}).

\begin{figure}[!ht]
\centering
\includegraphics[trim = 0 0 0 0, clip, width =150mm]{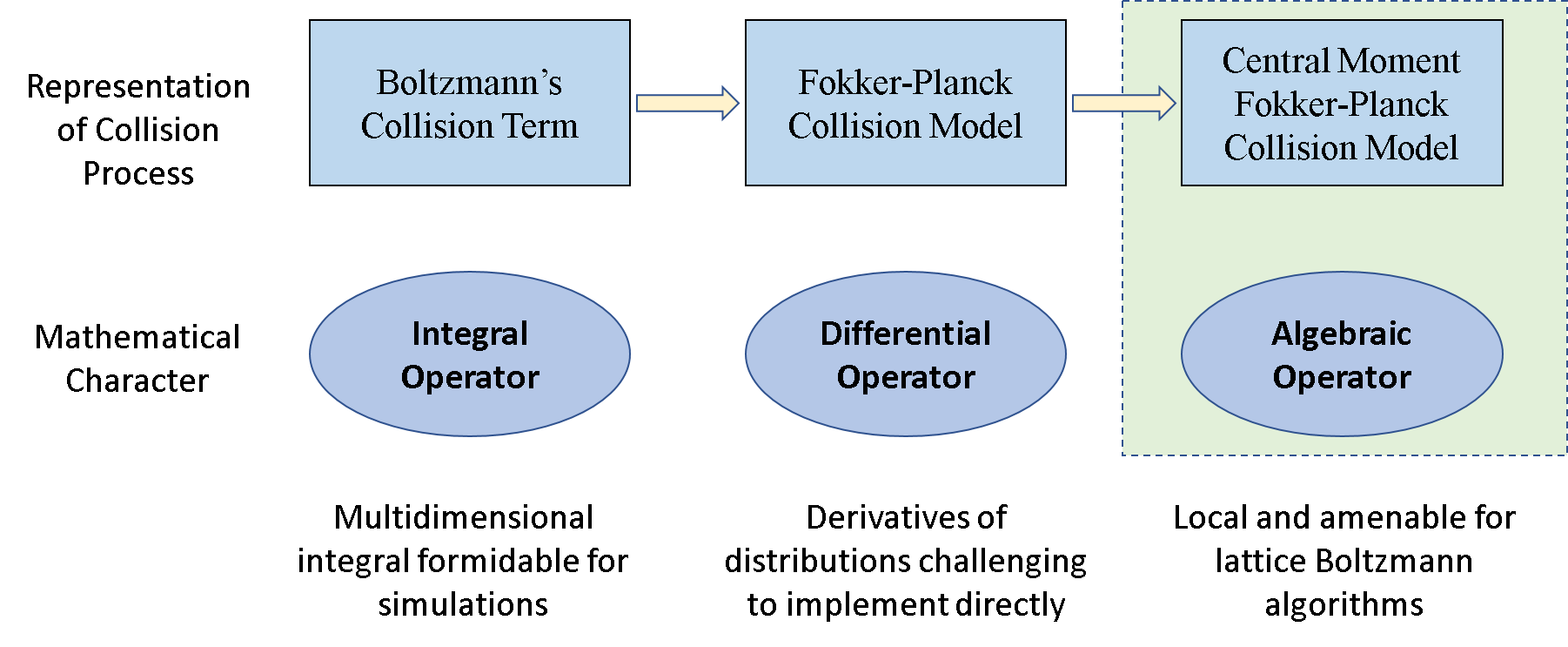}
\caption{Representation of collision processes at different levels of modeling description and the associated mathematical nature of the collision operator.}
\label{fig:MathCharacterFPOperator}
\end{figure}

In this work, we present a new central moment LBM constructed from the simplified formulation of the continuous Boltzmann equation with the FP collision model via a matching principle, i.e., the changes in different discrete central moments independently supported by a chosen lattice of the LB algorithm under collision are equal to those given by the central moments of the continuous FP-based collision model of the Boltzmann equation for effective flow simulations. Thus, we propose to construct a LBM that implements the FP collision model in terms of their resulting system of central moments rather than relying on the solution of a stochastic proxy formulation based on Langevin-type equation, or trying to solve the FP equation directly as given in the velocity space. Such an approach is reminiscent of and has some analogy to formulating and solving the problem of dispersion of a tracer species under the combined action of molecular diffusion and the driving nonuniform velocity field -- the so-called classical Taylor dispersion~\cite{taylor1953dispersion}; in such cases, it is known that solving the distribution of the tracer species directly depends on invoking various approximations and hence restrictive~\cite{taylor1953dispersion}, while recasting the problem in terms of the evolution equations of the various moments of the distribution of the tracer species avoids the need to use restrictive assumptions and hence yields more general solutions as shown by Aris~\cite{aris1956dispersion}. As such, in our case, the use of central moments avoids the need to approximate the cumbersome derivatives of the distribution functions with respect to the particle velocities present in the FP model, which can be represented exactly in terms of the central moments of lower orders, with an underlying local structure which is naturally amenable for highly stable LB implementations. Interestingly, as shown schematically in Fig.~\ref{fig:MathCharacterFPOperator}, from a mathematical perspective, the Boltzmann's collision term is an \emph{integral} operator, while its modeling via a FP kinetic model leads to a \emph{differential} operator, which when replaced with its central moments along with suitable refinements under the standard LB discretization leads a significantly simpler \emph{algebraic} operator that greatly facilitates algorithmic implementation.

Our formulation for the LBM based on FP-guided collision results in a new path for the evolution of different central moments, especially at higher orders, under collision and can be re-interpreted in terms of the relaxation of the various central moments to “equilibria” that depend only on the adjacent, lower order post-collision moments, appearing in the form of recurrence relations. Since the collision process is fundamentally stochastic and Markovian in nature as noted above, we designate such equilibria identified under the equivalent relaxation-type interpretation, which also involves scaling the drift coefficient by the order of the participating moment to define the relaxation rates, as \emph{Markovian central moment attractors}. This is distinct from the path of relaxation of central moments to the central moments of the Maxwellian under collision, since the Markovian central moment attractors depend on products of the lower order moments with components of a diffusion coefficient tensor, thereby allowing for considerations of their non-equilibrium effects. This is especially the case for evolving the fourth and higher order central moments under collision. As we will show in a later section, only in the special case where we assume the lower order moments appearing in the Markovian central moment attractors to be in equilibrium, will the latter coincide with those of the central moments of the Maxwellian. In addition, we will present and implement a method to include body forces in our formulation based directly on the Boltzmann's acceleration term by invoking the matching principle, i.e., by setting the discrete central moment changes due to body forces to be equal to those given by the central moments of the acceleration term of the continuous Boltzmann equation.

We will construct central moment LBMs using FP-guided collision, referred to as the FPC-LBM, in two (2D) and three (3D) dimensions using the standard D2Q9 and D3Q27 lattices, respectively, and present the implementation details of the resulting algorithms. In order to establish their consistency to the Navier-Stokes equations, we will present Chapman-Enskog analysis of the equivalent central moment system of the Boltzmann equation using the FP collision model in 2D and 3D. We will perform a detailed numerical study of our central moment LBM using FP-guided collision to demonstrate its accuracy and stability characteristics for various standard benchmark flow problems in 2D and 3D, and also present comparisons of its performance when compared to those using other collision models.

Finally, we note that for simulation of flows at relatively very low fluid viscosities, there is necessarily large disparities in the relaxation rates of the non-conserved second order and the higher order moments when the LB formulation is implemented using the multiple relaxation times, and if the central moments of the Maxwellian are used as equilibria, the contributions from such higher moments involves terms analogous to the non-equilibrium momentum fluxes (related to the strain rate tensor) and can dominate the corresponding physical contributions from the second-order non-equilibrium moments, which manifest as numerical hyperviscosities. On the other hand, this can be addressed by suitably prescribing the model parameters associated with the FP model, viz., the diffusion coefficient tensor in the Markovian central moment attractor. More specifically, the diffusion tensor parameter associated with the FP model is related to the variance of the distribution function, which corresponds to the different components of the second order central moments. When we consider the latter to be at equilibrium in evolving the conserved hydrodynamic variables via the relaxation of second order moments as required to recover the Navier-Stokes equations, while retaining them to be in general non-equilibrium states for evolving the higher moments, the resulting Markovian central moment attractors naturally become nonlinear functions of lower non-conserved moments, which, as shown later this work, can effectively avoid such numerical hyperviscosities, while also greatly enhancing numerical stability in simulations of flows at higher Reynolds numbers. Furthermore, to demonstrate the capability of our approach for accurate turbulence simulations, we will presents results obtained using the FPC-LBM for turbulent channel flow and compare them with state-of-the-art direct numerical simulations (DNS) data obtained previously using a Navier-Stokes based solver.

In summary, our present work is motivated by the fact that various collision models of the LBM are prone to numerical instabilities or are subject to numerical hyperviscosities when simulating flows at relatively low fluid viscosities or high Reynolds numbers. We address both these issues by constructing FPC-LBM in 2D and 3D based on refining a model from kinetic theory that results in a more robust and accurate formulation for flow simulations in such cases.

This paper is organized as follows. In the next section (Sec.~\ref{sec:FP2D}), we will present the derivation of a central moment formulation of the continuous FP collision operator of the Boltzmann equation in 2D, followed by the construction of the discrete Markovian central moment attractors using the D2Q9 lattice for the 2D FPC-LBM. Section~\ref{sec:FP3D} presents the corresponding 3D continuous central moment FP collision model and the respective discrete attractors using the D3Q27 lattice for the 3D FPC-LBM. Mathematical consistency analysis via the Chapman-Enskog expansions carried out directly using central moments in 2D and 3D are given in Appendices~\ref{sec:CEAnalysis2DFPC-LBM} and~\ref{sec:CEAnalysis3DFPC-LBM}, respectively; complete algorithmic implementation details of the 2D and 3D FPC-LBMs are discussed in Appendices~\ref{sec:2DFPC-LBM} and~\ref{sec:3DFPC-LBM}, respectively. Simulations using the FPC-LBMs for various standard benchmark flow problems that establish their accuracy, improvements in numerical stability when compared to using a variety of other existing collision models, and its ability in avoiding hyperviscosity effects are presented in Sec.~\ref{sec:ResultsandDiscussion}. Finally, Sec.~\ref{sec:summaryconclusions} briefly summarizes our main contributions and the key conclusions drawn from this work.

\section{Construction and Analysis of 2D Central Moment Fokker-Planck Collision Model of Boltzmann Equation and 2D FPC-LBM\label{sec:FP2D}}
We start with the continuous Boltzmann equation given by (see e.g.,~\cite{kremer2010introduction})
\begin{eqnarray}\label{boltzmann_equation}
\frac{\partial f}{\partial t}+ \bm{\xi}\cdot\bm{\nabla}f = \left(\frac{\delta f}{\delta t}\right)_{\!\!\!coll}^{\!\!\!Boltz}+\left( \frac{\delta f}{\delta t} \right)_{\!\!\!forcing},
\end{eqnarray}
where $f=f(\bm{x},\bm{\xi},t)$ is the single-particle density distribution function. In 2D, the position vector and the particle velocity are given by $\bm{x}=(x,y)$ and $\bm{\xi}=(\xi_x,\xi_y)$, respectively, with $\bm{\nabla}=(\partial_x,\partial_y)$. Considering athermal flows for simplicity, the fluid density $\rho$ and bulk velocity $\bm{u}=(u_x,u_y)$ are obtained through the leading zeroth and first moments of $f$ with respect to $\bm{\xi}$ given by
\begin{equation}
\rho = \int fd\bm{\xi}, \qquad \rho\bm{u}= \int f\bm{\xi}d\bm{\xi}.
\end{equation}
Here, the left side of Eq.~(\ref{boltzmann_equation}) accounts for the variation in $f$ due to particle advection or streaming, and if the fluid is subjected to an external force $\bm{F} = (F_x,F_y)$, the particles undergo acceleration whose effect on $f$ is represented through the term $\left(\frac{\delta f}{\delta t}\right)_{\!\!forcing}$ in its right side given by
\begin{equation}\label{eq:boltzmannforcingterm}
\left(\frac{\delta f}{\delta t}\right)_{\!\!forcing} = - \frac{\bm{F}}{\rho}\cdot\bm{\nabla}_\xi f,
\end{equation}
where $\bm{\nabla}_\xi=(\partial_{\xi_x},\partial_{\xi_y})$ is the gradient in the velocity space. The particles interact through collision that redistributes their velocities, which is accounted for on the rate of change in $f$ via the Boltzmann's collision term for binary collisions $\left(\frac{\delta f}{\delta t}\right)_{\!\!coll}$ given by~\cite{kremer2010introduction}
\begin{eqnarray}\label{collision_integral}
\left(\frac{\delta f}{\delta t}\right)_{\!\!\!coll}^{\!\!\!Boltz} =  \int g b(f^{'}f_{1}^{'} - f f_{1} ) \,      \mathrm{d}b \, \mathrm{d}\varepsilon \, \mathrm{d}\xi_i,
\end{eqnarray}
where the distribution functions of particle pairs before and after collision are given by $f \equiv f(\bm{x}, \bm{\xi}, t)$ , $f_{1} \equiv f(\bm{x}, \bm{\xi}_{1}, t)$, and $f' \equiv f(\bm{x}, \bm{\xi}', t)$ , $f_{1}^{'} \equiv f(\bm{x},\bm{\xi}_{1}^{'}, t)$. Here,  $ g = |\bm{\xi} - \bm{\xi}_{1} |$ is the relative speed, $\varepsilon$ is the azimuthal collision angle and b is the impact parameter. Equation (\ref{collision_integral}) is a multidimensional integral operator and represents one of the main difficulties with using the Boltzmann equation. Thus, it is often replaced by simpler models that represent certain essential features of the collision process. Varieties of collision models have been constructed over the years. One such popular model is the BGK model~\cite{bhatnagar1954model} given by
\begin{eqnarray}\label{BGK_collision}
\left(\frac{\delta f}{\delta t} \right)_{\!\!\!coll}^{\!\!\!BGK} = \omega_{BGK} (f^M-f),
\end{eqnarray}
where $f^M$ is the local Maxwell distribution function given by
\begin{eqnarray}\label{maxwell_distribution}
f^M = \frac{\rho}{(2\pi c_s^2)^{\frac{D}{2}}}\exp\left[\frac{-(\bm{\xi}-\bm{u})^2}{2c_s^2}\right],
\end{eqnarray}
and $\omega_{BGK}$ is the collision frequency. In Eq.~(\ref{maxwell_distribution}), $\rho$ is the density, $\bm{u}$ is the fluid velocity, $c_s$ is the speed of sound (generally related to temperature, but for athermal flows held as a constant), and D is the number of spatial dimensions. Equation (\ref{BGK_collision}) represents the effect of collisions as a relaxation toward an equilibrium state defined in Eq.~(\ref{maxwell_distribution}), and can be derived as a drastic approximation of the collision term (Eq.~(\ref{collision_integral})).

Another simplification of the Boltzmann's collision integral (Eq.~(\ref{collision_integral})) is the Fokker-Planck (FP) model proposed in Lebowitz (1960)\cite{lebowitz1960nonequilibrium}. While originally used to represent the drift and diffusion processes of a Brownian particle, it can be shown to a rigorous approximation of the collision term (Eq.~(\ref{collision_integral})) in the Boltzmann equation under the assumptions of the small velocity changes under collision and the particles undergo grazing collision, which is often the case for scattering processes involving inverse power interaction potentials that result in large impact parameters (see Cercignani~\cite{cercignani1988Boltzmann}, Gombosi~\cite{gombosi1994gaskinetic} and Liboff~\cite{liboff2003kinetic}). The Fokker-Planck collision model can be represented as~\cite{lebowitz1960nonequilibrium}
\begin{eqnarray}\label{FP_collision}
\left( \frac{\delta f}{\delta t}\right)_{\!\!\!coll}^{\!\!\!FP} = \omega_{FP} \left[\frac{\partial}{\partial \xi_i}((\xi_i - u_i)f) + D_{ij}' \frac{\partial^2 f}{\partial\xi_i \partial \xi_j} \right],
\end{eqnarray}
where $\omega_{FP}$ is a frequency characterizing collisions and $D'_{ij}$ is the diffusion tensor parameter related to the variance of the distribution function or its second order central moment, which will be discussed in more detail later. The effective diffusion rate coefficient $D_{ij}$ itself is related to the diffusion tensor parameter $D'_{ij}$ via the FP relaxation frequency as $D_{ij}=D'_{ij}\omega_{FP}$. Equation~(\ref{FP_collision}) represents the effect of collision in terms of drift and diffusion in the velocity space. The drift term models a process analogous to the dynamical friction and reflects the idea that collisions tend to gradually eliminate all gradients in the particle velocities of the fluid; the diffusion term represents the effect of diffusion of $f$ in the velocity space resulting in a broadening of the distribution function via molecular chaos~\cite{gombosi1994gaskinetic}. It is interesting to note that unlike the BGK model (Eq.~(\ref{BGK_collision})), the FP model (Eq.~(\ref{FP_collision})) preserves the quadratic nonlinearity of the Boltzmann's collision integral (Eq.~(\ref{collision_integral})) since $u_i$ and $D'_{ij}$ appearing in Eq.~(\ref{FP_collision}) could themselves depend on $f$. We note here that Eq.~(\ref{FP_collision}) is a generalization of the FP model developed in~\cite{lebowitz1960nonequilibrium} by considering a tensor for the diffusion parameter $D'_{ij}$ rather than a scalar $D'$.

We will now formulate a central moment representation of the continuous FP model (Eq.~(\ref{FP_collision})), which will then serve as a basis for constructing a corresponding discrete central moment lattice Boltzmann (LB) method for flow simulations. We start by defining the inner products of any two objects $a$ and $b$ as the integration over velocity space in 2D as
\begin{eqnarray}
\left<a,b\right>= \int\displaylimits_{-\infty}^{\infty} \int\displaylimits_{-\infty}^{\infty} a \; b\; \mathrm{d}\xi_x\; \mathrm{d}\xi_y.\nonumber
\end{eqnarray}
Now, we introduce the weights $W_{mn}$ of order ($m+n$) based on the integral powers of the components of the \emph{peculiar} velocity $\xi_i - u_i$ (i.e., the particle velocity relative to the bulk fluid velocity) as
\begin{eqnarray}\label{cm_weights}
W_{mn}=(\xi_x-u_x)^m(\xi_y-u_y)^n.
\end{eqnarray}
Then, we can define the central moment of the distribution function $f$ of order ($m+n$) as the inner product of $f$ and $W_{mn}$ as
\begin{eqnarray}\label{central_moments}
\Pi_{mn}=\left<f,W_{mn}\right>.\nonumber
\end{eqnarray}
For convenience, we rewrite the FP collision operator given in Eq.~(\ref{FP_collision}) in a shorthand notation as
\begin{eqnarray}\label{FP_short}
\left( \frac{\delta f}{\delta t}\right)_{\!\!\!coll}^{\!\!\!FP} = \omega_{FP}\left(\frac{\delta f}{\delta t}^{FP1}+\frac{\delta f}{\delta t}^{FP2}\right),
\end{eqnarray}
where
\begin{eqnarray}\label{dFP}
\frac{\delta f}{\delta t}^{FP1} = \frac{\partial}{\partial\xi_i}((\xi_i-u_i)f), \quad
\frac{\delta f}{\delta t}^{FP2} = D'_{ij}\frac{\partial^2 f}{\partial\xi_i\partial\xi_j} =  D'_{xx}\frac{\partial^2 f}{\partial \xi_x^2}+D'_{yy}\frac{\partial^2 f}{\partial \xi_y^2} + 2D'_{xy}\frac{\partial^2 f}{\partial \xi_x\partial \xi_y}.
\end{eqnarray}
Here, FP1 represents the `drift' term of the FP collision operator while FP2 corresponds to the `diffusion term'.

We then define the rate of change of the central moment of order ($m+n$) under collision via the FP model, by taking the inner product of each term in the right hand side of Eq.~(\ref{FP_short}) along with the attendant terms in Eq.~(\ref{dFP}), with the central moment weights of order $(m+n)$ given in Eq.~(\ref{cm_weights}), as in
\begin{eqnarray}\label{cmchanges}
\left(\frac{\delta \Pi_{mn}}{\delta t}\right)_{\!\!\!coll}^{\!\!\!FP} = \omega_{FP}\left[\left<\frac{\delta f}{\delta t}^{FP1},W_{mn}\right> + \left<\frac{\delta f}{\delta t}^{FP2},W_{mn}\right>\right]=\omega_{FP}\left[\frac{\delta \Pi_{mn}^{FP1}}{\delta t}+\frac{\delta \Pi_{mn}^{FP2}}{\delta t}\right].
\end{eqnarray}
Next, we take these inner products in Eq.~(\ref{cmchanges}) term by term, starting with the first term
\begin{eqnarray}\label{cmdrift}
\frac{\delta \Pi_{mn}^{FP1}}{\delta t} = \left<\frac{\partial}{\partial \xi_x}((\xi_x-u_x)f),W_{mn}\right> + \left<\frac{\partial}{\partial \xi_y}((\xi_y-u_y)f),W_{mn}\right>.
\end{eqnarray}
To simplify this last term, consider the following chain rule $\frac{\partial}{\partial \xi_i}((\xi_i-u_i)f)= f  + (\xi_i-u_i)\frac{\partial f}{\partial \xi_i}$ and then using this to rewrite Eq.~(\ref{cmdrift}), and absorbing the $(\xi_i - u_i)$ term into the respective central moment weight, $W_{mn}$, as
\begin{eqnarray}\label{cmdrift2}
\frac{\delta \Pi_{mn}^{FP1}}{\delta t} = \left<\frac{\partial f}{\partial \xi_x},W_{m+1,n}\right> + \left<\frac{\partial f}{\partial \xi_y},W_{m,n+1}\right> + 2\left<f,W_{mn}\right>.
\end{eqnarray}
We now further evaluate the right hand side of Eq.~(\ref{cmdrift2}) term by term, starting with the first term and rewriting it as
\begin{eqnarray}\label{cmdrift3}
\left<\frac{\partial f}{\partial \xi_x},W_{m+1,n,p}\right> = \left(\int\displaylimits_{-\infty}^{\infty}(\xi_x-u_x)^{m+1} \frac{\partial f}{\partial \xi_x}\mathrm{d}\xi_x\right)\left(\int\displaylimits_{-\infty}^{\infty}(\xi_y-u_y)^{n}\mathrm{d}\xi_y\right),
\end{eqnarray}
where the first term in the right hand side of  Eq.~(\ref{cmdrift3}) can be evaluated via integration by parts as
\begin{eqnarray}
\int\displaylimits_{-\infty}^{\infty}(\xi_x-u_x)^{m+1} \frac{\partial f}{\partial \xi_x}\mathrm{d}\xi_x= (\xi_x-u_x)^{m+1} f\biggr\rvert_{-\infty}^{\infty}- \int\displaylimits_{-\infty}^{\infty}(m+1)(\xi_x-u_x)^m f \mathrm{d}\xi_x.\nonumber
\end{eqnarray}
Assuming that $f$ decays faster than the increase due to any polynomials of $\xi_x - u_x$ , i.e., $(\xi_x-u_x)^{m+1} f = 0 $ as $ \xi_x \rightarrow \pm \infty$, the first term on the right hand side of Eq.~(\ref{cmdrift2}) becomes
\begin{eqnarray}
\left<\frac{\partial f}{\partial \xi_x},W_{m+1,n}\right> = -(m+1)\left(\int\displaylimits_{-\infty}^{\infty}(\xi_x-u_x)^{m} f\mathrm{d}\xi_x\right)\left(\int\displaylimits_{-\infty}^{\infty}(\xi_y-u_y)^{n}\mathrm{d}\xi_y\right),\nonumber
\end{eqnarray}
which is simplified to
\begin{eqnarray}\label{cmdrift4}
\left<\frac{\partial f}{\partial \xi_x},W_{m+1,n}\right> = -(m+1)\left<f,W_{mn}\right>= -(m+1)\Pi_{mn}.
\end{eqnarray}
Similarly, the second term on the right hand side of Eq.~(\ref{cmdrift2}) is evaluated the same way as
\begin{eqnarray}\label{cmdrift5}
\left<\frac{\partial f}{\partial \xi_y},W_{m,n+1}\right> = -(n+1)\Pi_{mn}.
\end{eqnarray}
Finally, using Eq.~(\ref{cmdrift4}) and Eq.~(\ref{cmdrift5}) back into  Eq.~(\ref{cmdrift2}) and simplifying, we obtain the following expression for the rate of change in the central moment of order $(m+n)$ under collision due to the drift term in the FP collision model as
\begin{eqnarray}\label{eq:cmdrift}
\frac{\delta \Pi_{mn}^{FP1}}{\delta t} = -(m+n)\Pi_{mn}.
\end{eqnarray}

Next, we consider the rate of change of central moment under collision due to the diffusion term in the FP collision model by evaluating the second term on the right hand side of Eq.~(\ref{cmchanges}) and using Eq.~(\ref{dFP}) as
\begin{eqnarray} \label{cmdiff}
\frac{\delta \Pi_{mn}^{FP2}}{\delta t} = \left< D'_{xx}\frac{\partial^2 f}{\partial \xi_x^2},W_{mn}\right> +  \left<D'_{yy}\frac{\partial^2 f}{\partial \xi_y^2},W_{mn}\right> + \left<2D'_{xy}\frac{\partial^2 f}{\partial\xi_x\partial\xi_y} ,W_{mn}    \right> .
\end{eqnarray}
The first term on the right hand side of Eq.~(\ref{cmdiff}) is evaluated as
\begin{eqnarray}\label{cmdiff2}
\left<D'_{xx}\frac{\partial^2 f}{\partial \xi_x^2},W_{mn}\right> = D'_{xx}\left[ \int\displaylimits_{-\infty}^{\infty}(\xi_x-u_x)^m \mathrm{d}\left(\frac{\partial f}{\partial \xi_x}\right)   \right] \left( \int\displaylimits_{-\infty}^{\infty}(\xi_y-u_y)^n \mathrm{d}\xi_y\right)
\end{eqnarray}
where the term inside the square brackets on the right hand side of Eq.~(\ref{cmdiff2}) is readily evaluated via integration by parts as
\begin{eqnarray}
\int\displaylimits_{-\infty}^{\infty}(\xi_x-u_x)^m \mathrm{d}\left(\frac{\partial f}{\partial \xi_x}\right) = (\xi_x-u_x)^m \frac{\partial f}{\partial \xi_x} \biggr\rvert_{-\infty}^{\infty}- \int\displaylimits_{-\infty}^{\infty}m(\xi_x-u_x)^{m-1}\left(\frac{\partial f}{\partial \xi_x}\right)\mathrm{d}\xi_x.\nonumber
\end{eqnarray}
As before, assuming $(\xi_x - u_x)^m \frac{\partial f}{\partial \xi_x} = 0$ as $\xi_x \rightarrow \pm \infty$, it follows that
\begin{eqnarray}
\left<D'_{xx}\frac{\partial^2 f}{\partial \xi_x^2},W_{mn}\right> =  D'_{xx}\left(-m\int\displaylimits_{-\infty}^{\infty}(\xi_x-u_x)^{m-1}\left(\frac{\partial f}{\partial \xi_x}\right)\mathrm{d}\xi_x\right) \left( \int\displaylimits_{-\infty}^{\infty}(\xi_y-u_y)^n \mathrm{d}\xi_y\right),\nonumber
\end{eqnarray}
which then yields
\begin{eqnarray}\label{cmdiffDxxdxxf}
\left<D'_{xx}\frac{\partial^2 f}{\partial \xi_x^2},W_{mn}\right> = -mD'_{xx}\left<\frac{\partial f}{\partial x},W_{m-1,n}\right>.
\end{eqnarray}
Subsequently, invoking the identity in Eq.~(\ref{cmdrift4}), we get
\begin{eqnarray}
\left<\frac{\partial f}{\partial \xi_x},W_{m-1,n}\right> = -(m-1)\left<f,W_{m-2,n}\right>= -(m-1)\Pi_{m-2,n},\nonumber
\end{eqnarray}
and hence substituting this last equation in  Eq.~(\ref{cmdiffDxxdxxf}) results in the following reduced expression for the latter
\begin{eqnarray}
\left<D'_{xx}\frac{\partial^2 f}{\partial \xi_x^2},W_{mn}\right>= m(m-1)D'_{xx}\Pi_{m-2,n}.
\end{eqnarray}
Similarly, the second and third terms in the right hand side of Eq.~(\ref{cmdiff}) follow from the above considerations as
\begin{eqnarray}
\left<D'_{yy}\frac{\partial^2 f}{\partial \xi_y^2},W_{mn}\right>  = n(n-1)D'_{yy}\Pi_{m,n-2},\quad
\left<D'_{xy}\frac{\partial^2 f}{\partial\xi_x \partial\xi_y},W_{mn}\right>  = mnD'_{xy}\Pi_{m-1,n-1}.
\end{eqnarray}
Thus, the total rate of change of the central moment of order $(m+n)$ under collision due to the diffusion of the particle distribution function given in Eq.~(\ref{cmdiff}) follows by combining the last two equations as
\begin{eqnarray}\label{eq:cmdiffusion}
\frac{\delta \Pi_{mn}^{FP2}}{\delta t} = m(m-1)D'_{xx}\Pi_{m-2,n} +n(n-1)D'_{yy}\Pi_{m,n-2} + 2mnD'_{xy}\Pi_{m-1,n-1}
\end{eqnarray}

Finally, combining Eqs.~(\ref{eq:cmdrift}) and (\ref{eq:cmdiffusion}), the \emph{net} rate of change of central moment of order $(m+n)$ due to the drift and diffusion processes based on the Fokker-Planck collision model is given by
\begin{eqnarray}\label{eq:cmnetdriftdiffusion}
\left(\frac{\delta \Pi_{mnp}}{\delta t}\right)_{\!\!\!coll}^{\!\!\!FP} &=& \omega_{FP}\left[-(m+n)\Pi_{mn}+ m(m-1)D'_{xx}\Pi_{m-2,n}\right.\nonumber\\
&&\left.\qquad\qquad\qquad\qquad\quad+n(n-1)D'_{yy}\Pi_{m,n-2}+ 2mnD'_{xy}\Pi_{m-1,n-1}\right].\nonumber
\end{eqnarray}
Since $\omega_{FP}$ is a free parameter it can be rescaled by absorbing the prefactor $(m+n)$ of $\Pi_{mn}$ appearing inside the square bracket of the right side of the last equation and setting it as a renormalized relaxation frequency $\omega_{mn}$ given by $\omega_{mn}=(m+n)\omega_{FP}$. In effect, we can then recast the central moment FP collision model as a \emph{relaxation} of $\Pi_{mn}$ to its ``equilibrium state" $\Pi_{mn}^{Mv}$, which we term as the \emph{Markovian central moment attractor} of order $(m+n)$, at a rate $\omega_{mn}$. Such an attractor can be equivalently obtained from the stationary condition on the net effect of collision due to drift and diffusion given by
\begin{eqnarray}
  \left[\frac{\delta \Pi_{mn}^{FP1}}{\delta t}+\frac{\delta \Pi_{mn}^{FP2}}{\delta t}\right] &=& 0 \quad \mbox{when}\quad \Pi_{mn} = \Pi_{mn}^{Mv}.
\end{eqnarray}
Then, the refined central moment-based FP rate equation under collision can be finally expressed as
\begin{eqnarray}\label{FPcollision}
\left( \frac{\delta}{\delta t} \Pi_{mnp}\right)_{\!\!\!coll}^{\!\!\!FP} = \omega_{mn}\left[ \Pi_{mn}^{Mv}-\Pi_{mn}\right],
\end{eqnarray}
where
\begin{eqnarray}\label{eq:Mvcmattractor2D}
\Pi_{mn}^{Mv} = \frac{m(m-1)}{m+n}D'_{xx}\Pi_{m-2,n} + \frac{n(n-1)}{m+n}D'_{yy}\Pi_{m,n-2}+ \frac{2mn}{m+n}D'_{xy}\Pi_{m-1,n-1}.
\end{eqnarray}
Few important remarks are in order here. Clearly, $\Pi_{mn}^{Mv}$ involves a recurrence relationship to evolve a higher central moment of order $(m+n)$ in terms of products of lower central moments of order $(m+n-2)$ and the diffusion tensor parameter $D'_{ij}$. In other words, the attractor is a nonlinear function of not only the conserved hydrodynamic moments but also that of the various lower order non-conserved (kinetic) central moments since as discussed next, in general, the components of $D'_{ij}$ themselves are related to the second order central moments. Moreover, as seen in Eq.~(\ref{FPcollision}) and also emphasized earlier schematically in Fig.~\ref{fig:MathCharacterFPOperator}, the mathematical character of the central moment formulation of the FP collision operator is \emph{algebraic} and \emph{local} rather than being differential or integral in nature, which lends itself to an effective numerical implementation using the LBM framework as discussed later in this section. Finally, since $\omega_{mn}$ can differ depending on the order of the moment, viz., $(m+n)$, the formulation given in Eqs.~(\ref{FPcollision}) and (\ref{eq:Mvcmattractor2D}) naturally allows the use of multiple relaxation times or rates for relaxing different central moments to their respective Marvokian attractors. Such a generalization is analogous to adapting the original BGK model that uses a single relaxation time to using multiple relaxation times that relax different moments to their equilibria based on the Maxwell distribution as commonly adopted in previous LB approaches. However, the central moment FP collision model as developed here is conceptually different as it arises from modeling the drift-diffusion processes leading to a more general form for its ``equilibria" as noted above.

\subsection{Choice of diffusion tensor parameter}\label{subsec:choiceDij2D}
The selection of the diffusion tensor parameter $D'_{ij}$ is a key aspect in the numerical implementation of the central moment FP collision model. Before discussing it, let's introduce the following convenient notation:
\begin{eqnarray}
D'_{20}=D'_{xx}, \quad D'_{02} = D'_{yy}, \quad D'_{11}=D'_{xy}.
\end{eqnarray}
In general, $D'_{ij}$ is related to the respective components of the variance (or the spread) of the distribution function $f$ or its second order central moments. Now, if we evaluate the second order central moment attractors given in Eq.~(\ref{eq:Mvcmattractor2D}) and use the notation in the last equation, we get $\Pi_{20}^{Mv}=\Pi_{00} D'_{20}$, $\Pi_{02}^{Mv}=\Pi_{00} D'_{02}$, and $\Pi_{11}^{Mv}=\Pi_{00} D'_{11}$, which upon inverting yield $D'_{20}=\Pi_{20}^{Mv}/\Pi_{00}$, $D'_{02}=\Pi_{02}^{Mv}/\Pi_{00}$, and $D'_{11}=\Pi_{11}^{Mv}/\Pi_{00}$. As shown in the Chapman-Enskog analysis of the continuous kinetic equation with the central moment FP collision model as developed above (see Appendix~\ref{sec:CEAnalysis2DFPC-LBM}) that for correctly recovering the conserved momentum fields, the leading components of second order moments $\Pi_{20}^{(0)}=\Pi_{20}^{Mv}$, $\Pi_{02}^{(0)}=\Pi_{02}^{Mv}$ and $\Pi_{11}^{(0)}=\Pi_{11}^{Mv}$ should be isotropic with the diagonal parts related to the pressure field $P=\rho c_s^2$, i.e., $\Pi_{20}^{(0)}=\Pi_{02}^{(0)}=\rho c_s^2$ and $\Pi_{11}^{(0)}=0$. Then, in view of the above, and noting that $\Pi_{00}=\rho$, it follows that for evolving the conserved modes, which in turn require the computation of the rate of change of second order moments, we require the diffusion coefficient tensor to be chosen based on the \emph{equilibrium} second order central moments, i.e., $D'_{20}=\Pi_{20}^{(0)}/\Pi_{00}=c_s^2$, $D'_{02}=\Pi_{02}^{(0)}/\Pi_{00}=c_s^2$, and $D'_{11}=\Pi_{11}^{(0)}/\Pi_{00}=0$.

On the other hand, for the physically important and challenging case involving the simulation of high Reynolds number flows, which are associated with fluids with relatively low viscosities that depend on the relaxation frequencies of the second order moments (see Appendix~\ref{sec:CEAnalysis2DFPC-LBM}), we generally have $1/\omega_{II}\ll 1/\omega_{h}$, where $\omega_{II}$ is taken as the relaxation frequency of the second order moments (i.e., $(m+n)=2$) and $\omega_{h}$ correspond to that of higher order moments (i.e., $(m+n)>2$). In other words, the second order moments are fast modes while the higher order ones are slower to relax. In such situations, when evolving the higher moments, the fast second order moments would have already approached their post-collision states which are not in equilibrium but contain their non-equilibrium contributions as well, which should be accounted for in their associated Markovian attractors that depend on $D'_{ij}$. Hence, in such situations, to relax the higher order moments with $(m+n)>2$ under collisions, we prescribe $D'_{ij}$ to be equal to $\Pi_{ij}/\Pi_{00}$ or $D'_{20}=\Pi_{20}/\rho$, $D'_{02}=\Pi_{02}/\rho$, and $D'_{11}=\Pi_{11}/\rho$. One may interpret such modification to the diffusion tensor parameter for evolving the higher order moments when compared to the second order moments as a form of \emph{renormalization}.

Then, in summary, the diffusion tensor parameter appearing in the Markovian central moment attractor in Eq.~(\ref{eq:Mvcmattractor2D}) for evolving the moment $\Pi_{mn}$ of order $(m+n)$ under collision is selected as follows:
\begin{eqnarray}
&& D'_{20} = c_s^2, \qquad D'_{02} = c_s^2, \qquad D'_{11} = 0 \qquad \mbox{for}\quad (m+n) \le 2,\nonumber\\
&& D'_{20} = \frac{\Pi_{20}}{\rho}, \quad D'_{02} = \frac{\Pi_{02}}{\rho}, \quad D'_{11} = \frac{\Pi_{11}}{\rho} \quad \mbox{for} \quad (m+n) > 2.
\end{eqnarray}

\subsection{Selected continuous Markovian central moment attractors in 2D}
From the above considerations, we can then evaluate the continuous Markovian central moment attractors given in Eq.~(\ref{eq:Mvcmattractor2D}) for the components of moments that are independently supported by the D2Q9 lattice, which are required for the LB algorithm discussed in this section. Then, we get
\begin{eqnarray}\label{eq:MvCM2D}
&&\Pi_{00}^{Mv} = \rho, \quad \Pi_{10}^{Mv} =0, \quad \Pi_{01}^{Mv} = 0, \nonumber\\
&&\Pi_{20}^{Mv} = c_s^2\rho, \quad \Pi_{02}^{Mv} = c_s^2\rho, \nonumber\\
&&\Pi_{21}^{Mv} = 0, \quad \Pi_{12}^{Mv} = 0, \nonumber\\
&&\Pi_{22}^{Mv}=\frac{1}{\rho}\left[\Pi_{20}\Pi_{02}+2\Pi_{11}^2\right].
\end{eqnarray}
Interestingly, the attractor for the fourth order moment is now a nonlinear function of the second order central moments.

\subsection{Central moment of Boltzmann's acceleration term due to body force in 2D}
In addition, where there is a body force, it results in a rate of change contribution to the distribution function through $\left(\frac{\delta f}{\delta t}\right)_{\!\!\!forcing}$ in the continuous Boltzmann equation (Eq.~(\ref{boltzmann_equation})). We
define the central moment of the rate of change of the distribution function due to the body force of order ($m+n$) as
\begin{eqnarray}\label{cm_forces}
\Gamma_{mn}= \left<  \left(\frac{\delta f}{\delta t}\right)_{\!\!\!forcing} , W_{mn} \right>,
\end{eqnarray}
which can be evaluated by using Eq.~(\ref{eq:boltzmannforcingterm}) in Eq.~(\ref{cm_forces}) to get
\begin{eqnarray}
\Gamma_{mn} = - \frac{F_x}{\rho} \left< \frac{\partial f}{\partial \xi_x}, W_{mn} \right>  - \frac{F_y}{\rho} \left< \frac{\partial f}{\partial \xi_y}, W_{mn} \right>.
\end{eqnarray}
Then, from Eqs.~(\ref{cmdrift4}) and~(\ref{cmdrift5}), it follows that $\left< \frac{\partial f}{\partial \xi_x}, W_{mn} \right> = -m\Pi_{m-1,n}$ and $\left< \frac{\partial f}{\partial \xi_y}, W_{mn} \right> = -n\Pi_{m,n-1}$. Hence, the rate of change of the central moment of order ($m+n$) due to the body force is given by
\begin{eqnarray}\label{forcing:2D}
\Gamma_{mn} = m \frac{F_x}{\rho}\Pi_{m-1,n} + n \frac{F_y}{\rho}\Pi_{m,n-1},
\end{eqnarray}
which relates it to $\bm{F}=(F_x,F_y)$ and to the central moment components of $f$ of a lower order (reduced by a degree of 1) and is exact (see~\cite{premnath2012inertial}).

A Chapman-Enskog analysis of the 2D central moment-based continuous Boltzmann equation with the above FP collision model that is directly based on an expansion in terms of the central moments to establish its consistency with the 2D Navier-Stokes equations is given in Appendix~\ref{sec:CEAnalysis2DFPC-LBM}.

\subsection{Construction of 2D FPC-LBM}
We will now discretize the Boltzmann equation (see Eq.~(\ref{boltzmann_equation})) with using the FP guided collision operator given in Eq.~(\ref{FPcollision}) along with the forcing term (see Eq.~(\ref{eq:boltzmannforcingterm})) as follows: first we replace the continuous particle velocity $\bm{\xi}$ with a discrete particle velocity set $\bm{e}_{\alpha}$, with each direction corresponding to a lattice link, such that we have $f_\alpha(\bm{x},t)=f(\bm{x},\bm{e}_{\alpha},t)$, and then integrate the resulting discrete velocity Boltzmann equation along the particle characteristics over a time step $\delta_t$ that exactly spans a distance equal to the magnitude of $\bm{e}_{\alpha}\delta_t$ which then further discretizes both the space and time (see He \emph{et al}~\cite{he1998,he1999}). In 2D, the standard D2Q9 lattice is used whose discrete particle velocities $\bm{e}_{\alpha}$ have the following Cartesian components:
\begin{eqnarray*}
\bigr| \mathbf{e}_x\bigr> = (0,1,0,-1,0,1,-1,-1,1)^{\dag}, \label{ex}\\
\bigr|\mathbf{e}_y\bigr> = (0,0,1,0,-1,1,1,-1,-1)^{\dag}, \label{ey}
\end{eqnarray*}
where, henceforth, $\dag$ denotes the transpose operator and we use the Dirac's notation of $\bigr|\cdot\bigr>$ and $\bigr<\cdot\bigr|$ for denoting the column and row vectors, respectively. Then the resulting LBM can be generically represented as a two-step process involving a \emph{collision step} which is then followed by a \emph{streaming step}:
\begin{subequations}
\begin{eqnarray}
  \tilde{f}_\alpha(\bm{x},t) &=& f_\alpha(\bm{x},t)+\Omega_\alpha(\bm{x},t), \quad \alpha = 0,1,2,\ldots, q \label{eq:LBMsteps-collide}\\
  f_\alpha(\bm{x},t+\delta_t) &=& \tilde{f}_\alpha(\bm{x}-\bm{e}_{\alpha}\delta_t,t),\label{eq:LBMsteps-stream}
\end{eqnarray}
\end{subequations}
where $q=9$ and $\Omega_\alpha(\bm{x},t)$ represents the cumulative changes in the distribution function $f_\alpha$ due to collision and the effect of body force during a time step $\delta_t$. In practice, such changes will be obtained in terms of central moments via the FP-guided collision and source term updates derived for the continuous Boltzmann equation projected to the countable independent moments represented by the D2Q9 lattice, which will then be mapped back to those in terms of the distribution functions. To accomplish this, we first define the \emph{discrete central moments} and \emph{discrete raw moments} of order ($m+n$) of the distribution function $f_{\alpha}$, its corresponding Markovian attractor $f_{\alpha}^{Mv}$, and the source term $S_{\alpha}$ for the body force, respectively, as follows:
\begin{subequations}
\begin{eqnarray}
\left( \begin{array}{c}\Ks{mn} \\[3mm] \Ks{mn}^{Mv} \\[3mm] \Sig_{mn} \end{array} \right)  &=& \sum_{\alpha = 0}^{8} \left( \begin{array}{c}f_{\alpha} \\[3mm] f_{\alpha}^{Mv} \\[3mm] S_{\alpha} \end{array} \right)  ( e_{\alpha x} - u_{x})^m  ( e_{\alpha y} - u_{y})^n, \;\;\mbox{and}\\[0.2in]
\left( \begin{array}{c}\Kp_{mn} \\[3mm] \kappa^{Mv'}_{mn} \\[3mm]  \Sig_{mn}' \end{array} \right)  &=& \sum_{\alpha = 0}^{8} \left( \begin{array}{c}f_{\alpha} \\[3mm] f_{\alpha}^{Mv}\\[3mm]  S_{\alpha} \end{array} \right)  e_{\alpha x}^m  e_{\alpha y}^n .
\end{eqnarray}
\end{subequations}
Then, the countable independent central moments and raw moments for the D2Q9 lattice respectively are given by
\begin{subequations}\label{eqn:30}
\begin{eqnarray}
\mc= ( \Ks{00},\Ks{10}, \Ks{01}, \Ks{20}, \Ks{02}, \Ks{11}, \Ks{21}, \Ks{12}, \Ks{22} ),
\end{eqnarray}
\begin{eqnarray}
\m = ( \Kp_{00}, \Kp_{10},\Kp_{01}, \Kp_{20}, \Kp_{02}, \Kp_{11},\Kp_{21}, \Kp_{12},\Kp_{22} ),
\end{eqnarray}
\end{subequations}
and, similarly, those for the attractors and sources can be listed. Let us also define a 9-dimensional vector $\mathbf{f}$ containing the distribution functions via
\begin{eqnarray}
\mathbf{f} = (f_0 , f_1, f_2, \dots , f_8)^{\dag},
\end{eqnarray}
and one can similarly group the attractors $f_{\alpha}^{Mv}$ and sources $S_{\alpha}$ in respective vectors.

To perform the collision step in terms of relaxations of different central moments to their respective attractors along with source term updates, we need to first successively map the distribution functions $\mathbf{f}$ to raw moments $\m$ and then to central moments $\mc$; then the post-collision central moments need to be transformed back to raw moments and then to the distribution functions to complete the collision step. To achieve these, we introduce the mappings between the raw moments and the distribution functions as
\begin{equation}\label{eq:map-df-rm}
\m = \PP\mathbf{f},  \; \; \; \; \; \; \; \; \;  \; \; \; \; \; \; \; \; \; \mathbf{f} = \PP^{-1}\m,
\end{equation}
where $\PP$ is a matrix dependent on the basis vectors for moments, i.e., via the components of the monomials $\left< \bm{e}_x^m  \bm{e}_y^n \right|$ given by
\begin{equation}\label{eq:dfrm}
\tensr{P} =\bigr[ \; \bigr|\bm{1}\bigr>,    \;\;     \bigr|\bm{e}_x \bigr>, \;\;\bigr|\bm{e}_y\bigr>, \;\; \bigr|\bm{e}_x^2\bigr>, \;\; \bigr|\bm{e}_y^2\bigr>,  \;\; \bigr|\bm{e}_x \bm{e}_y\bigr>, \;\; \bigr|\bm{e}_x^2 \bm{e}_y\bigr>, \;\; \bigr|\bm{e}_x \bm{e}_y^2\bigr>, \;\;  \bigr|\bm{e}_x^2 \bm{e}_y^2\bigr> \;     \bigr]^\dag,
\end{equation}
where $ \left|\mathbf{1}\right> $ is a 9-dimensional unit vector given by $\bigr|\bm{1}\bigr> =\left| \; \bm{e}_x^0 \bm{e}_y^0\; \right> = (1,1,1,1,1,1,1,1,1)^{\dag}$. In addition, we represent the transformations between central moments and raw moments as
\begin{equation}\label{eq:rmcm}
\mc = \F \m,  \; \; \; \; \; \; \; \; \; \; \; \; \; \; \; \; \; \; \m = \Fi \mc,
\end{equation}
where both the frame transformation matrix $\F$ and its inverse $\F^{-1}$ are lower triangular matrices dependent on the fluid velocity components $\bm{u} = (u_x,u_y)$ and can be readily evaluated through collecting the coefficients of the binomial expansions and their inversions (see e.g.,~\cite{yahia2021central}).

Before we describe our 2D FPC-LBM, we first obtain the expressions for the discrete central moments of the Markovian attractor $\Ks{mn}^{Mv}$ and the source term $\Sig_{mn}$ due to force field by matching the corresponding continuous versions as given in Eq.~(\ref{eq:Mvcmattractor2D}) (or Eq.~(\ref{eq:MvCM2D})) and Eq.~(\ref{forcing:2D}), respectively, i.e., $\Ks{mn}^{Mv}=\Pi_{mn}^{Mv}$ and
$\Sig_{mn}=\Gamma_{mn}$. Thus, the discrete central moments of the Markovian attractors for the D2Q9 lattice read as
\begin{gather}\label{D2Q9attractors}
\Ks{00}^{Mv} = \rho, \; \; \;  \; \; \; \Ks{10}^{Mv}=0, \; \; \;  \; \; \; \Ks{01}^{Mv} = 0, \\[3mm]
\Ks{20}^{Mv} = c_s^2 \rho, \; \; \; \; \; \;  \Ks{02}^{Mv} =c_s^2 \rho, \; \; \;  \; \; \;  \Ks{11}^{Mv} = 0, \nonumber\\[3mm]
\Ks{21}^{Mv} = 0, \; \; \;  \; \; \;   \Ks{12}^{Mv} = 0, \nonumber \\[3mm]
\boxed{\Ks{22}^{ Mv} = \frac{1}{\rho}\left(\tilde{\K}_{20}\tilde{\K}_{02} +2 \tilde{\K}_{11}\tilde{\K}_{11}\right) }.\nonumber
\end{gather}
It can be seen that the fourth order attractor depends on the sum of the products of the current states of the second order central moments. In practical implementations, we will consider the post-collision state of the central moments $\Ks{20}$, $\Ks{02}$, and  $\Ks{11}$  (designated by $\tilde{\K}_{20}$, $\tilde{\K}_{02}$, and  $\tilde{\K}_{11}$) in evaluating the fourth order Markovian attractor. Moreover, the contributions of the body force to the various central moments can be listed as follows:
\begin{gather}\label{eq:sourceCM2D}
\Sig_{00} = 0, \; \; \;  \; \; \; \Sig_{10} = F_x,\; \; \; \; \; \;\Sig_{01} = F_y,\\[3mm]
\Sig_{20} = 2\frac{F_x}{\rho} \Ks{10} = 0, \; \; \; \; \; \; \Sig_{02} =2\frac{F_y}{\rho} \Ks{01} = 0, \; \; \; \; \; \;\Sig_{11} = \frac{F_x}{\rho} \Ks{10}+\frac{F_y}{\rho} \Ks{01} = 0, \nonumber\\[3mm]
\Sig_{21} = 2\frac{F_x}{\rho} \Ks{11}+\frac{F_y}{\rho} \Ks{20}, \; \; \; \; \; \; \Sig_{12} = \frac{F_x}{\rho} \Ks{02}+\frac{F_y}{\rho} \Ks{11},\nonumber \\[3mm]
\Sig_{22} = 2 \left( \frac{F_x}{\rho} \Ks{12} + \frac{F_y}{\rho} \Ks{21} \right). \nonumber
\end{gather}

The complete algorithmic details of the implementation of the 2D FPC-LBM using the D2Q9 lattice are given in Appendix~\ref{sec:2DFPC-LBM}. The solution method can be briefly summarized as follows. During every time step, the discrete distribution functions are first mapped to a set of raw moments and then to central moments. Then, the collision step is performed by relaxing various central moments to their equilibria based on the Markovian attractors of the Fokker-Planck collision model discussed earlier. Once the post-collision central moments are computed, they are transformed back to corresponding raw moments and then to distribution functions. Subsequently, the streaming step is performed by lock-step advection of the post-collision distribution functions along the respective lattice directions. Finally, using these updated distribution functions, the hydrodynamic fields, including the local density and velocity fields are obtained by via taking their kinetic moments, which completes one time step of the FPC-LBM.

\section{Construction and Analysis of 3D Central Moment Fokker-Planck Collision Model of Boltzmann Equation and 3D FPC-LBM\label{sec:FP3D}}
We will now extend our formulation discussed in the previous section to 3D. Here, we consider the Boltzmann equation, its acceleration term, and the FP collision operator given in Eqs.~(\ref{boltzmann_equation}), (\ref{eq:boltzmannforcingterm}), and (\ref{FP_collision}), respectively, by taking the phase space coordinates, their derivatives, and the body force as $\bm{x}=(x,y,z)$, $\bm{\xi}=(\xi_x,\xi_y,\xi_z)$, $\bm{\nabla}=(\partial_x,\partial_y,\partial_z)$, $\bm{\nabla}_\xi=(\partial_{\xi_x}, \partial_{\xi_y}, \partial_{\xi_z})$, and $\bm{F}=(F_x,F_y,F_z)$. Then, to obtain the rate of change of central moments of $f$ due to collision guided by the Fokker-Planck model in 3D, we follow the same procedure as before, by taking the inner product of each term in the Fokker-Planck collision model with the central moment weights. Here, the inner product of two objects, $a$ and $b$, is now defined in 3D as
\begin{eqnarray}
\left<a,b\right>= \int\displaylimits_{-\infty}^{\infty} \int\displaylimits_{-\infty}^{\infty} \int\displaylimits_{-\infty}^{\infty} a \; b\; \mathrm{d}\xi_x\; \mathrm{d}\xi_y\; \mathrm{d}\xi_z
\end{eqnarray}
and the corresponding central moment weights of order ($m+n+p$) as
\begin{eqnarray}\label{cm_weights3D}
W_{mnp}=(\xi_x-u_x)^m(\xi_y-u_y)^n(\xi_z-u_z)^p.
\end{eqnarray}
Then, the continuous central moment of order ($m+n+p$) is written as
\begin{eqnarray}
\Pi_{mnp}=\left<f,W_{mnp}\right>.
\end{eqnarray}

As before, we rewrite the FP collision operator given in Eq.~(\ref{FP_collision}) conveniently in a shorthand notation as shown in Eq.~(\ref{FP_short}), where $\frac{\delta f}{\delta t}^{FP1}$ is given in Eq.~(\ref{dFP}) and $\frac{\delta f}{\delta t}^{FP2}$ now modifies in 3D to
\begin{eqnarray}\label{dFP2for3D}
\frac{\delta f}{\delta t}^{FP2} = D'_{xx}\frac{\partial^2 f}{\partial \xi_x^2}+D'_{yy}\frac{\partial^2 f}{\partial \xi_y^2}+D'_{zz}\frac{\partial^2 f}{\partial \xi_z^2} + 2D'_{xy}\frac{\partial^2 f}{\partial \xi_x\partial \xi_y} + 2D'_{xz}\frac{\partial^2 f}{\partial \xi_x\partial \xi_z}+ 2D'_{yz}\frac{\partial^2 f}{\partial \xi_y\partial \xi_z},
\end{eqnarray}
which expresses the total rate of change of $f$ under collision due to its diffusion in different possible directions in the velocity space, and whose magnitude is determined by diffusion tensor parameter $D'_{ij}$. For ease of presentation, we rewrite the latter by following a slightly more general and compact notation:
\begin{eqnarray}
D'_{200}=D'_{xx}, \quad D'_{020}=D'_{yy}, \quad D'_{002}=D'_{zz}, \quad D'_{110}=D'_{xy}, \quad D'_{011}=D'_{yz}, \quad D'_{101}=D'_{xz}.
\end{eqnarray}
Then, as in Eq.~(\ref{cmchanges}), we define the rate of change of the central moment of order ($m+n+p$) under collision via the FP model, by taking the inner product of each term in the right hand side of Eq.~(\ref{FP_short}) using the attendant terms $\frac{\delta f}{\delta t}^{FP1}$ in Eq.~(\ref{dFP}) and $\frac{\delta f}{\delta t}^{FP2}$ in Eq.~(\ref{dFP2for3D}) with the central moment weights of order $(m+n+p)$ given in Eq.~(\ref{cm_weights3D}) as
\begin{eqnarray}\label{cmchanges3D}
\left(\frac{\delta \Pi_{mnp}}{\delta t}\right)_{\!\!\!coll}^{\!\!\!FP} = \omega_{FP}\left[\left<\frac{\delta f}{\delta t}^{FP1},W_{mnp}\right> + \left<\frac{\delta f}{\delta t}^{FP2},W_{mnp}\right>\right]=\omega_{FP}\left[\frac{\delta \Pi_{mnp}^{FP1}}{\delta t}+\frac{\delta \Pi_{mnp}^{FP2}}{\delta t}\right].
\end{eqnarray}
Now, following the derivation given in the previous section, the central moments of the derivatives of the distribution function in different directions in the velocity space can be expressed as follows:
\begin{eqnarray}\label{eq:CMfgradients3D}
\left<\frac{\partial f}{\partial \xi_x}, W_{mnp}\right> = - m\Pi_{m-1,n,p}, \left<\frac{\partial f}{\partial \xi_y}, W_{mnp}\right> = - n\Pi_{m,n-1,p}, \left<\frac{\partial f}{\partial \xi_z}, W_{mnp}\right> = - p\Pi_{m,n,p-1}.
\end{eqnarray}
Using these successively in pairs, we get the central moments of the cross-diffusion terms in the velocity space as
\begin{eqnarray}
\left<\frac{\partial^2 f}{\partial \xi_x \xi_y}, W_{mnp}\right> &=& m n \Pi_{m-1,n-1,p}, \quad \left<\frac{\partial^2 f}{\partial \xi_y \xi_z}, W_{mnp}\right> = n p \Pi_{m,n-1,p-1}, \nonumber \\
\left<\frac{\partial^2 f}{\partial \xi_x \xi_z}, W_{mnp}\right> &=& m p \Pi_{m-1,n,p-1}.
\end{eqnarray}
Also, the central moments of the diffusion terms in the principal directions in the velocity space read as
\begin{eqnarray}
\left<\frac{\partial^2 f}{\partial \xi_x^2}, W_{mnp}\right> &=& m (m-1) \Pi_{m-2,n,p}, \quad \left<\frac{\partial^2 f}{\partial \xi_y^2}, W_{mnp}\right> = n (n-1) \Pi_{m,n-2,p}, \nonumber \\
\left<\frac{\partial^2 f}{\partial \xi_z^2}, W_{mnp}\right> &=& p (p-1) \Pi_{m,n,p-2},
\end{eqnarray}
Using the above identities, the drift and diffusion contributions to the rate of change of central moment of order ($m+n+p$) under collision can be written as follows:
\begin{eqnarray}\label{cmdrift3D}
\frac{\delta \Pi_{mnp}^{FP1}}{\delta t} &=& \left<\frac{\partial f}{\partial \xi_x},W_{m+1,n,p}\right> + \left<\frac{\partial f}{\partial \xi_y},W_{m,n+1,p}\right> + \left<\frac{\partial f}{\partial \xi_z},W_{m,n,p+1}\right> + 3\left<f,W_{mnp}\right>\nonumber\\
&=& -(m+n+p) \Pi_{mnp}.
\end{eqnarray}
and
\begin{eqnarray}
\frac{\delta \Pi_{mnp}^{FP2}}{\delta t} &=& m(m-1)D'_{200}\Pi_{m-2,n,p} + n(n-1)D'_{020}\Pi_{m,n-2,p} + p(p-1)D'_{002}\Pi_{m,n,p-2} + \nonumber\\
&& 2mnD'_{110}\Pi_{m-1,n-1,p} + 2npD'_{011}\Pi_{m,n-1,p-1} + 2mpD'_{101}\Pi_{m-1,n,p-1}.
\end{eqnarray}
As such, from these last two equations, it can be seen that the drift represents a depletion process, while the diffusion is a gain process under collision. Combining them and since the characteristic Fokker-Planck inverse time scale $\omega_{FP}$ is a free parameter, we can rescale it using $(m+n+p)$
as $\omega_{mnp}=(m+n+p)\omega_{FP}$, such that it can then be used to effectively control the rate of relaxation of the central moment $\Pi_{mnp}$ to its attractor designated as $\Pi_{mnp}^{Mv}$. Then, the central moment FP collision operator can be compactly rewritten in 3D as
\begin{equation}\label{FPcoll3d}
\left(\frac{\delta}{\delta t} \Pi_{mnp}\right)^{\!\!\! FP}_{\!\!\! coll} = \omega_{mnp} \left[ \Pi_{mnp}^{Mv}-\Pi_{mnp} \right],
\end{equation}
where the attendant Markovian central moment attractor reads as
\begin{eqnarray}\label{eq:Mvcmattractor3D}
\Pi_{mnp}^{ Mv} &=&  D'_{200}\frac{m(m-1)}{(m+n+p)}\Pi_{m-2,n,p} + D'_{020}\frac{n(n-1)}{(m+n+p)}\Pi_{m,n-2,p} +   \\
                    && D'_{002}\frac{p(p-1)}{(m+n+p)}\Pi_{m,n,p-2} + 2D'_{110}\frac{mn}{(m+n+p)}\Pi_{m-1,n-1,p} +\nonumber \\
                    && 2D'_{011}\frac{np}{(m+n+p)}\Pi_{m,n-1,p-1} + 2D'_{101}\frac{mp}{(m+n+p)}\Pi_{m-1,n,p-1}. \nonumber
\end{eqnarray}
We emphasize that all the key remarks made regarding the Markovian central moment attractor for the 2D case in the paragraph below Eq.~(\ref{eq:Mvcmattractor2D}) is applicable for the above Eq.~(\ref{eq:Mvcmattractor3D}) as well by replacing ($m+n$) with ($m+n+p$) along with Eq.~(\ref{FPcoll3d}) by replacing $\omega_{mn}$ with $\omega_{mnp}$.

\subsection{Choice of diffusion tensor parameter}
The choice of the diffusion tensor parameter $D'_{ij}$, which may also be referred to as the second order Kramers-Moyal expansion coefficients~\cite{pawula1967approximation}, plays a crucial role in the performance of the numerical simulations that use the central moment FP collision formulation. A detailed discussion of the considerations involved in their selection in the 2D case is given earlier in Sec.~\ref{subsec:choiceDij2D}, which also readily extends for the present 3D model given above. Following this, for evolving moments of second order (i.e., $(m+n+p) = 2$) that is used to update the hydrodynamic fields, we require $D'_{200}=\Pi_{200}^{(0)}/\Pi_{000}^{(0)}$, $D'_{020}=\Pi_{020}^{(0)}/\Pi_{000}^{(0)}$, $D'_{002}=\Pi_{002}^{(0)}/\Pi_{000}^{(0)}$, $D'_{110}=\Pi_{110}^{(0)}/\Pi_{000}^{(0)}$, $D'_{101}=\Pi_{101}^{(0)}/\Pi_{000}^{(0)}$, and $D'_{011}=\Pi_{011}^{(0)}/\Pi_{000}^{(0)}$, where $\Pi_{mnp}^{(0)}=\Pi_{mnp}^{Mv}=c_s^2\rho$ for $(mnp)=(200), (020)$ and $(002)$ and  $\Pi_{mnp}^{(0)}=\Pi_{mnp}^{Mv}=0$ for $(mnp)=(110), (101)$ and $(011)$; on the other hand, for evolving higher moments (i.e., $(m+n+p) > 2$), we prescribe $D'_{200}=\Pi_{200}/\Pi_{000}$, $D'_{020}=\Pi_{020}/\Pi_{000}$, $D'_{002}=\Pi_{002}/\Pi_{000}$, $D'_{110}=\Pi_{110}/\Pi_{000}$, $D'_{101}=\Pi_{101}/\Pi_{000}$, and $D'_{011}=\Pi_{011}/\Pi_{000}$. Then, using $\Pi_{000}=\Pi_{000}^{(0)}=\rho$, in summary, the diffusion tensor parameter appearing in the Markovian central moment attractor in Eq.~(\ref{eq:Mvcmattractor3D}) for evolving the moment $\Pi_{mnp}$ of order $(m+n+p)$ under collision is selected as follows:
\begin{eqnarray}\label{eq:diffusion-tensor-parameter-3D}
&& D'_{200} = D'_{020} = D'_{002} = c_s^2, \quad \quad\quad\quad\quad\quad\quad\;\;\mbox{for}\quad (m+n+p) \le 2, \nonumber\\
&& D'_{110} =  D'_{101} =  D'_{011} = 0,  \nonumber \\
&& D'_{200} = \frac{\Pi_{200}}{\rho}, D'_{020} = \frac{\Pi_{020}}{\rho}, D'_{002} = \frac{\Pi_{002}}{\rho}, \quad \mbox{for}\quad (m+n+p) > 2, \\
&& D'_{110} = \frac{\Pi_{110}}{\rho}, D'_{101} = \frac{\Pi_{101}}{\rho}, D'_{011} = \frac{\Pi_{011}}{\rho}. \nonumber
\end{eqnarray}

\subsection{Selected continuous Markovian central moment attractors in 3D}
Based on the above selection of the diffusion tensor parameter, in anticipation of constructing the FPC-LBM in 3D later in this section we now evaluate the continuous Markovian central moment attractor given in Eq.~(\ref{eq:Mvcmattractor3D}) for the components independently supported by the D3Q27 lattice. Up to the third moments, which directly appear in the hydrodynamics in the Chapman-Enskog analysis, they read as
\begin{eqnarray}
&&\Pi_{000}^{Mv} = \rho, \quad \Pi_{100}^{Mv} = \Pi_{010}^{Mv} = \Pi_{001}^{Mv} = 0,
\end{eqnarray}
\begin{eqnarray}
&&\Pi_{110}^{Mv} = \Pi_{101}^{Mv} = \Pi_{011}^{Mv} = 0, \quad \Pi_{200}^{Mv} = \Pi_{020}^{Mv} = \Pi_{002}^{Mv} = c_s^2 \rho,\nonumber
\end{eqnarray}
\begin{eqnarray}
&&\Pi_{120}^{Mv} = \Pi_{102}^{Mv} = \Pi_{210}^{Mv} = \Pi_{012}^{Mv} = \Pi_{201}^{Mv} = \Pi_{021}^{Mv} = \Pi_{111}^{Mv} = 0.\nonumber
\end{eqnarray}
In addition, the higher order continuous central moment attractors from the fourth to the sixth order, which complete all the 26 independent moments that can be evolved when using the D3Q27 lattice, are given by
\begin{eqnarray}
\Pi_{220}^{Mv}=\frac{1}{\rho}\left[\Pi_{200}\Pi_{020}+2\Pi_{110}^2\right], \qquad
\Pi_{202}^{Mv}=\frac{1}{\rho}\left[\Pi_{200}\Pi_{002}+2\Pi_{101}^2\right], \nonumber
\end{eqnarray}
\vspace{-3mm}
\begin{eqnarray}
\Pi_{022}^{Mv}=\frac{1}{\rho}\left[\Pi_{020}\Pi_{002}+2\Pi_{011}^2\right], \qquad
\Pi_{211}^{Mv}=\frac{1}{\rho}\left[\Pi_{200}\Pi_{011}+2\Pi_{110}\Pi_{101}\right], \nonumber
\end{eqnarray}
\vspace{-3mm}
\begin{eqnarray}
\Pi_{121}^{Mv}=\frac{1}{\rho}\left[\Pi_{020}\Pi_{101}+2\Pi_{110}\Pi_{011}\right], \qquad
\Pi_{112}^{Mv}=\frac{1}{\rho}\left[\Pi_{002}\Pi_{110}+2\Pi_{101}\Pi_{011}\right], \nonumber
\end{eqnarray}
\vspace{-3mm}
\begin{eqnarray}
\Pi_{122}^{ Mv}=\frac{2}{5\rho}\left[\Pi_{020}\Pi_{102}+\Pi_{002}\Pi_{120}+4\Pi_{011}\Pi_{111}+2(\Pi_{101}\Pi_{021}+\Pi_{110}\Pi_{012})\right], \nonumber
\end{eqnarray}
\vspace{-3mm}
\begin{eqnarray}
\Pi_{212}^{ Mv} = \frac{2}{5\rho}(\Pi_{200}\Pi_{012}+\Pi_{002}\Pi_{210} + 4\Pi_{101}\Pi_{111}+2(\Pi_{110}\Pi_{102}+\Pi_{011}\Pi_{201})), \nonumber
\end{eqnarray}
\vspace{-3mm}
\begin{eqnarray}
\Pi_{221}^{ Mv} = \frac{2}{5\rho}(\Pi_{200}\Pi_{021}+\Pi_{020}\Pi_{201} + 4\Pi_{110}\Pi_{111}+2(\Pi_{011}\Pi_{210}+\Pi_{101}\Pi_{120})), \nonumber
\end{eqnarray}
\begin{eqnarray}
\Pi_{222}^{Mv}\! \! \!&=& \! \! \! \frac{1}{3\rho}[\Pi_{200}\Pi_{022}+\Pi_{020}\Pi_{202}+\Pi_{002}\Pi_{220}\nonumber \\
 & & \qquad \qquad+ 4(\Pi_{110}\Pi_{112}+\Pi_{101}\Pi_{121}+\Pi_{011}\Pi_{211})].\label{eq:continouous-Markovian-CM-3D}
\end{eqnarray}
It can be seen from the above set of equations that the Markovian central moment attractors of 4th and higher orders involve various combinations of the quadratic products of the lower order central moments, analogous, but differing in detail, to those appearing in the transformations involved between central moments and cumulants before performing collision using the latter in the cumulant LBM~\cite{geier2015cumulant}. On the other hand, the factorized LBM~\cite{geier2009factorized} prescribes products of lower moments for higher moment attractors in a rather \emph{ad hoc} manner with substantially fewer terms as the order of the moment increases. Thus, in terms of complexity, we may regard the central moment FP collision model to be intermediate between the factorized and cumulant formulations of the LBM while derivable from refining a collision model of the Boltzmann equation that also has various other applications in statistical mechanics and beyond as noted in the introduction.

Another interesting point to note here is that the representation of the \emph{collision step} as a set of relaxations to quadratically nonlinear combinations of lower order central moments implies that it \emph{cannot be written in a matrix form}, just like the cumulant LBM, even if the mappings needed prior to and following collision be expressed in that way; by contrast, in other existing LB models, typically a matrix formulation is contrived to represent the effect of collisions, which effectively implies considering relaxations to central moment equilibria that are linear functions of non-conserved lower order moments. As such, there is no a priori physical reason to express the effect of collisions in a matrix form noting that the Boltzmann's collision integral term itself is \emph{quadratically nonlinear} in terms of the distribution functions and the present central moment FP formulation broadly reflects such nonlinearity in the attractors.

\subsection{Central moment of Boltzmann's acceleration term due to body force in 3D}
To account for the effect of any external body force $\bm{F}=(F_x,F_y,F_z)$, as in the 2D case discussed earlier, we first introduce the central moment of the rate of change of the distribution function due to the body force of order ($m+n+p$) as
\begin{eqnarray}\label{cm_forces3D}
\Gamma_{mnp}= \left<  \left(\frac{\delta f}{\delta t}\right)_{\!\!\!forcing} , W_{mnp} \right>.
\end{eqnarray}

Evaluating this last equation using Eq.~(\ref{eq:boltzmannforcingterm}) along with the identities for the central moments of the derivatives of $f$ in the velocity space given in Eq.~(\ref{eq:CMfgradients3D}), we obtain the rate of change of the central moment of order ($m+n+p$) due to the body force given by
\begin{eqnarray}\label{forcing:3D}
\Gamma_{mnp} = m \frac{F_x}{\rho}\Pi_{m-1,n,p} + n \frac{F_y}{\rho}\Pi_{m,n-1,p}+p \frac{F_z}{\rho}\Pi_{m,n,p-1},
\end{eqnarray}
which generalizes the expression derived earlier for the 2D case to 3D~\cite{premnath2012inertial}.

For completeness, consistency of the above central moment-based FP collision model to the 3D Navier-Stokes equations is shown using a Chapman-Enskog analysis in Appendix~\ref{sec:CEAnalysis3DFPC-LBM}.

\subsection{Construction of 3D FPC-LBM}
We will now construct a FPC-LBM in 3D on the D3Q27 lattice, with particle velocities $\mathbf{e_{\alpha}}$, where $\alpha = 1,2,\cdots,26$, whose cartesian components are given by
\begin{eqnarray*}
\bigr|\mathbf{e}_x\bigr>  &=& (0,1,-1,0,0,0,0,1,-1,1,-1,1,-1,1,-1,0,0,0,0,1,-1,1,-1,1,-1,1,-1)^\dag,\nonumber\\
\bigr|\mathbf{e}_y\bigr>  &=& (0,0,0,1,-1,0,0,1,1,-1,-1,0,0,0,0,1,-1,1,-1,1,1,-1,-1,1,1,-1,-1)^\dag,\nonumber\\
\bigr|\mathbf{e}_z\bigr>  &=& (0,0,0,0,0,1,-1,0,0,0,0,1,1,-1,-1,1,1,-1,-1,1,1,1,1,-1,-1,-1,-1)^\dag.\nonumber
\end{eqnarray*}
Following the discretization process as outlined in the 2D case, we arrive a LB scheme with the standard collide-and-stream steps generically represented in Eqs.~(\ref{eq:LBMsteps-collide}) and (\ref{eq:LBMsteps-stream}), where $q = 27$. To express the collision
step in terms of changes to various central moments via their relaxations to attractors as prescribed by the FP collision model along
with those due to the body force for the D3Q27 lattice, we first define the \emph{discrete central moments} and \emph{discrete raw moments} of order ($m+n+p$) of the distribution function $f_{\alpha}$, its corresponding Markovian attractor $f_{\alpha}^{Mv}$, and the source term $S_{\alpha}$ for the body force, respectively, as follows:
\begin{subequations}
\begin{eqnarray}
\left( \begin{array}{c}\Ks{mnp} \\[3mm] \Ks{mnp}^{Mv} \\[3mm] \Sig_{mnp} \end{array} \right)  &=& \sum_{\alpha = 0}^{26} \left( \begin{array}{c}f_{\alpha} \\[3mm] f_{\alpha}^{Mv} \\[3mm] S_{\alpha} \end{array} \right)  ( e_{\alpha x} - u_{x})^m  ( e_{\alpha y} - u_{y})^n( e_{\alpha z} - u_{z})^p, \;\;\mbox{and}\\[0.2in]
\left( \begin{array}{c}\Kp_{mnp} \\[3mm] \kappa^{Mv'}_{mnp} \\[3mm]  \Sig'_{mnp} \end{array} \right)  &=& \sum_{\alpha = 0}^{26} \left( \begin{array}{c}f_{\alpha} \\[3mm] f_{\alpha}^{Mv}\\[3mm]  S_{\alpha} \end{array} \right)  e_{\alpha x}^m  e_{\alpha y}^n e_{\alpha z}^p.
\end{eqnarray}
\end{subequations}
Then, the countable independent central moments and raw moments for the D3Q27 lattice are, respectively, given by
\begin{subequations}
\begin{align}
\mc= ( & \Ks{000}, \Ks{100}, \Ks{010}, \Ks{001}, \Ks{110}, \Ks{101}, \Ks{011}, \Ks{200}, \Ks{020}, \Ks{002},\Ks{120}, \Ks{102}, \Ks{210},\Ks{012}, \nonumber \\[5pt] &  \Ks{201}, \Ks{021}, \Ks{111}, \Ks{220}, \Ks{202}, \Ks{022},  \Ks{211}, \Ks{121}, \Ks{112}, \Ks{122}, \Ks{212}, \Ks{221}, \Ks{222} ),
\end{align}
and
\begin{align}
\m = ( & \Kps{000}, \Kps{100}, \Kps{010}, \Kps{001}, \Kps{110}, \Kps{101}, \Kps{011}, \Kps{200}, \Kps{020}, \Kps{002}, \Kps{120}, \Kps{102}, \Kps{210}, \Kps{012}, \nonumber\\[5pt]  &  \Kps{201}, \Kps{021}, \Kps{111}, \Kps{220}, \Kps{202}, \Kps{022}, \Kps{211}, \Kps{121}, \Kps{112}, \Kps{122}, \Kps{212}, \Kps{221}, \Kps{222} ),
\end{align}
\end{subequations}
and, similarly, one can write those for the attractors and sources. In addition, we can also express a 27-dimensional vector $\mathbf{f}$ containing the distribution functions via
\begin{equation}\label{eqn:81}
\mathbf{f} = ( f_{ 0}, f_{ 1}, f_{ 2}, \cdots, f_{ 26}).
\end{equation}

To express the collision step in terms of relaxations of different central moments to their respective attractors along with source term updates, we have previously introduced the mappings between distribution functions $\mathbf{f}$ to raw moments $\m$ (see Eq.~(\ref{eq:map-df-rm})) and those between the raw moments $\m$ and central moments $\mc$(see Eq.~(\ref{eq:rmcm})). The projection matrix $\tensr{P}$ appearing in the former case for the D3Q27 lattice can be written as
\begin{eqnarray}
\tensr{P}=\bigr[\bigr|\bm{1}\bigr>,\;\;
\bigr|\bm{ e}_x \bigr>,\;\;
\bigr|\bm{e}_y\bigr>,\;\;
\bigr|\bm{e}_z\bigr>,\;\;
\bigr|\bm{e}_x \bm{e}_y\bigr>,\;\;
\bigr|\bm{e}_x \bm{e}_z\bigr>,\;\;
\bigr|\bm{e}_y \bm{e}_z\bigr>,\;\;
\bigr|\bm{e}_x^2\bigr>,\;\;
\bigr|\bm{e}_y^2\bigr>,\;\;
\bigr|\bm{e}_z^2\bigr>,\;\;
\bigr|\bm{e}_x \bm{e}_y^2\bigr>,\;\;
\bigr|\bm{e}_x \bm{e}_z^2\bigr>,\;\;\nonumber\\[2mm]
\bigr|\bm{e}_x^2 \bm{e}_y\bigr>,\;\;
\bigr|\bm{e}_y \bm{e}_z^2\bigr>,\;\;
\bigr|\bm{e}_x^2 \bm{e}_z\bigr>,\;\;
\bigr|\bm{e}_y^2 \bm{e}_z\bigr>,\;\;
\bigr|\bm{e}_x \bm{e}_y \bm{e}_z\bigr>,\;\;
\bigr|\bm{e}_x^2 \bm{e}_y^2\bigr>,\;\;
\bigr|\bm{e}_x^2 \bm{e}_z^2\bigr>,\;\;
\bigr|\bm{e}_y^2 \bm{e}_z^2\bigr>,\;\;
\bigr|\bm{e}_x^2  \bm{e}_y \bm{e}_z\bigr>,\;\;\nonumber\\[2mm]
\bigr|\bm{e}_x \bm{e}_y^2 \bm{e}_z\bigr>,\;\;
\bigr|\bm{e}_x \bm{e}_y \bm{e}_z^2\bigr>,\;\;
\bigr|\bm{e}_x \bm{e}_y^2 \bm{e}_z^2\bigr>,\;\;
\bigr|\bm{e}_x^2 \bm{e}_y \bm{e}_z^2\bigr>,\;\;
\bigr|\bm{e}_x^2 \bm{e}_y^2 \bm{e}_z\bigr>,\;\;
\bigr|\bm{e}_x^2 \bm{e}_y^2 \bm{e}_z^2\bigr>\nonumber
\bigr]^\dag
\end{eqnarray}
in which $\left| \mathbf{1} \right>$ is a 27-dimensional vector with unit elements forming a basis for the zeroth moment given by
$\left| \mathbf{1} \right> = (1,1,1,1,1,1,1,1,1,1,1,1,1,1,1,1,1,1,1,1,1,1,1,1,1,1,1)^{\dag}$; the frame transformation matrix $\F$
(as well as its inverse $\Fi$) appearing in the latter case depend on the local fluid velocity components $\bm{u} = (u_x,u_y,u_z)$ and
are low triangular matrices in structure that can be readily obtained via extracting the coefficients of the binomial expansions of
central moments in terms of raw moments (see e.g.,~\cite{yahia2021three}).

The discrete Markovian central moment attractors $\kappa_{mnp}^{Mv}$ needed in the construction of the 3D FPC-LBM are obtained, as in the 2D case, by matching the corresponding continuous versions given in Eq.~(\ref{eq:continouous-Markovian-CM-3D}), i.e., $\kappa_{mnp}^{Mv} = \Pi_{mnp}^{ Mv}$ for all independent countable moments supported by the D3Q27 lattice. Thus, we get
\[ \Ks{ 000}^{ Mv} = \rho, \; \; \; \; \; \; \Ks{ 100}^{ Mv} = 0, \; \; \; \; \; \; \Ks{ 010}^{ Mv} = 0, \; \; \; \; \; \; \Ks{ 001}^{ Mv} = 0, \\[3mm]\]
\[ \Ks{ 110}^{ Mv} = 0, \; \; \; \; \; \;\Ks{ 101}^{ Mv} = 0, \; \; \; \; \; \;\Ks{ 011}^{ Mv} = 0, \; \; \; \; \; \; \Ks{ 200}^{ Mv} = \rho c_s^2, \; \; \; \; \; \;\Ks{ 020}^{ Mv} = \rho c_s^2, \; \; \; \; \; \;\Ks{ 002}^{ Mv} =\rho c_s^2,\\[3mm] \]
\[\Ks{ 120}^{ Mv} = 0, \; \; \; \; \; \; \Ks{ 102}^{ Mv} =  0, \; \; \; \; \; \;\Ks{ 210}^{ Mv} = 0, \; \; \; \; \; \;\Ks{ 012}^{ Mv} = 0,  \; \; \; \; \; \;\Ks{ 201}^{ Mv} = 0, \; \; \; \; \; \;\Ks{ 021}^{ Mv} = 0, \; \; \; \; \; \;\Ks{ 111}^{ Mv} = 0, \\[3mm] \]
\[\boxed{\KMv{220} = \frac{1}{\rho} (\Kts{200}\Kts{020}+2\Kts{110}\Kts{110})},\; \; \;  \; \; \;
\boxed{\KMv{202} = \frac{1}{\rho} (\Kts{200}\Kts{002}+2\Kts{101}\Kts{101})}, \\[3mm]\]
\[\boxed{\KMv{022} = \frac{1}{\rho}(\Kts{020}\Kts{002}+2\Kts{011}\Kts{011})},\\[3mm]\]
\[ \boxed{\KMv{211} = \frac{1}{\rho}(\Kts{200}\Kts{011}+2\Kts{110}\Kts{101})},\; \; \;  \; \; \;
\boxed{\KMv{121} = \frac{1}{\rho}(\Kts{020}\Kts{101}+2\Kts{110}\Kts{011})},\\[3mm]\]
\[\boxed{\KMv{112} = \frac{1}{\rho}(\Kts{002}\Kts{110}+2\Kts{011}\Kts{101})},\\[3mm]\]
\[\boxed{\KMv{122} = \frac{2}{5\rho}(\Kts{020}\Kts{102}+\Kts{002}\Kts{120} + 4\Kts{011}\Kts{111}+2(\Kts{101}\Kts{021}+\Kts{011}\Kts{012}))},\\[3mm]\]
\[\boxed{\KMv{212} = \frac{2}{5\rho}(\Kts{200}\Kts{012}+\Kts{002}\Kts{210} + 4\Kts{101}\Kts{111}+2(\Kts{110}\Kts{102}+\Kts{011}\Kts{201}))},\\[3mm]\]
\[\boxed{\KMv{221} = \frac{2}{5\rho}(\Kts{200}\Kts{021}+\Kts{020}\Kts{201} + 4\Kts{110}\Kts{111}+2(\Kts{011}\Kts{210}+\Kts{101}\Kts{120}))},\\[3mm]\]
\begin{equation}\label{eqn:84}
\boxed{\KMv{222} = \frac{1}{3\rho}(\Kts{200}\Kts{022}+\Kts{020}\Kts{202}+\Kts{002}\Kts{220}+4(\Kts{110}\Kts{112}+\Kts{101}\Kts{121}+\Kts{011}\Kts{211}))}.\nonumber
\end{equation}
Thus, the above form a chain of attractors, especially for the fourth and higher central moments, each of which depends on certain combinations of the most recent (current) state of the adjacent and relevant lower moment, differing by a degree of two. In practical implementations, such lower central moments needed in the attractors are already in the post-collision states and the updated values will be used to prescribe the attractors for the higher central moments when they relax under collision. In addition, the source terms to the various central moments expressing the contributions from the presence of any body force is expressed from matching with their continuous counterparts given in Eq.~(\ref{forcing:3D}), which can be compactly written as
\begin{eqnarray}\label{eq:cmupdatesourceD3Q27}
\sigma_{mnp} = m \frac{F_x}{\rho}\kappa_{m-1,n,p} + n \frac{F_y}{\rho}\kappa_{m,n-1,p}+p \frac{F_z}{\rho}\kappa_{m,n,p-1}.
\end{eqnarray}

The implementation details of the 3D FPC-LBM using the D3Q27 lattice are given in Appendix~\ref{sec:3DFPC-LBM}.

\section{Results and discussion\label{sec:ResultsandDiscussion}}
We now demonstrate numerically the accuracy and stability properties of the new LB schemes based on central moment Fokker-Planck guided collision model, viz., the FPC-LBMs developed in the previous sections through a series of simulations of benchmark flow problems in two and three dimensions respectively. In two dimensions using the D2Q9 lattice, we show the accuracy of the FPC-LBM using a common test case, viz., the lid-driven square cavity flow, by comparing simulation results with those found in literature. Next, the stability of the FPC-LBM is demonstrated in two dimensions with the double periodic shear layer simulations, which give a visual representation of stability. Specifically, in this test case, two shear layers with an initial perturbation begin to roll up on themselves. When the grid resolution is relatively too coarse, secondary vortices begin to form at the thinnest parts of each layer for some of the LBM formulations, especially the SRT and MRT formulations. On the other hand, the central moment based LBM collision models are found to more robustly deal with such instabilities and this is shown visually when the secondary vortices mentioned above do not form for the same simulation parameters as those used with a less stable method. Furthermore, we show a direct comparison between two central moment collision models, the FPC-LBM and the MCM-LBM, which uses  Maxwellian central moments for equilibria, by simulating this flow problem at relatively high Mach numbers as well as beyond the usual limits of the relaxation parameter for the bulk viscosity when the simulation is generally accurate, and demonstrate the superior numerical performance of the former compared to the latter.

Then, in three dimensions, the lid-driven cubic cavity flow benchmark is used to indicate both accuracy and stability characteristics of FPC-LBM using the D3Q27 lattice. The orthogonal crossing shear wave test is then used to show how the FPC-LBM can deal with numerical hyperviscosity effects that arise in simulations of flows with extremely small physical fluid viscosities. For these cases, when appropriate, we show comparisons of the FPC-LBM with various other existing collision models in LBM. These include the single relaxation time (SRT)-LBM, the non-orthogonal raw moments-based multiple relaxation time (MRT)-LBM, the cascaded collision model with non-orthogonal moment basis using Maxwellian central moments for equilibria referred to as the MCM-LBM, the factorized LBM, and the cumulant LBM. Lastly, we show the capability of FPC-LBM in accurately and robustly resolving complex flows by simulating the canonical turbulent channel flow and comparing the resulting statistics against a recent direct numerical simulations (DNS) data available in the literature.

\subsection{Two-dimensional lid-driven square cavity flow: An accuracy study}
The benchmark known as lid-driven cavity flow consists of a square cavity with a viscous fluid inside; the cavity has a lid moving in the horizontal or $x$ direction with constant velocity $U$. In turn, the moving lid acts to shear the fluid which then forms vortical structures due to interaction with the surrounding stationary walls of the cavity. Furthermore, as the Reynolds number is increased, the characteristic velocity of the flow is then increased which causes counter-rotating vortices of varying sizes to form in the corners of the cavity. Here, we perform an accuracy study by comparing the results obtained using the FPC-LBM with those of Ghia et al (1982)~\cite{ghia1982high} for the cases of Reynolds numbers of $\mbox{Re} = 1000, 3200, 5000$, and $7500$. Here, the Reynolds number is defined as $\mbox{Re} = U_o L_o/\nu$, where $U_o$ is the characteristic plate velocity or $U$, $L_o$ is the characteristic length of one side of the square cavity, and $\nu$ is the kinematic viscosity of the fluid. The relaxation parameters related to the kinematic viscosity of the fluid, which depends on the choice of $\mbox{Re}$, are set according to Eq.~(\ref{eqn:transportcoeff2DFPCLBM}) (or Eq.~(\ref{eqn:transportcoeff3DFPCLBM}) in the 3D case), and, unless otherwise stated, that related to the bulk viscosity $\zeta$ and the others for relaxing third and higher moments are set to unity ; the speed of sound $c_s$ is taken to be $1/\sqrt{3}$ in all the simulations reported in this work. We used a grid resolution of $500\times 500$ and performed all our simulations at various Reynolds numbers noted above until they reach steady state by ensuring that the 2-norm of the residual error is less than $1\times 10^{-15}$.

\begin{figure}[ht!]
\centering
\begin{subfigure}{.48\textwidth}
\includegraphics[trim =35 0 70 10, clip, width =65mm]{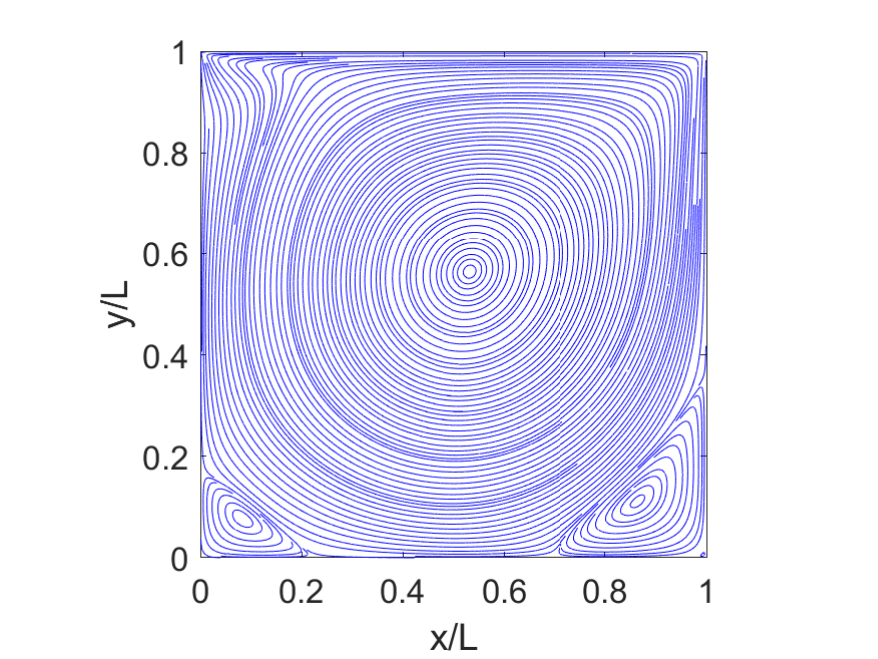}
\caption{$Re=1000$}
\end{subfigure}
\begin{subfigure}{.48\textwidth}
\includegraphics[trim = 35 0 70 10, clip, width = 65mm]{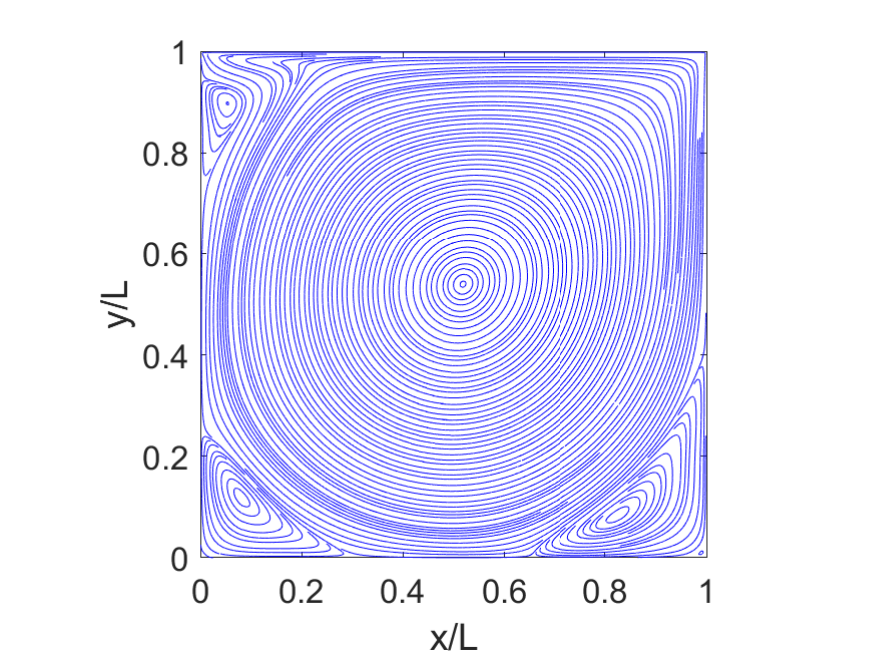}
\caption{$Re=3200$}
\end{subfigure}

\vspace{3mm}
\begin{subfigure}{.48\textwidth}
\includegraphics[trim =35 0 70 10, clip, width = 65mm]{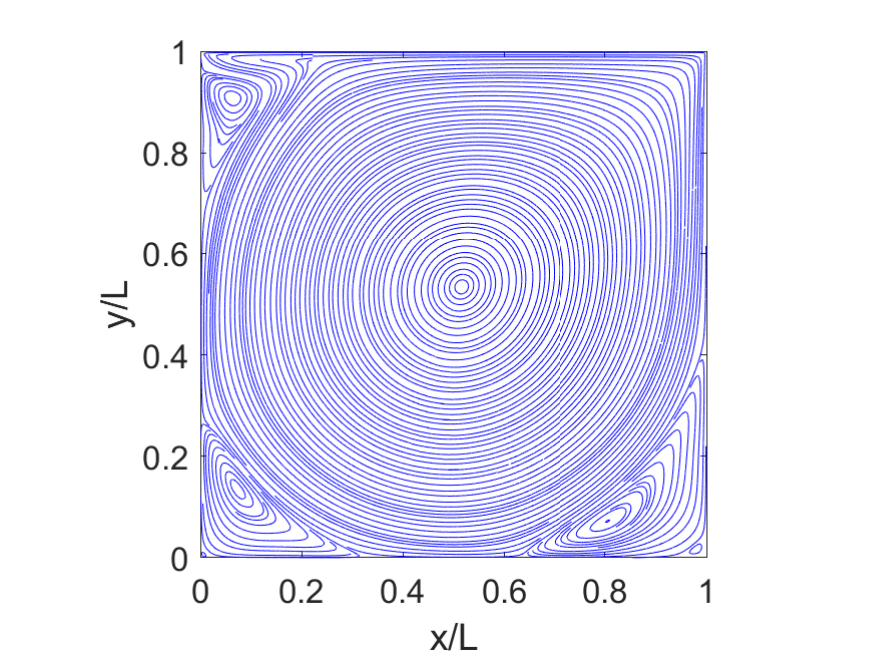}
\caption{$Re=5000$}
\end{subfigure}
\begin{subfigure}{.48\textwidth}
\includegraphics[trim =35 0 70 10, clip, width = 65mm]{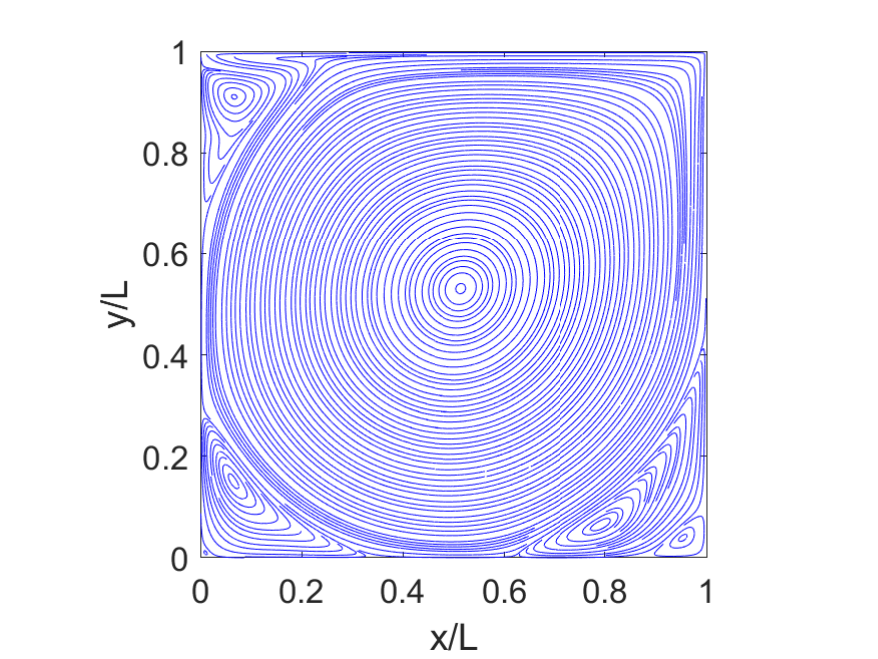}
\caption{$Re=7500$}
\end{subfigure}
\caption{Streamlines for two-dimensional lid-driven square cavity flow computed using the FPC-LBM at Reynolds numbers of $\mbox{Re}$ = 1000, 3200, 5000, and 7500. The formation of secondary and tertiary vortices is consistent with those in the benchmark results of Ghia \emph{et al} (1982)~\cite{ghia1982high} for each Reynolds number shown here.}
\label{LDC_Streamlines2d}
\end{figure}

\begin{figure}[ht!]
\centering
\begin{subfigure}{.48\textwidth}
\includegraphics[trim = 0 0 0 0, clip, width = 70mm]{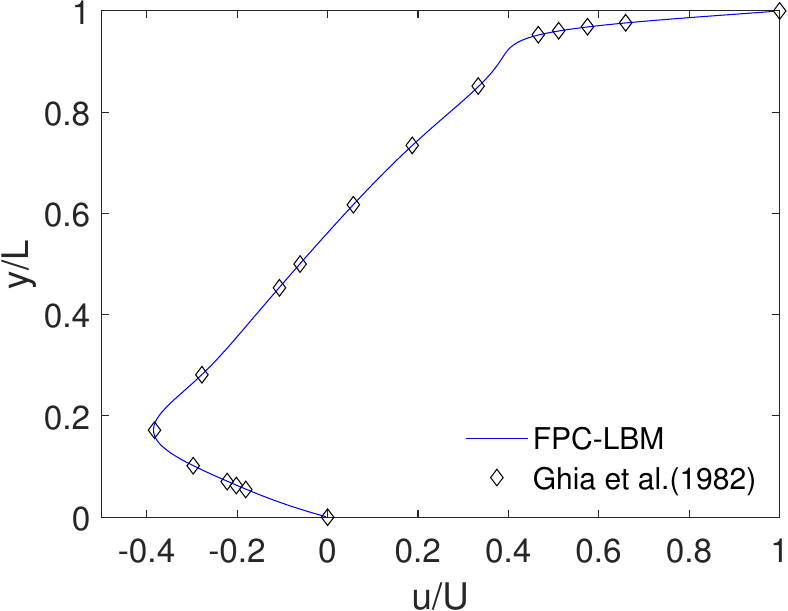}
\caption{$Re=1000$}
\end{subfigure}
\begin{subfigure}{.48\textwidth}
\includegraphics[trim = 0 0 0 0, clip, width = 70mm]{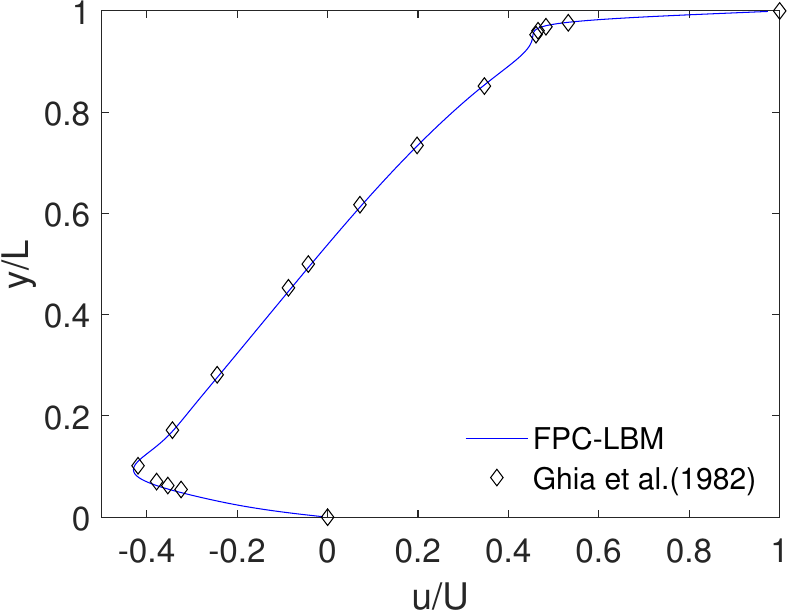}
\caption{$Re=3200$}
\end{subfigure}

\begin{subfigure}{.48\textwidth}
\includegraphics[trim = 0 0 0 -5, clip, width =70mm]{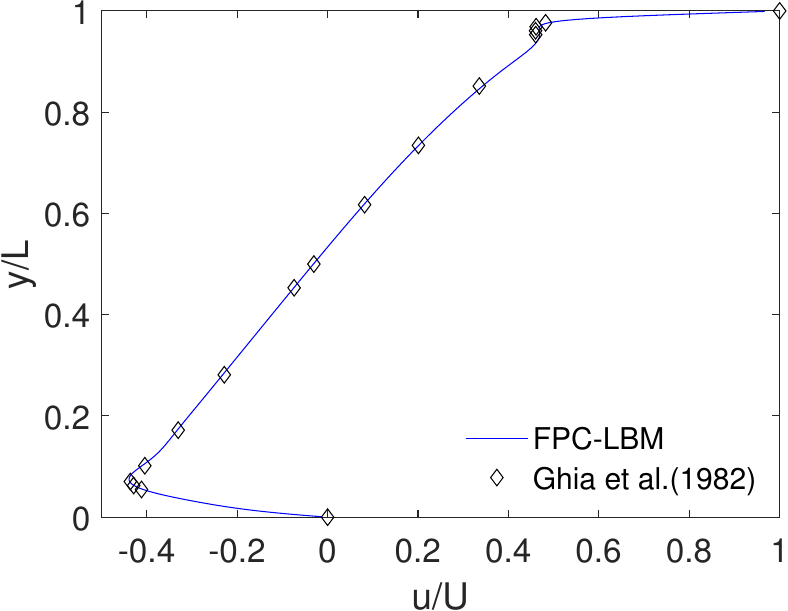}
\caption{$Re=5000$}
\end{subfigure}
\begin{subfigure}{.48\textwidth}
\includegraphics[trim = 0 0 0 -5, clip, width =70mm]{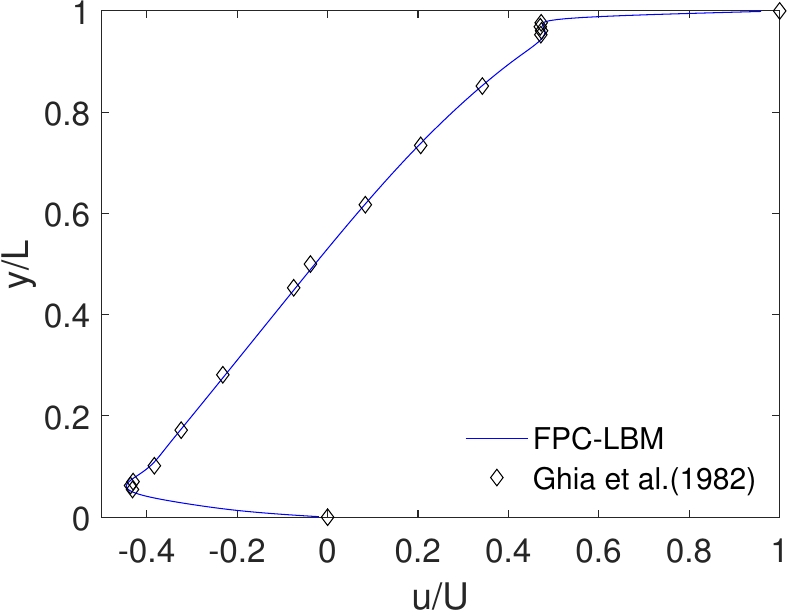}
\caption{$Re=7500$}
\end{subfigure}
\caption{Comparisons of the horizontal velocity component along the vertical centerline in a two-dimensional lid-driven square cavity flow at different Reynolds numbers computed using the FPC-LBM with the reference results of Ghia \emph{et al} (1982)~\cite{ghia1982high}. (a) $\mbox{Re}=1000$,   (b) $\mbox{Re}=3200$, (c) $\mbox{Re}=5000$, (d) $\mbox{Re}=7500$.}
\label{LDC2d_VerticalCL}
\end{figure}

\begin{figure}[ht!]
\centering
\begin{subfigure}{.48\textwidth}
\includegraphics[trim = 0 0 0 0, clip, width = 70mm]{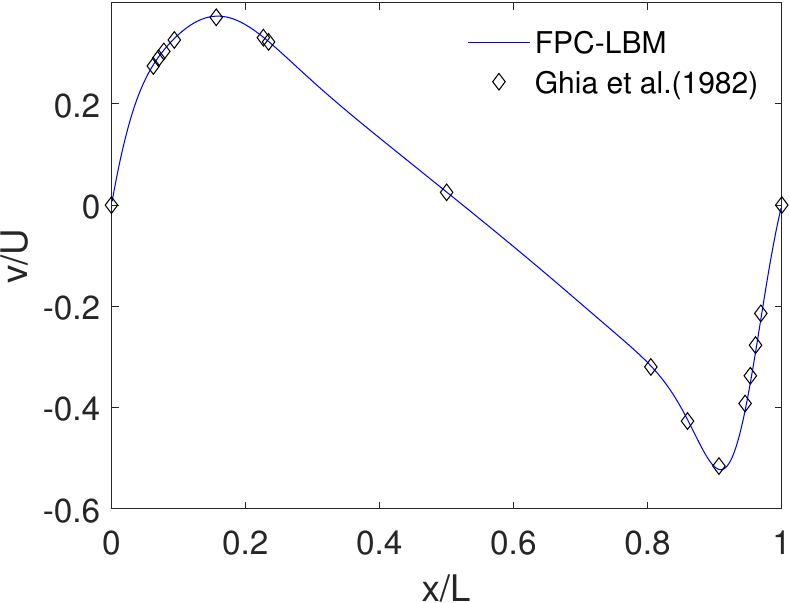}
\caption{$Re=1000$}
\end{subfigure}
\begin{subfigure}{.48\textwidth}
\includegraphics[trim = 0 0 0 0, clip, width = 70mm]{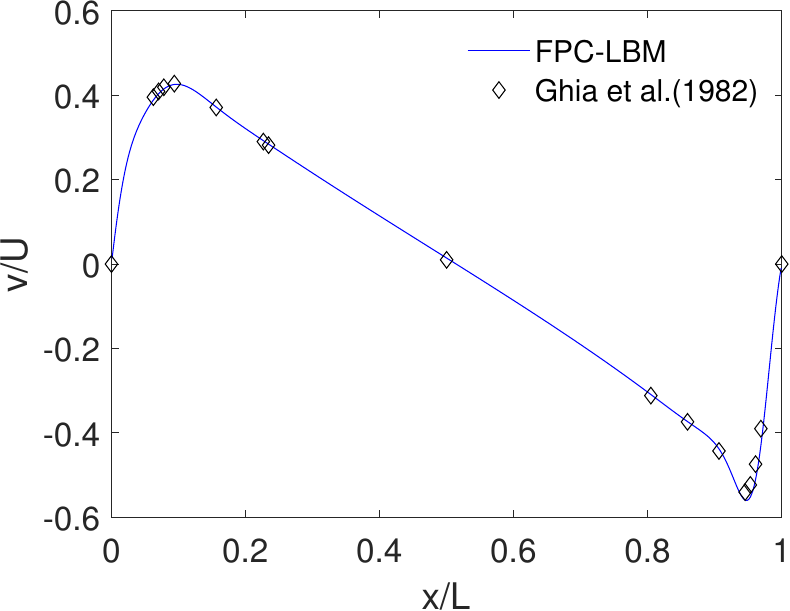}
\caption{$Re=3200$}
\end{subfigure}

\begin{subfigure}{.48\textwidth}
\includegraphics[trim = 0 0 0 -5, clip, width = 70mm]{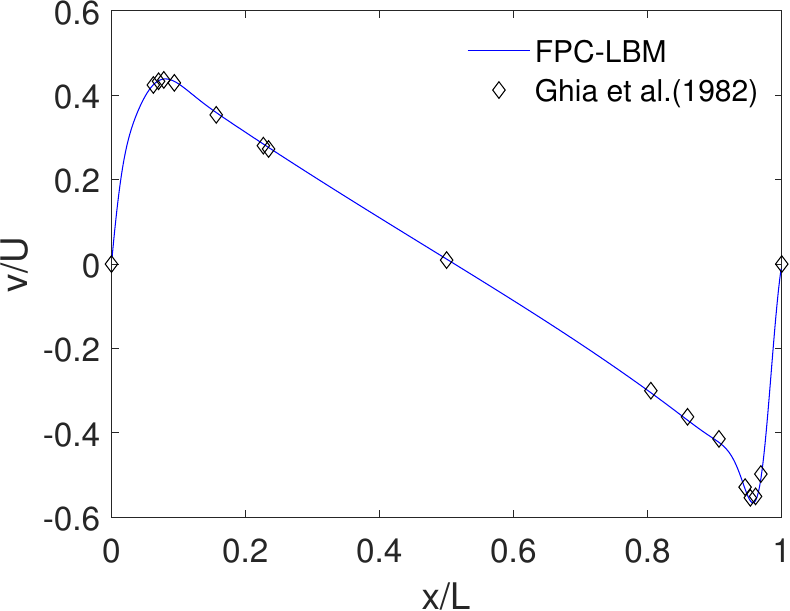}
\caption{$Re=5000$}
\end{subfigure}
\begin{subfigure}{.48\textwidth}
\includegraphics[trim = 0 0 0 -5, clip, width = 70mm]{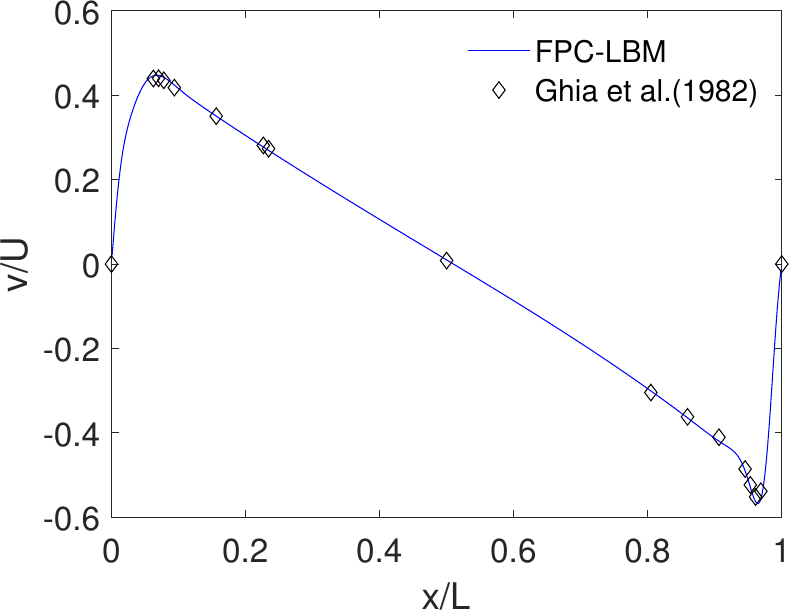}
\caption{$Re=7500$}
\end{subfigure}
\caption{Comparisons of the vertical velocity component along the horizontal centerline in a two-dimensional lid-driven square cavity flow at different Reynolds numbers computed using the FPC-LBM with the reference results of Ghia \emph{et al} (1982)~\cite{ghia1982high}. (a) $\mbox{Re}=1000$,   (b) $\mbox{Re}=3200$, (c) $\mbox{Re}=5000$, (d) $\mbox{Re}=7500$.}
\label{LDC2d_HorizontalCL}
\end{figure}

Shown in Figure~\ref{LDC_Streamlines2d} are the streamlines that come from the results using the FPC-LBM, and are found to be consistent with those found in literature~\cite{ghia1982high}. However, to make a direct comparison, we find the location of the ($x,y$) coordinates of the center of each of the corner vortices and compare them with the reference data reported by Ghia et al (1982)~\cite{ghia1982high} for all the cases of $\mbox{Re}$ noted above. These results are tabulated in Table~\ref{LDC_XYVortex}, side by side with our comparison case, and are found to be in relatively very good quantitative agreement. Moreover, the centerline horizontal and vertical components of the velocity profiles computed using the FPC-LBM for all the above four representative $\mbox{Re}$ are in excellent agreement with the reference results~\cite{ghia1982high} (see Figs.~\ref{LDC2d_VerticalCL} and \ref{LDC2d_HorizontalCL})
\begin{table}[ht]
\centering
\begin{tabular}{|c|c|c|c|c|c|}
\hline
Vortex    & Model    & $\mbox{Re}=1000$     &$\mbox{Re}=3200$   &$\mbox{Re}=5000$  &$\mbox{Re}=7500$ \\[ 0.05in]
\hline
PV&  FPC-LBM     & $(0.5306, 0.5650)$  & $ (0.5185 , 0.5395 )$ & $(0.5158, 0.5344)$ & $(0.5138, 0.5318)$ \\[ 0.05in]
PV& Ghia et. al.   & $(0.5313 , 0.5625)$  & $(0.5165 , 0.5469)$ & $(0.5117 , 0.5352)$ & $(0.5117 , 0.5322)$ \\[ 0.05in]
\hline
BR1&   FPC-LBM  & $(0.8646, 0.1115)$  & $( 0.8249, 0.0834)$ & $(0.8054 , 0.0720)$ & $( 0.7913, 0.0643)$ \\[ 0.05in]
BR1 &Ghia et al.  & $(0.8594 , 0.1094)$  & $(0.8125 , 0.0859)$  & $(0.8086 , 0.0742)$ & $(0.7813 , 0.0625)$  \\[ 0.05in]
\hline
BR2 & FPC-LBM   & $(0.9929, 0.0057)$  & $(0.9903 , 0.0083 )$ & $( 0.9795 , 0.0172)$ & $(0.9533 , 0.0401)$ \\[ 0.05in]
BR2 & Ghia et al. & $(0.9922 , 0.0078)$  & $(0.9844 , 0.0078)$ & $(0.9805 , 0.0195)$ & $(0.9492 , 0.0430)$ \\[ 0.05in]
\hline
BL& FPC-LBM        & $(0.0830, 0.0771)$   & $(0.0810 , 0.1191)$ & $( 0.0736,0.1344 )$ & $(0.0649 , 0.1510)$ \\[ 0.05in]
BL& Ghia et al.        & $(0.0859 , 0.0781)$   & $(0.0859 , 0.1094)$ & $(0.0703 , 0.1367)$ & $(0.0645 , 0.1504)$ \\[ 0.05in]
\hline
TL &FPC-LBM        & $N/A$                            & $(0.0574 , 0.8974)$ & $( 0.0635, 0.9089)$ & $(0.0669 , 0.9102)$ \\[ 0.05in]
TL &Ghia et al.     & $N/A $                           & $(0.0547 , 0.8984)$ & $(0.0625 , 0.9102)$ & $(0.0664 , 0.9141)$ \\[ 0.05in]
\hline
\end{tabular}
\caption{Comparison of the location of the center of various vortices in a 2D lid-driven square cavity flow for different Reynolds numbers. Computed results are obtained using the FPC-LBM with a grid resolution of $500\times500$ and are compared with those of Ghia \emph{et al}(1982)~\cite{ghia1982high}. PV – Primary Vortex, BR1 – Bottom Right 1, BR2 - Bottom Right 2, BL - Bottom Lefft, TL - Top Left.}
\label{LDC_XYVortex}
\end{table}


\subsection{Doubly periodic shear layers: Numerical performance study}
The benchmark commonly known as doubly periodic shear layers~\cite{minion1997performance} is employed here to indicate a level of stability of the FPC-LBM as compared to other collision models such as MCM-LBM , which uses the Maxwellian based equilibria, i.e., $\kappa_{mn}^{eq}=\Pi^{(0)}_{mn}=\left<f^M, W_{mn}\right>$, where $f^M$ is given in Eq.~(\ref{maxwell_distribution}). Since the SRT-LBM
and MRT-LBM are already known to be less stable compared to central moment-based LB formulations, our focus will be on comparing the different possible central moment-based LB scheme for simulating this flow problem. The key idea is that a pair of shear layers in a periodic box will roll up on themselves if given an initial perturbation in the direction perpendicular to the shear layers. Then, looking at the resulting vorticity field at a dimensionless time equal to one, we observe if the flow field is fully resolved. When using a relatively fine enough grid resolution, the layers will each be shown to roll up in a single vortex at the location of the initial perturbation for any LBM collision operator being used, which is considered to be a fully resolved flow field. However, if the discretized grid is made to be relatively coarse, then spurious secondary vortices can form at the thinnest parts of each layer due to the presence of numerical artifacts associated with the numerical method being used, which could ultimately destabilize the simulation. Furthermore, the Mach number is used here to enforce the characteristic speed of the shear layers, which can trigger numerical instabilities when made progressively larger for a chosen grid resolution, and one of the goals is to determine which one of the central moment LB formulation sustains stable simulations relative to the other in such cases.

The following equation is used to initialize the flow field $\bm{u}(t=0) = (u_x,u_y)$ by non-dimensionalizing the spatial coordinates with the side length of the periodic square domain, $L_o$:
\begin{eqnarray}
&u_x& = \begin{cases}
U_o \tanh(4(y-0.25)/w) \;\;\;\; y\leq 0.5, \\[3mm]
U_o \tanh(4(0.75-y)/w) \;\;\;\; y>0.5,\\
       \end{cases} \\[3mm]
&u_y& = \delta \sin(2\pi(x+0.25)), \nonumber
\end{eqnarray}
where $\delta$ is related to the magnitude of the initial perturbation and $w$ is related to the thickness of the shear layers; here, we use $\delta = 0.05$ and $w = 0.05$. In all of the following simulations, we set the Reynolds number to $\mbox{Re}=30000$, where the Reynolds number for this problem is defined by $\mbox{Re} = U_o L_o /\nu$. The Mach number associated in the simulations is defined as $\mbox{Ma} = U_o/c_s$, where $c_s = 1/\sqrt{3}$, and the characteristic time scale reads as $T=L_o/U_o$. Here, we compare the different collision models by running simulations over a range of grid resolutions as well as Mach numbers, and compare the resulting flow fields at the dimensionless time $t$, obtained by scaling time using $T$, of 1.0.

\begin{figure}[ht]
\centering
\vspace{3mm}
\begin{subfigure}{.32\textwidth}
\includegraphics[trim = 0 0 0 0,clip, width = 45mm]{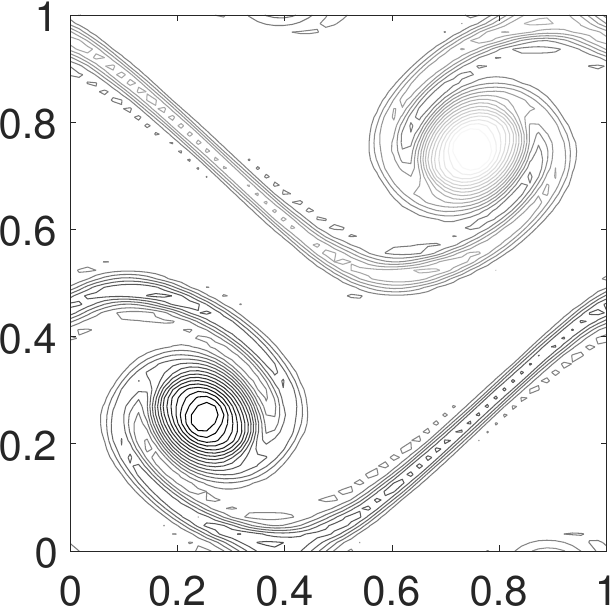}
\caption{$\mbox{Ma}=0.05$,   $L^2 = 64^2$}
\end{subfigure}
\begin{subfigure}{.32\textwidth}
\includegraphics[trim = 0 0 0 0,clip, width = 45mm]{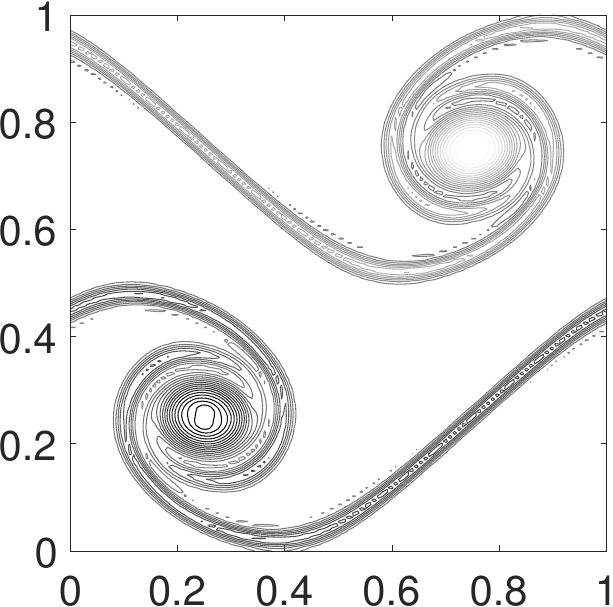}
\caption{$\mbox{Ma}=0.05$,   $L^2 = 128^2$}
\end{subfigure}
\begin{subfigure}{.32\textwidth}
\includegraphics[trim = 0 0 0 0,clip, width = 45mm]{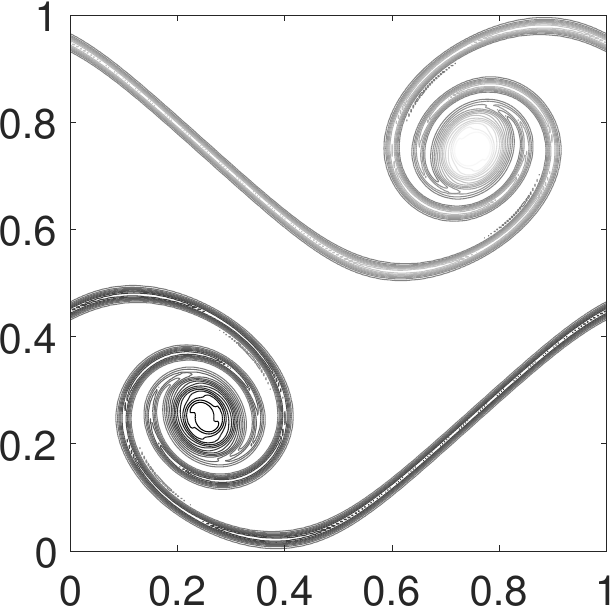}
\caption{$\mbox{Ma}=0.05$,   $L^2 = 256^2$}
\end{subfigure}
\vspace{3mm}
\begin{subfigure}{.32\textwidth}
\includegraphics[trim = 0 0 0 0,clip, width = 45mm]{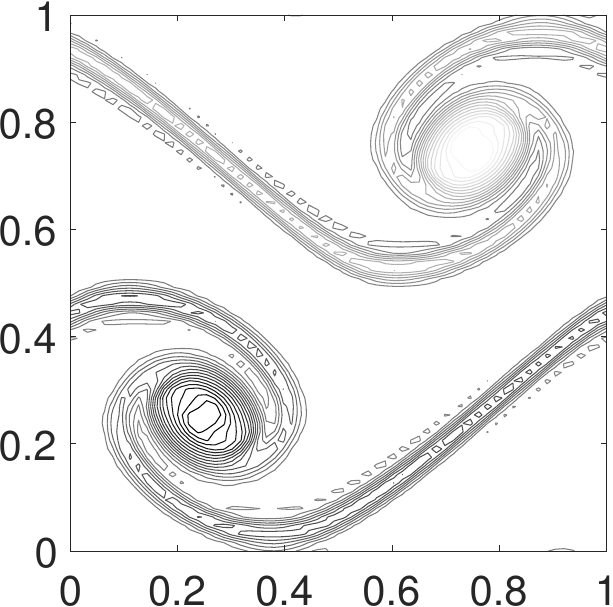}
\caption{$\mbox{Ma}=0.2$,   $L^2 = 64^2$}
\end{subfigure}
\begin{subfigure}{.32\textwidth}
\includegraphics[trim = 0 0 0 0,clip, width = 45mm]{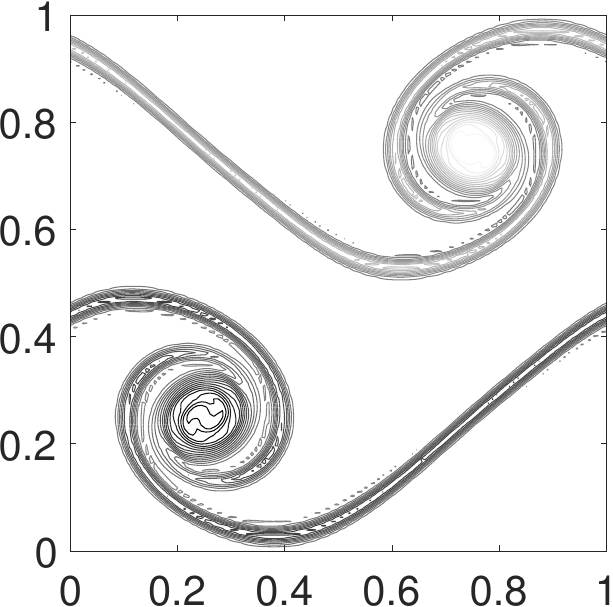}
\caption{$\mbox{Ma}=0.2$,   $L^2 = 128^2$}
\end{subfigure}
\begin{subfigure}{.32\textwidth}
\includegraphics[trim = 0 0 0 0,clip, width = 45mm]{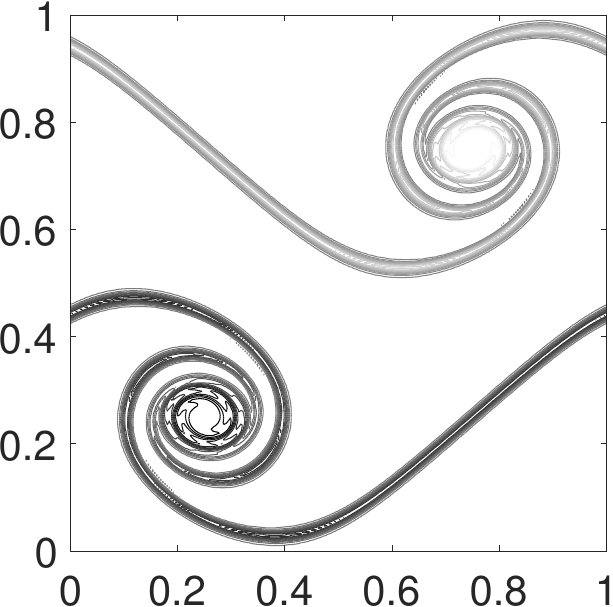}
\caption{$\mbox{Ma}=0.2$,   $L^2 = 256^2$}
\end{subfigure}
\vspace{3mm}
\begin{subfigure}{.32\textwidth}
\includegraphics[trim = 0 0 0 0,clip, width = 45mm]{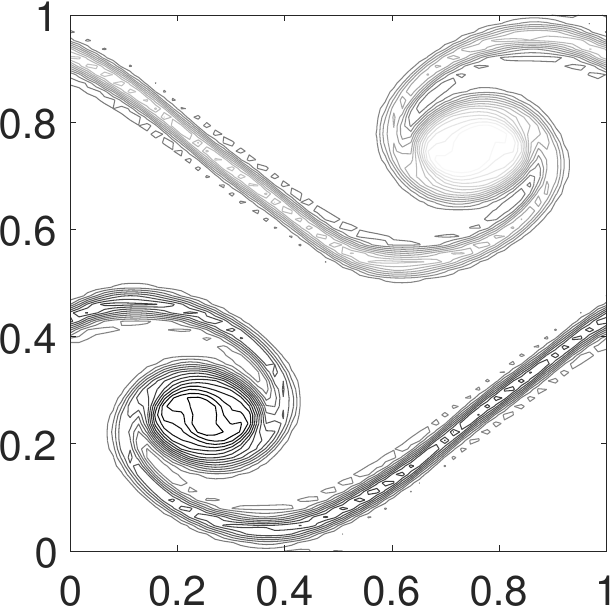}
\caption{$\mbox{Ma}=0.3$,   $L^2 = 64^2$}
\end{subfigure}
\begin{subfigure}{.32\textwidth}
\includegraphics[trim = 0 0 0 0,clip, width =45mm]{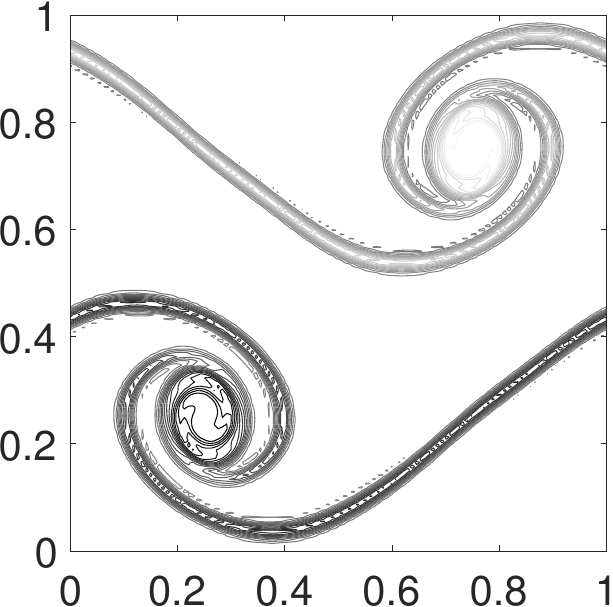}
\caption{$\mbox{Ma}=0.3$,   $L^2 =128^2$}
\end{subfigure}
\begin{subfigure}{.32\textwidth}
\includegraphics[trim = 0 0 0 0,clip, width = 45mm]{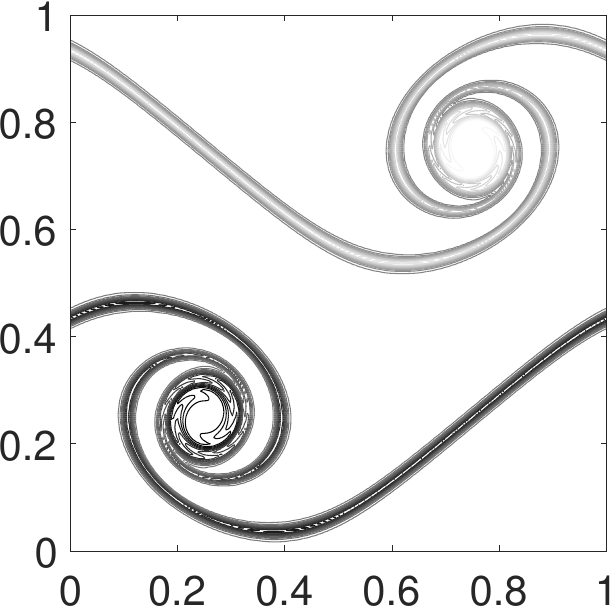}
\caption{$\mbox{Ma}=0.3$,   $L^2 =256^2$}
\end{subfigure}
\caption{Vorticity contours of doubly periodic shear layers that roll up due to an applied perturbation at $t = 1$ for different sets of lower Mach numbers of $0.05$, $0.2$ and $0.3$ (along rows) and at grid resolutions of $64^2$, $128^2$, and $256^2$ (along columns) computed using the Maxwellian equilibria based MCM-LBM. The MCM-LBM is seen to sufficiently capture the physics of this case for all of the grid resolutions and Mach numbers considered here as it has not caused the formation of any spurious secondary vortices.}
\label{DPSL_MCM1}
\end{figure}

\begin{figure}[ht]
\centering
\vspace{3mm}
\begin{subfigure}{.32\textwidth}
\includegraphics[trim = 0 0 0 0,clip, width =45mm]{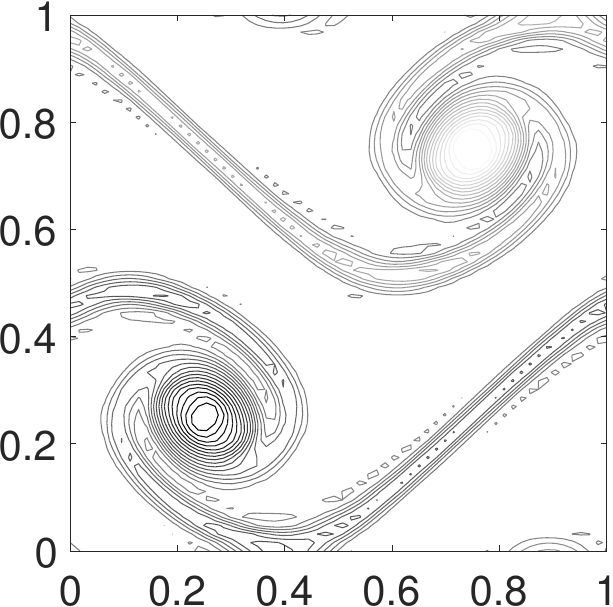}
\caption{$\mbox{Ma}=0.05$,   $L^2 = 64^2$}
\end{subfigure}
\begin{subfigure}{.32\textwidth}
\includegraphics[trim = 0 0 0 0,clip, width = 45mm]{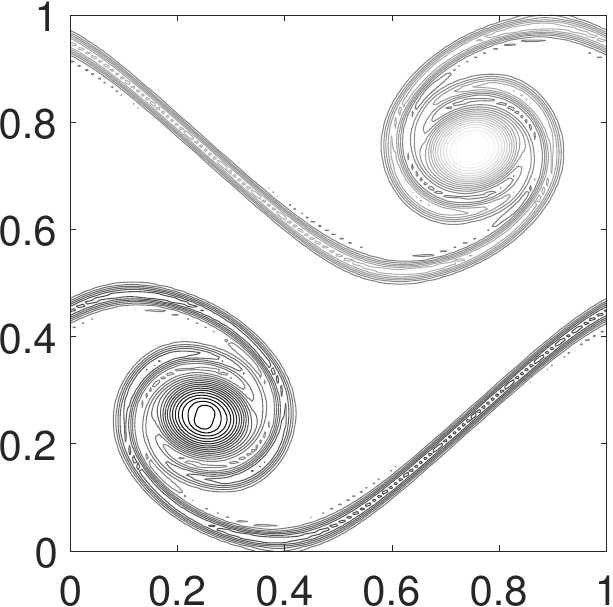}
\caption{$\mbox{Ma}=0.05$,   $L^2 = 128^2$}
\end{subfigure}
\begin{subfigure}{.32\textwidth}
\includegraphics[trim = 0 0 0 0,clip, width = 45mm]{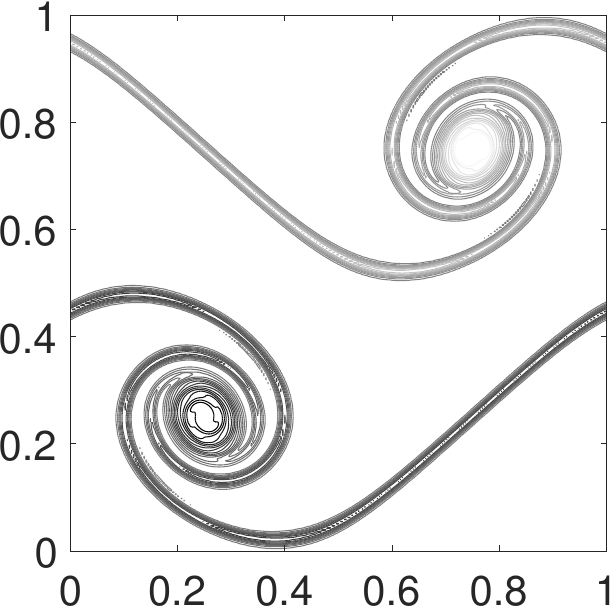}
\caption{$\mbox{Ma}=0.05$,   $L^2 = 256^2$}
\end{subfigure}
\vspace{3mm}
\begin{subfigure}{.32\textwidth}
\includegraphics[trim = 0 0 0 0,clip, width = 45mm]{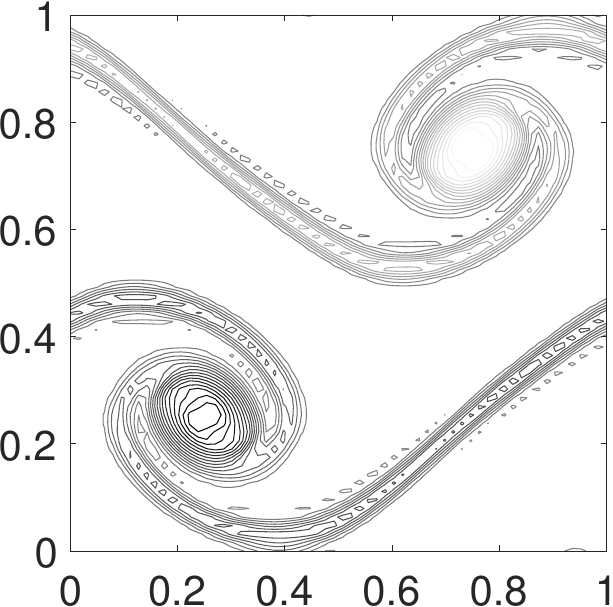}
\caption{$\mbox{Ma}=0.2$,   $L^2 = 64^2$}
\end{subfigure}
\begin{subfigure}{.32\textwidth}
\includegraphics[trim = 0 0 0 0,clip, width = 45mm]{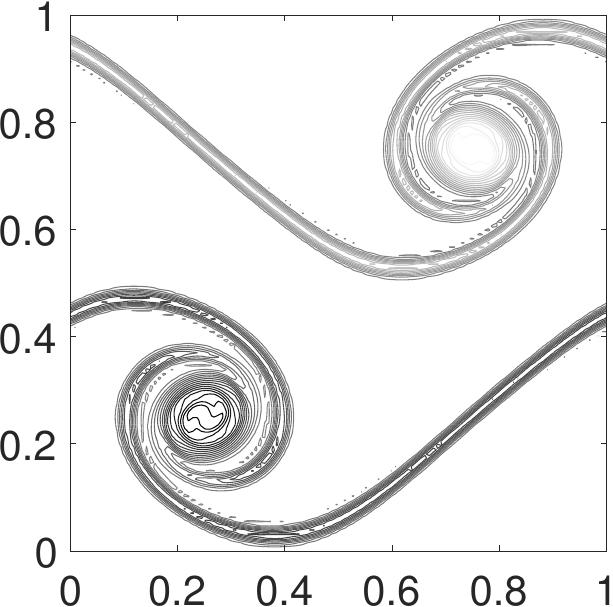}
\caption{$\mbox{Ma}=0.2$,   $L^2 = 128^2$}
\end{subfigure}
\begin{subfigure}{.32\textwidth}
\includegraphics[trim = 0 0 0 0,clip, width = 45mm]{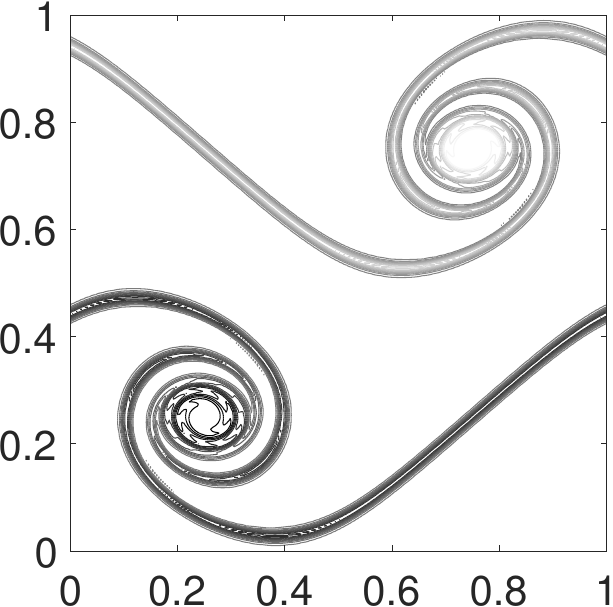}
\caption{$\mbox{Ma}=0.2$,   $L^2 = 256^2$}
\end{subfigure}
\vspace{3mm}
\begin{subfigure}{.32\textwidth}
\includegraphics[trim = 0 0 0 0,clip, width = 45mm]{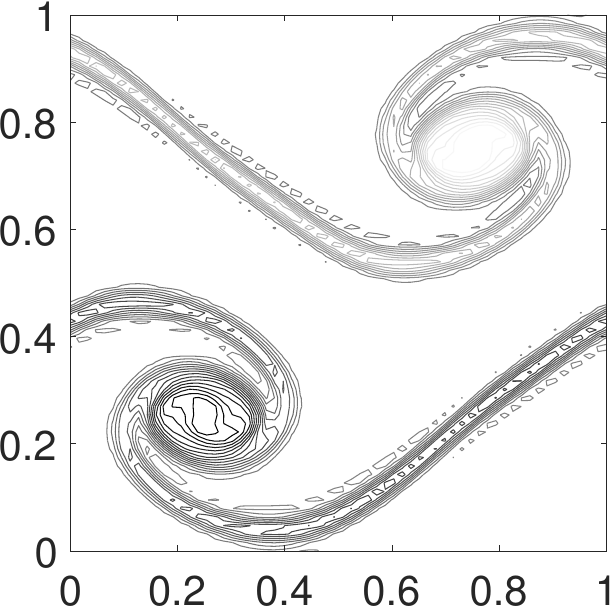}
\caption{$\mbox{Ma}=0.3$,   $L^2 = 64^2$}
\end{subfigure}
\begin{subfigure}{.32\textwidth}
\includegraphics[trim = 0 0 0 0,clip, width = 45mm]{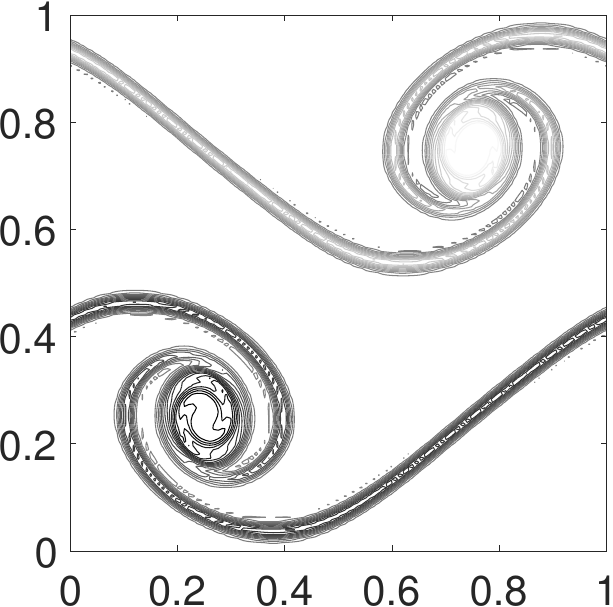}
\caption{$\mbox{Ma}=0.3$,   $L^2 =128^2$}
\end{subfigure}
\begin{subfigure}{.32\textwidth}
\includegraphics[trim = 0 0 0 0,clip, width = 45mm]{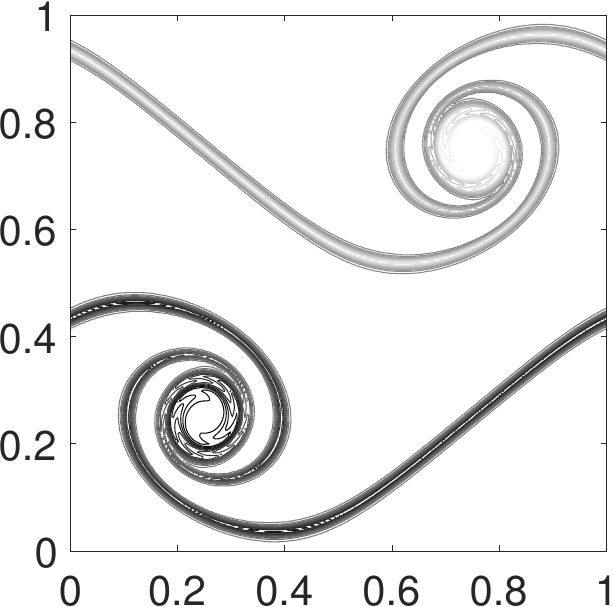}
\caption{$\mbox{Ma}=0.3$,   $L^2 =256^2$}
\end{subfigure}
\caption{Vorticity contours of doubly periodic shear layers that roll up due to an applied perturbation at $t = 1$ for different sets of lower Mach numbers of $0.05$, $0.2$ and $0.3$ (along rows) and at grid resolutions of $64^2$, $128^2$, and $256^2$ (along columns) computed using the Fokker-Planck equilibria based FPC-LBM. The FPC-LBM is seen to sufficiently capture the physics of this case for all of the grid resolutions and Mach numbers considered here as it has not caused the formation of any spurious secondary vortices. The FPC-LBM and MCM-LBM results are almost indistinguishable from one another for these cases indicating the robustness of central moment collision models in general and that we must consider more extreme cases to see significant differences between them.}
\label{DPSL_FPC1}
\end{figure}

In what follows, we make a direct comparison of the resulting vorticity fields between the FPC-LBM and the MCM-LBM. The results shown in Figs.~\ref{DPSL_MCM1} and \ref{DPSL_FPC1} consist of baseline cases using the MCM-LBM and FPC-LBM, respectively, using the Mach numbers of $\mbox{Ma}= 0.05, 0.2$, and $0.3$ along with grid resolutions of $L^2 = 64^2, 128^2$, and $256^2$. As such, the Mach numbers are generally in the lower range and hence all of the cases were found to be numerically stable without the formation of any spurious secondary vortices, and the results from both the collision models are almost indistinguishable. This generally shows the robustness of central moment collision models and also indicates that we must consider more extreme parametric conditions to see significant differences between them.

\begin{figure}[ht]
\centering
\vspace{3mm}
\begin{subfigure}{.32\textwidth}
\includegraphics[trim = 0 0 0 0,clip, width = 45mm]{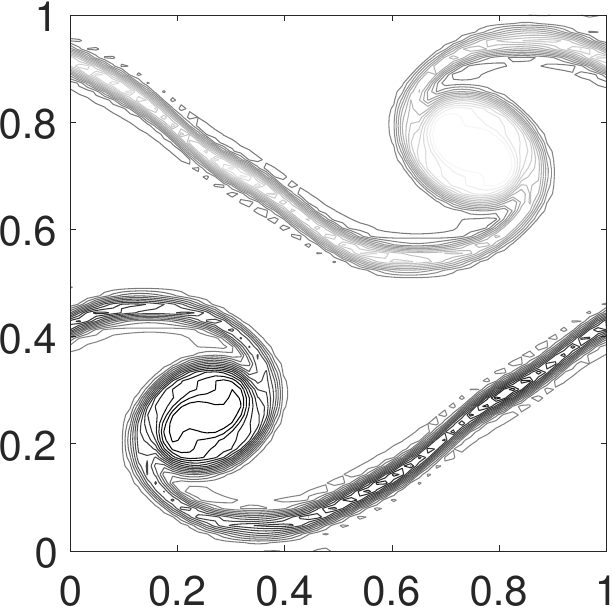}
\caption{$\mbox{Ma}=0.4$,   $L^2 = 64^2$}
\end{subfigure}
\begin{subfigure}{.32\textwidth}
\includegraphics[trim = 0 0 0 0,clip, width = 45mm]{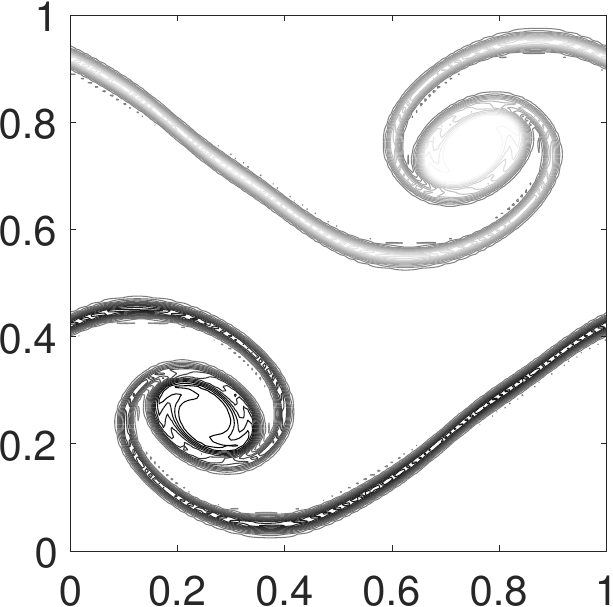}
\caption{$\mbox{Ma}=0.4$,   $L^2 = 128^2$}
\end{subfigure}
\begin{subfigure}{.32\textwidth}
\includegraphics[trim = 0 0 0 0,clip, width = 45mm]{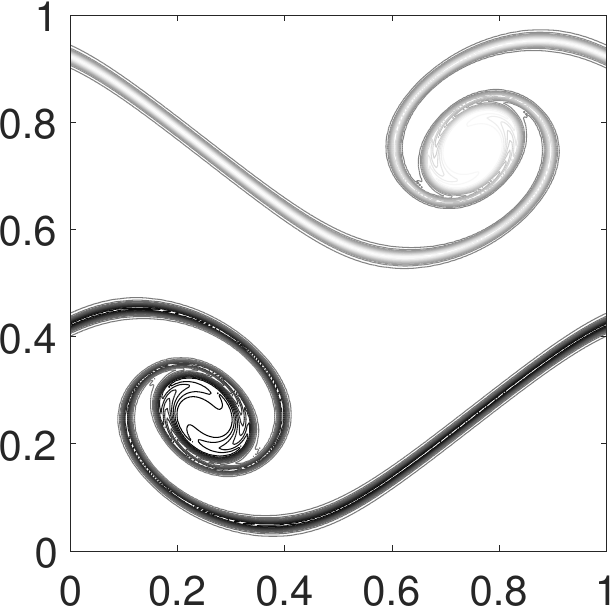}
\caption{$\mbox{Ma}=0.4$,   $L^2 = 256^2$}
\end{subfigure}
\vspace{3mm}
\begin{subfigure}{.32\textwidth}
\includegraphics[trim = 0 0 0 0,clip, width = 45mm]{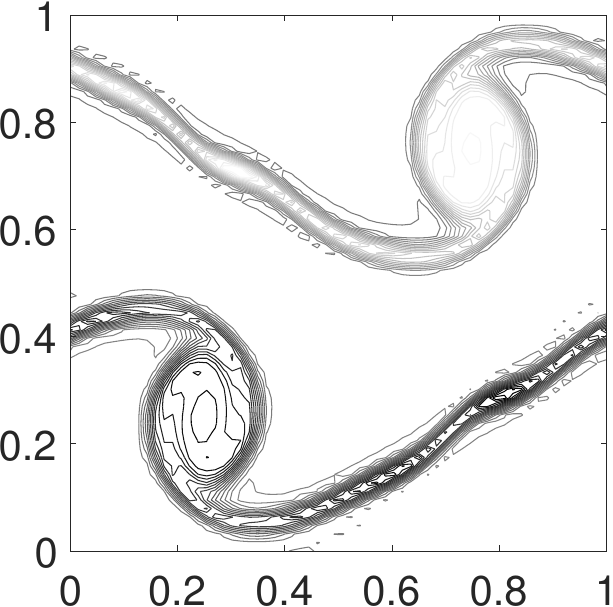}
\caption{$\mbox{Ma}=0.5$,   $L^2 = 64^2$}
\end{subfigure}
\begin{subfigure}{.32\textwidth}
\includegraphics[trim = 0 0 0 0,clip, width = 45mm]{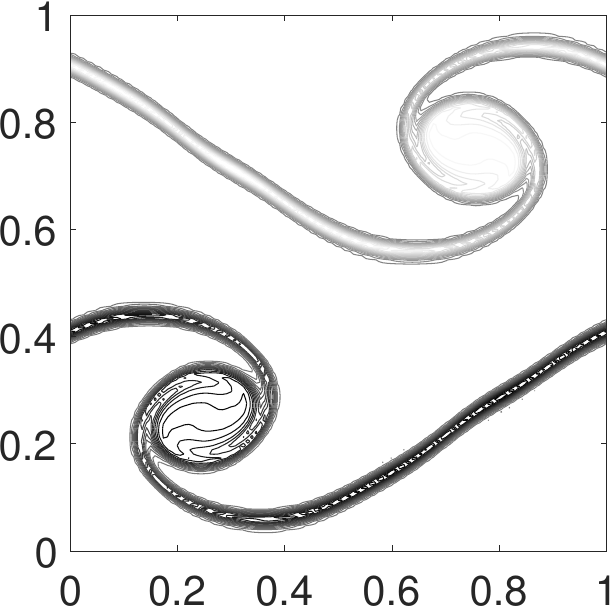}
\caption{$\mbox{Ma}=0.5$,   $L^2 = 128^2$}
\end{subfigure}
\begin{subfigure}{.32\textwidth}
\includegraphics[trim = 0 0 0 0,clip, width = 45mm]{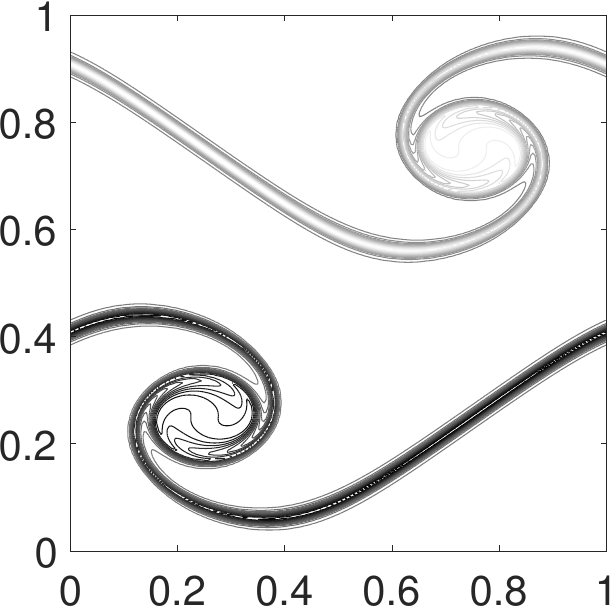}
\caption{$\mbox{Ma}=0.5$,   $L^2 = 256^2$}
\end{subfigure}
\vspace{3mm}
\begin{subfigure}{.32\textwidth}
\includegraphics[trim = 0 0 0 0,clip, width = 45mm]{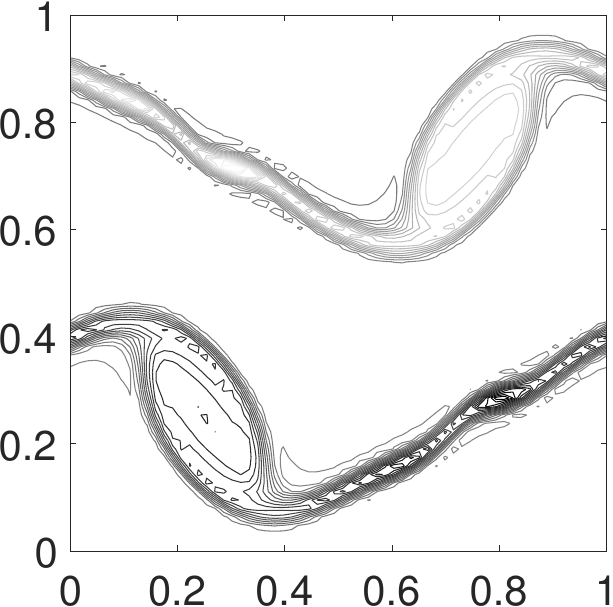}
\caption{$\mbox{Ma}=0.6$,   $L^2 = 64^2$}
\end{subfigure}
\begin{subfigure}{.32\textwidth}
\includegraphics[trim = 0 0 0 0,clip, width = 45mm]{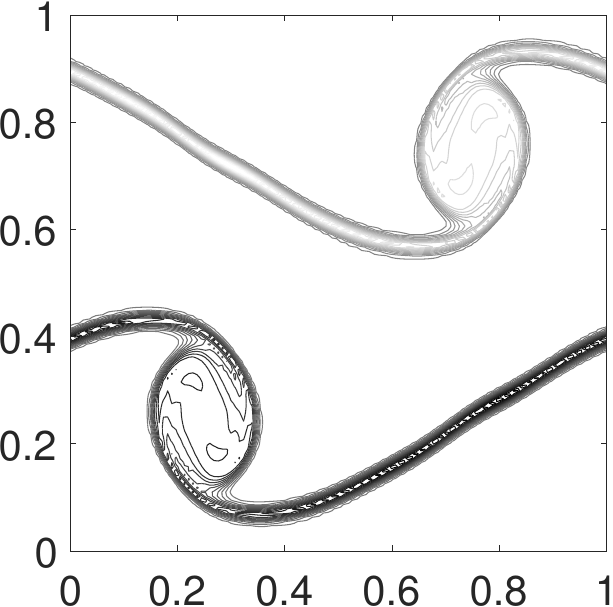}
\caption{$\mbox{Ma}=0.6$,   $L^2 =128^2$}
\end{subfigure}
\begin{subfigure}{.32\textwidth}
\includegraphics[trim = 0 0 0 0,clip, width = 45mm]{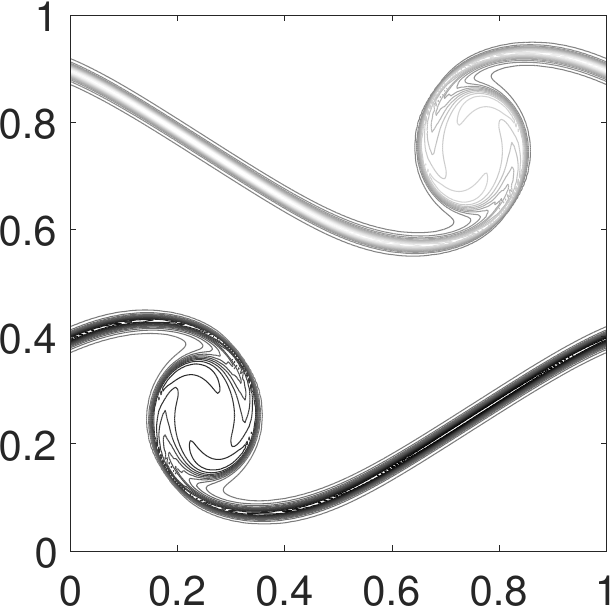}
\caption{$\mbox{Ma}=0.6$,   $L^2 =256^2$}
\end{subfigure}
\caption{Vorticity contours of doubly periodic shear layers that roll up due to an applied perturbation at $t = 1$ for different sets of higher Mach numbers of $0.4$, $0.5$ and $0.6$ (along rows) and at grid resolutions of $64^2$, $128^2$, and $256^2$ (along columns) computed using the Maxwellian equilibria based MCM-LBM. It is seen that at larger Mach numbers and coarse grid resolutions, the MCM-LBM becomes unstable with the formation of spurious secondary vortices which form on the layers for the cases of $\mbox{Ma}=0.4, 0.5,$ and $0.6$ at $L^2 = 64^2$.}
\label{DPSL_MCM2}
\end{figure}

\begin{figure}[ht]
\centering
\vspace{3mm}
\begin{subfigure}{.32\textwidth}
\includegraphics[trim = 0 0 0 0,clip, width = 45mm]{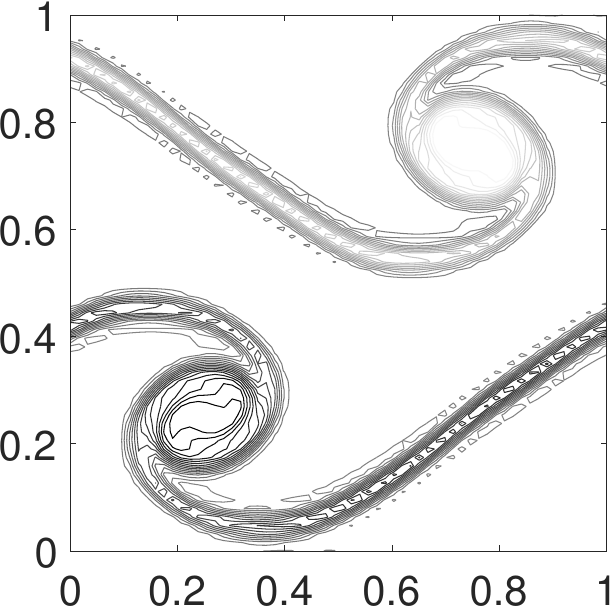}
\caption{$\mbox{Ma}=0.4$,   $L^2 = 64^2$}
\end{subfigure}
\begin{subfigure}{.32\textwidth}
\includegraphics[trim = 0 0 0 0,clip, width = 45mm]{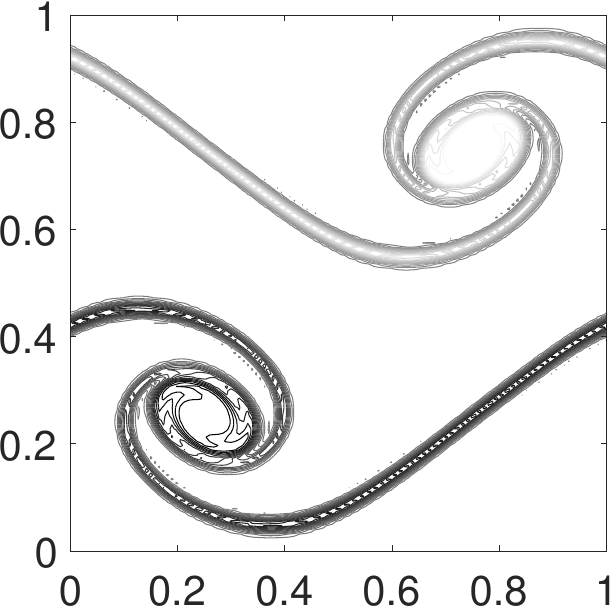}
\caption{$\mbox{Ma}=0.4$,   $L^2 = 128^2$}
\end{subfigure}
\begin{subfigure}{.32\textwidth}
\includegraphics[trim = 0 0 0 0,clip, width = 45mm]{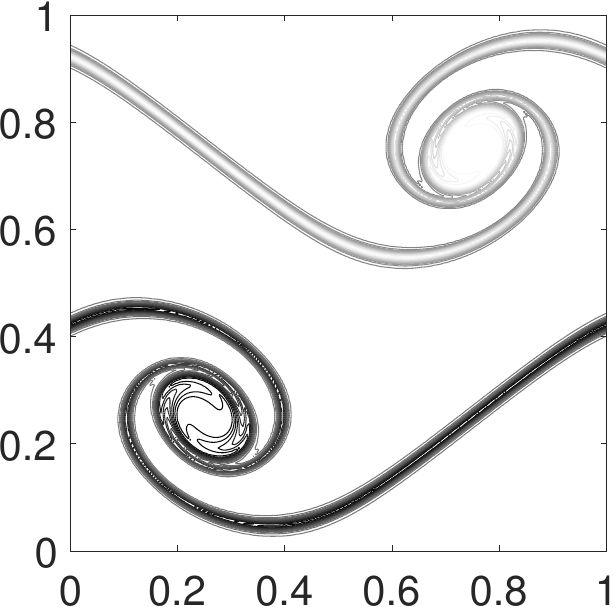}
\caption{$\mbox{Ma}=0.4$,   $L^2 = 256^2$}
\end{subfigure}
\vspace{3mm}
\begin{subfigure}{.32\textwidth}
\includegraphics[trim = 0 0 0 0,clip, width = 45mm]{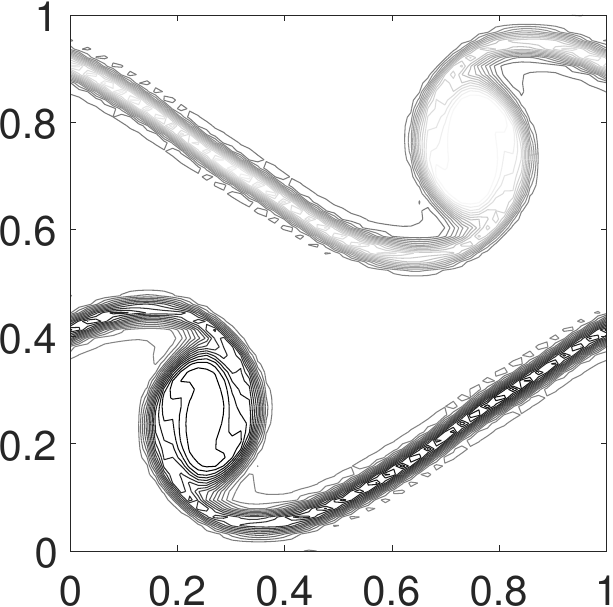}
\caption{$\mbox{Ma}=0.5$,   $L^2 = 64^2$}
\end{subfigure}
\begin{subfigure}{.32\textwidth}
\includegraphics[trim = 0 0 0 0,clip, width = 45mm]{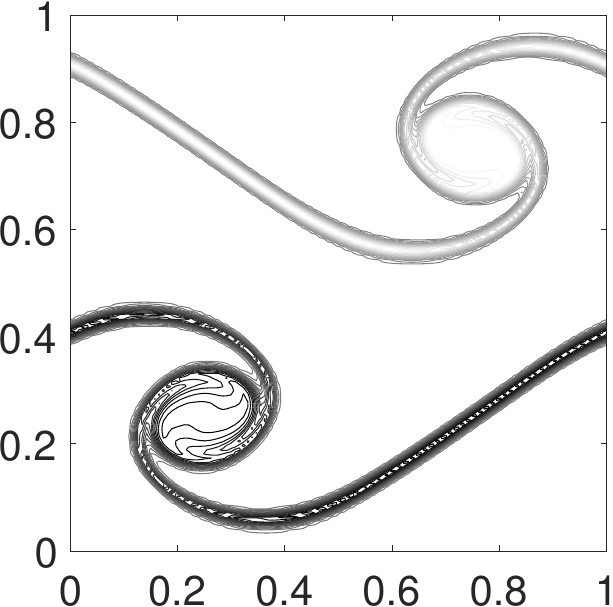}
\caption{$\mbox{Ma}=0.5$,   $L^2 = 128^2$}
\end{subfigure}
\begin{subfigure}{.32\textwidth}
\includegraphics[trim = 0 0 0 0,clip, width = 45mm]{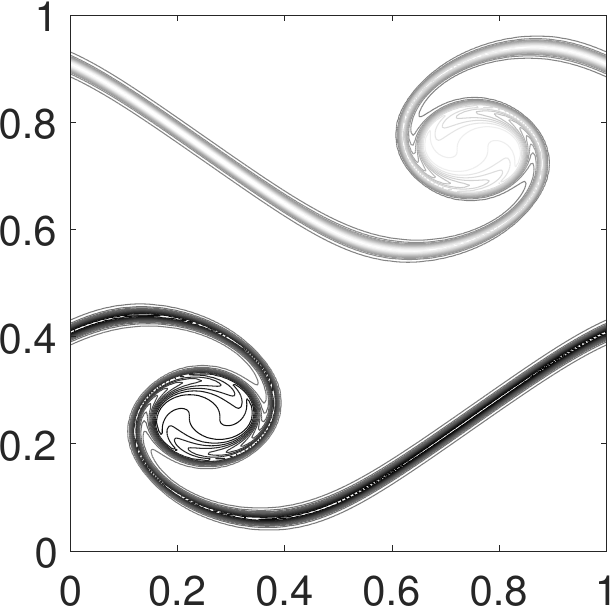}
\caption{$\mbox{Ma}=0.5$,   $L^2 = 256^2$}
\end{subfigure}
\vspace{3mm}
\begin{subfigure}{.32\textwidth}
\includegraphics[trim = 0 0 0 0,clip, width = 45mm]{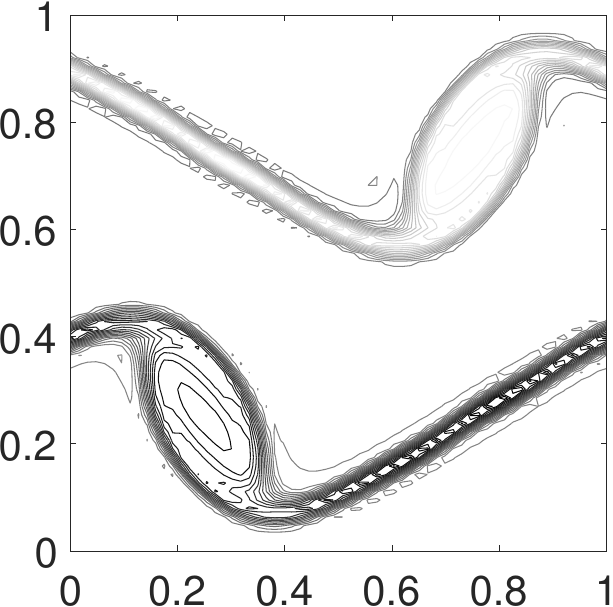}
\caption{$\mbox{Ma}=0.6$,   $L^2 = 64^2$}
\end{subfigure}
\begin{subfigure}{.32\textwidth}
\includegraphics[trim = 0 0 0 0,clip, width = 45mm]{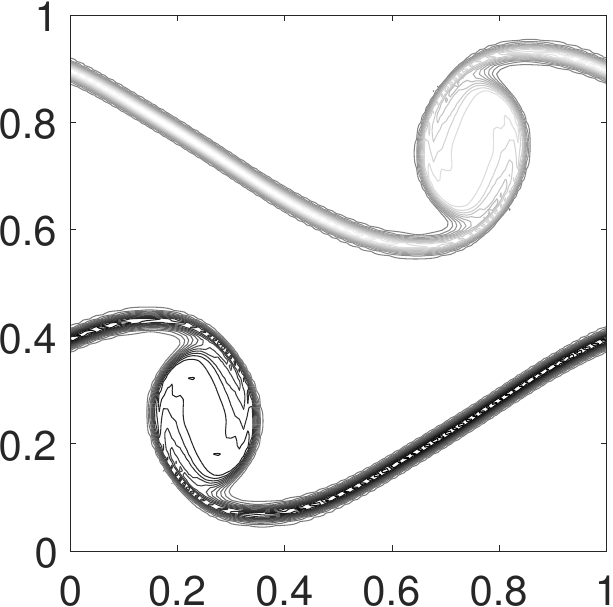}
\caption{$\mbox{Ma}=0.6$,   $L^2 =128^2$}
\end{subfigure}
\begin{subfigure}{.32\textwidth}
\includegraphics[trim = 0 0 0 0,clip, width = 45mm]{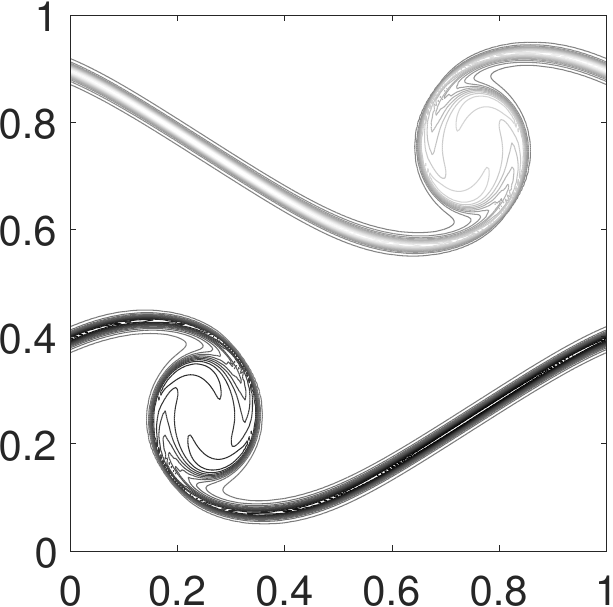}
\caption{$\mbox{Ma}=0.6$,   $L^2 =256^2$}
\end{subfigure}
\caption{Vorticity contours of doubly periodic shear layers that roll up due to an applied perturbation at $t = 1$ for different sets of higher Mach numbers of $0.4$, $0.5$ and $0.6$ (along rows) and at grid resolutions of $64^2$, $128^2$, and $256^2$ (along columns) computed using the Maxwellian equilibria based MCM-LBM.Vorticity contours at $t=1$. It is seen that the FPC-LBM remains stable and no spurious vortices are formed for the same cases of higher Mach numbers and grid resolutions as those shown previously for the MCM-LBM which did not remain stable (see Fig.~\ref{DPSL_MCM2}). This indicates that the FPC-LBM is a more robust collision model when compared to the MCM-LBM for simulations at coarse grid resolutions and at relatively large Mach numbers.}
\label{DPSL_FPC2}
\end{figure}

In this regard, we perform two more comparison studies that attempt this by first using larger Mach numbers and second by using a relatively large increase in bulk viscosity that exceeds the point of having positive effects. The results from the larger Mach number cases of $\mbox{Ma} = 0.4, 0.5$, and $0.6$, and using the same three different grid resolutions as before are presented in Fig.~\ref{DPSL_MCM2} using MCM-LBM and Fig.~\ref{DPSL_FPC2} using FPC-LBM. Evidently, from Fig.~\ref{DPSL_MCM2}, the MCM-LBM simulations show the formation of spurious secondary vortices for all cases of these higher $\mbox{Ma}$ for the coarsest grid resolution of $64^2$, and these artifacts progressively become more prominent as $\mbox{Ma}$ increases. On the other hand, the results from the FPC-LBM (see Fig.~\ref{DPSL_FPC2}) show no such spurious secondary vortices for those same cases, which indicates that the FPC-LBM can sustain stable simulations without any noticeable artifacts even at large Mach numbers and coarse grid resolutions.

\begin{figure}[ht]
\centering
\vspace{3mm}
\begin{subfigure}{.32\textwidth}
\includegraphics[trim = 0 0 0 0,clip, width = 45mm]{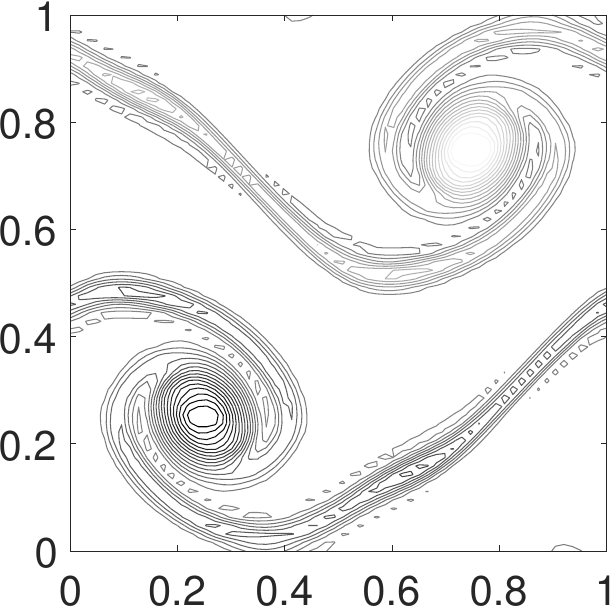}
\caption{$\mbox{Ma}=0.05$,   $L^2 = 64^2$}
\end{subfigure}
\begin{subfigure}{.32\textwidth}
\includegraphics[trim = 0 0 0 0,clip, width = 45mm]{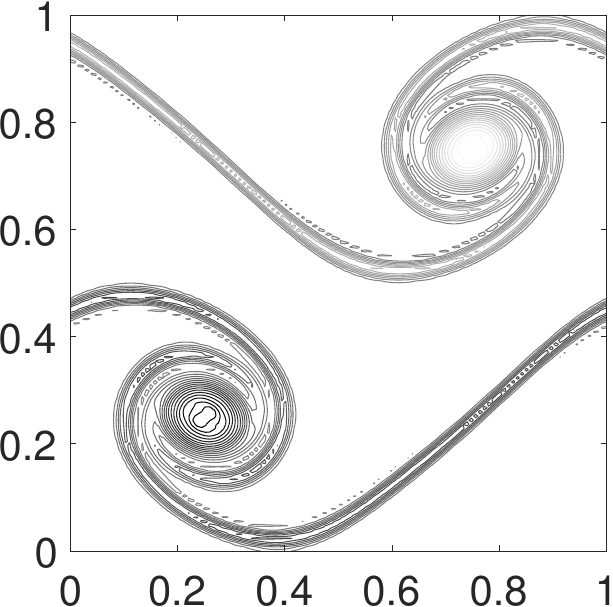}
\caption{$\mbox{Ma}=0.05$,   $L^2 = 128^2$}
\end{subfigure}
\begin{subfigure}{.32\textwidth}
\includegraphics[trim = 0 0 0 0,clip, width = 45mm]{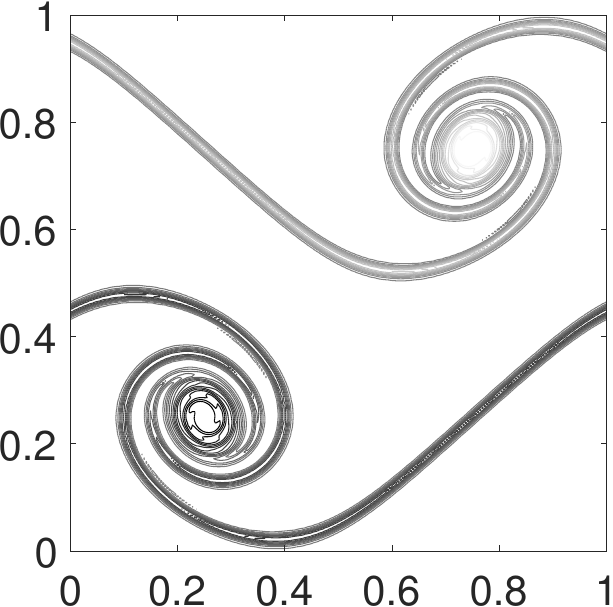}
\caption{$\mbox{Ma}=0.05$,   $L^2 = 256^2$}
\end{subfigure}
\vspace{3mm}
\begin{subfigure}{.32\textwidth}
\includegraphics[trim = 0 0 0 0,clip, width = 45mm]{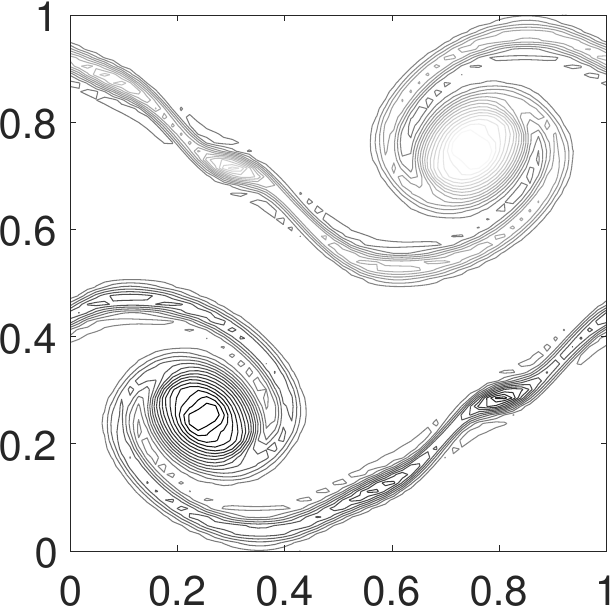}
\caption{$\mbox{Ma}=0.2$,   $L^2 = 64^2$}
\end{subfigure}
\begin{subfigure}{.32\textwidth}
\includegraphics[trim = 0 0 0 0,clip, width = 45mm]{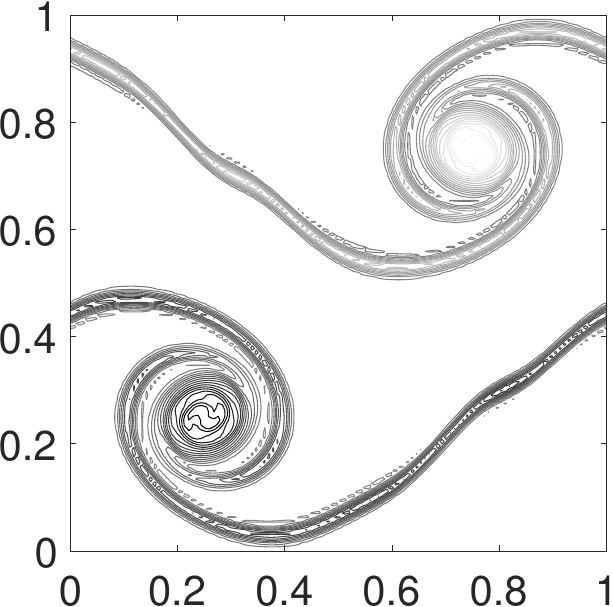}
\caption{$\mbox{Ma}=0.2$,   $L^2 = 128^2$}
\end{subfigure}
\begin{subfigure}{.32\textwidth}
\includegraphics[trim = 0 0 0 0,clip, width = 45mm]{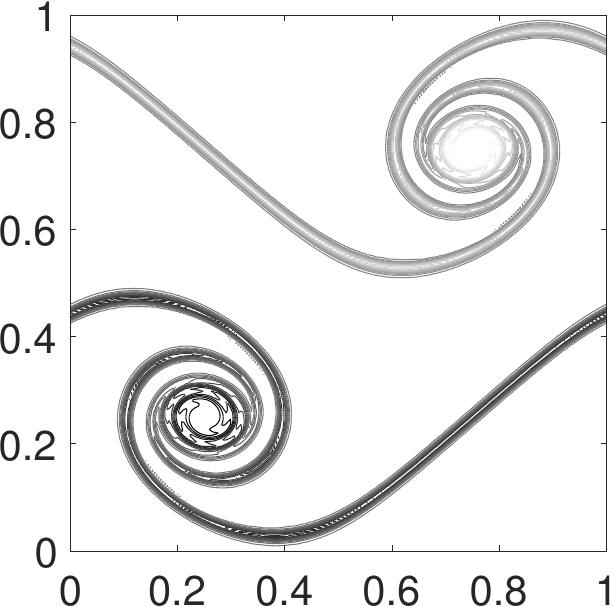}
\caption{$\mbox{Ma}=0.2$,   $L^2 = 256^2$}
\end{subfigure}
\vspace{3mm}
\begin{subfigure}{.32\textwidth}
\includegraphics[trim = 0 0 0 0,clip, width = 45mm]{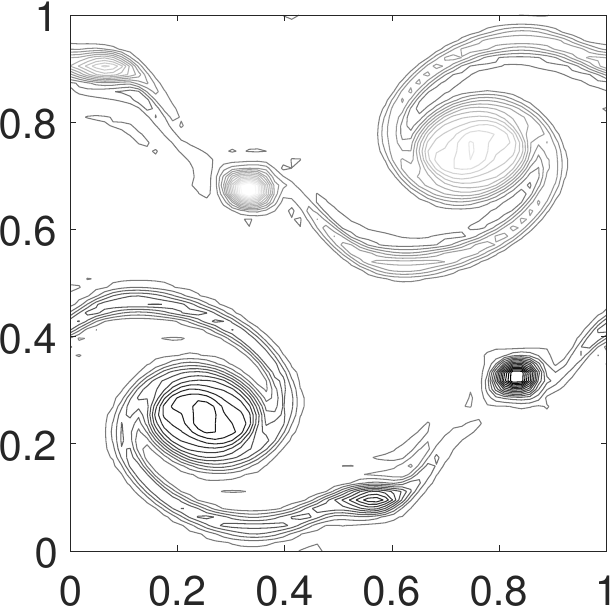}
\caption{$\mbox{Ma}=0.3$,   $L^2 = 64^2$}
\end{subfigure}
\begin{subfigure}{.32\textwidth}
\includegraphics[trim = 0 0 0 0,clip, width = 45mm]{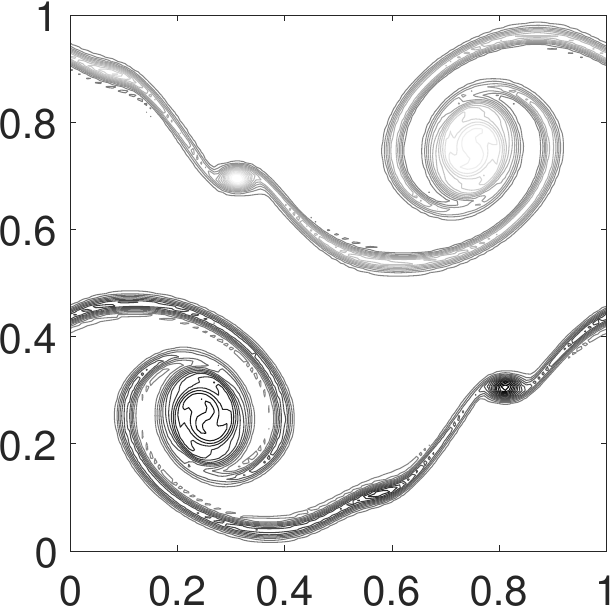}
\caption{$\mbox{Ma}=0.3$,   $L^2 =128^2$}
\end{subfigure}
\begin{subfigure}{.32\textwidth}
\includegraphics[trim = 0 0 0 0,clip, width = 45mm]{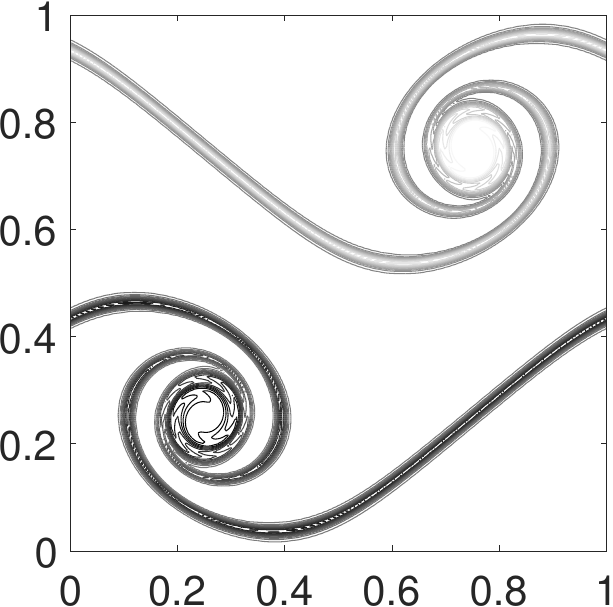}
\caption{$\mbox{Ma}=0.3$,   $L^2 =256^2$}
\end{subfigure}
\caption{Vorticity contours of doubly periodic shear layers that roll up due to an applied perturbation at $t = 1$ for different sets of lower Mach numbers of $0.05$, $0.2$ and $0.3$ (along rows) and at grid resolutions of $64^2$, $128^2$, and $256^2$ (along columns) at an extremely large bulk viscosity by setting the relaxation parameter associated with the bulk viscosity to $\omega_3 = 0.35$ and computed using the Maxwellian equilibria based MCM-LBM.  It is seen that MCM-LBM becomes unstable under an extreme increase in bulk viscosity, especially for coarse grid resolutions which become progressively worse as the Mach number is increased. By increasing the relaxation time
$\tau_3 = 1/\omega_3$ associated with the bulk viscosity to $2.857$  from being $\sim 1.0$ or smaller, the range of beneficial limits is exceeded, and instead the simulations begin to numerically destabilize with MCM-LBM.}
\label{DPSL_MCM3}
\end{figure}

\begin{figure}[ht]
\centering
\vspace{3mm}
\begin{subfigure}{.32\textwidth}
\includegraphics[trim = 0 0 0 0,clip, width = 45mm]{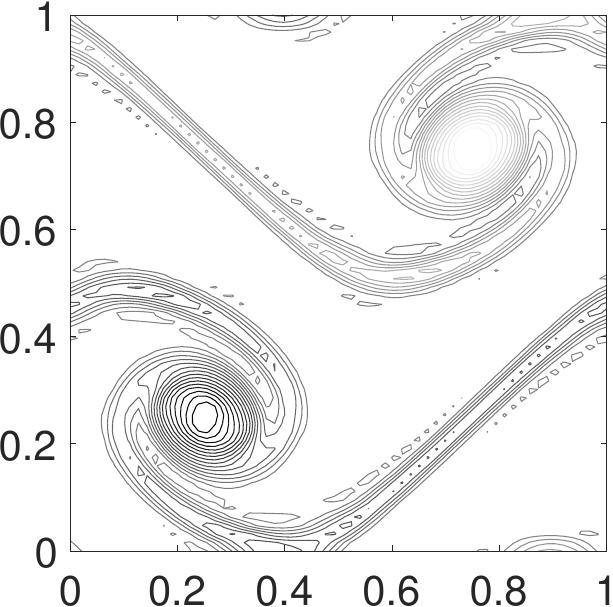}
\caption{$\mbox{Ma}=0.05$,   $L^2 = 64^2$}
\end{subfigure}
\begin{subfigure}{.32\textwidth}
\includegraphics[trim = 0 0 0 0,clip, width = 45mm]{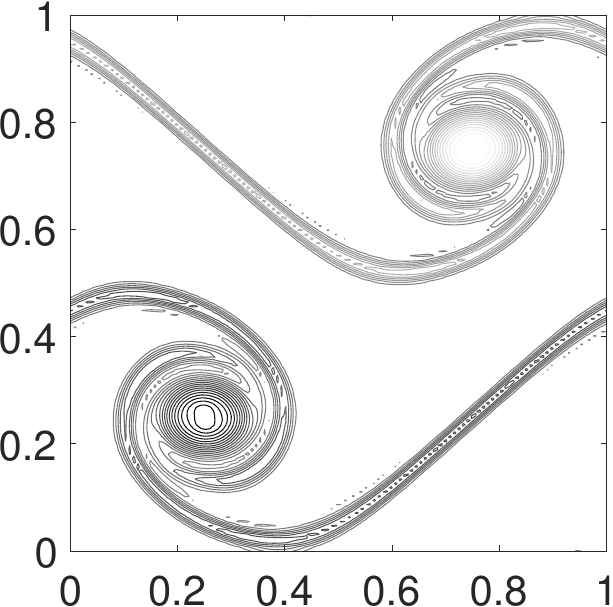}
\caption{$\mbox{Ma}=0.05$,   $L^2 = 128^2$}
\end{subfigure}
\begin{subfigure}{.32\textwidth}
\includegraphics[trim = 0 0 0 0,clip, width = 45mm]{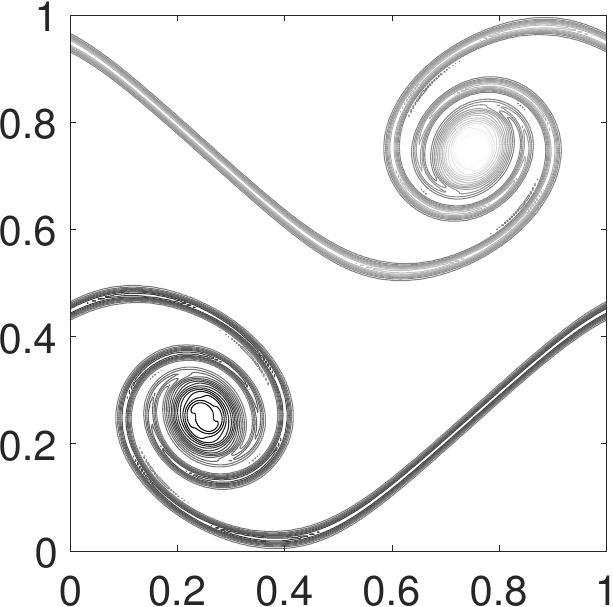}
\caption{$\mbox{Ma}=0.05$,   $L^2 = 256^2$}
\end{subfigure}
\vspace{3mm}
\begin{subfigure}{.32\textwidth}
\includegraphics[trim = 0 0 0 0,clip, width = 45mm]{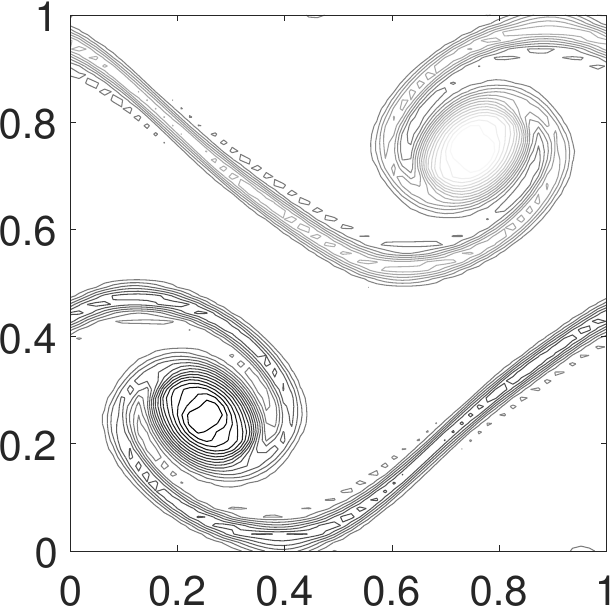}
\caption{$\mbox{Ma}=0.2$,   $L^2 = 64^2$}
\end{subfigure}
\begin{subfigure}{.32\textwidth}
\includegraphics[trim = 0 0 0 0,clip, width = 45mm]{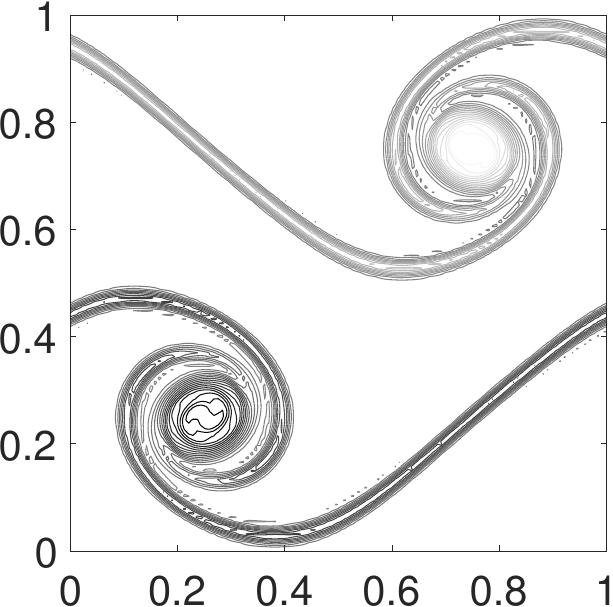}
\caption{$\mbox{Ma}=0.2$,   $L^2 = 128^2$}
\end{subfigure}
\begin{subfigure}{.32\textwidth}
\includegraphics[trim = 0 0 0 0,clip, width = 45mm]{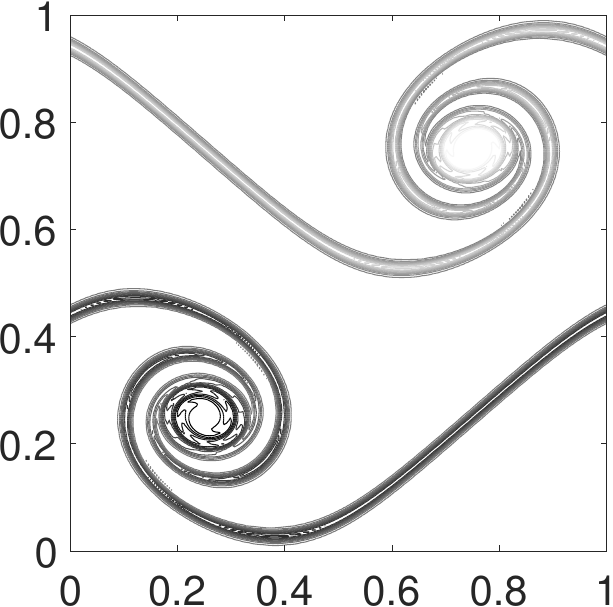}
\caption{$\mbox{Ma}=0.2$,   $L^2 = 256^2$}
\end{subfigure}
\vspace{3mm}
\begin{subfigure}{.32\textwidth}
\includegraphics[trim = 0 0 0 0,clip, width = 45mm]{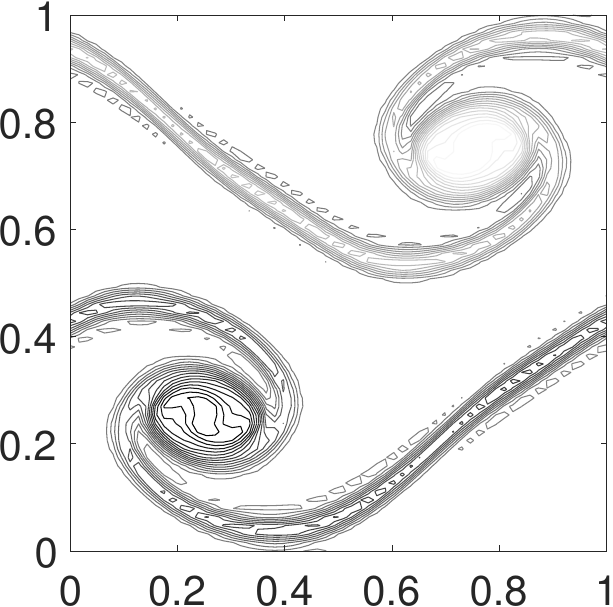}
\caption{$\mbox{Ma}=0.3$,   $L^2 = 64^2$}
\end{subfigure}
\begin{subfigure}{.32\textwidth}
\includegraphics[trim = 0 0 0 0,clip, width = 45mm]{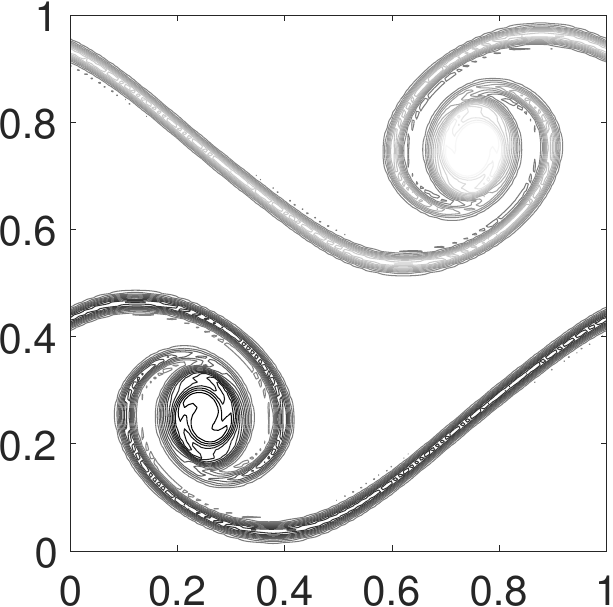}
\caption{$\mbox{Ma}=0.3$,   $L^2 =128^2$}
\end{subfigure}
\begin{subfigure}{.32\textwidth}
\includegraphics[trim = 0 0 0 0,clip, width = 45mm]{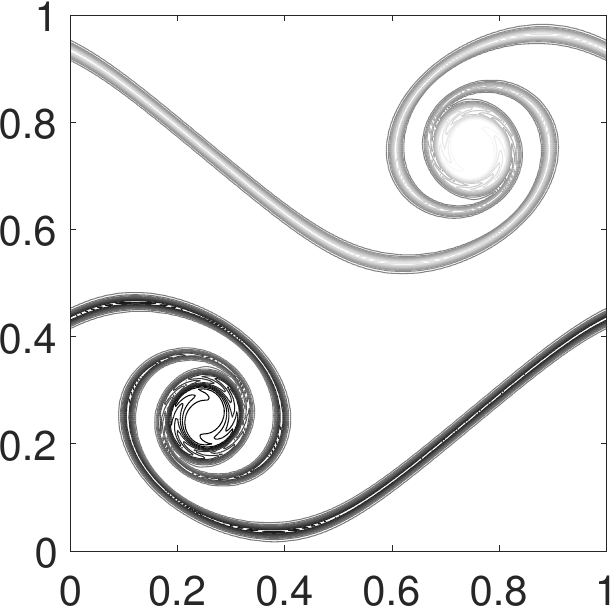}
\caption{$\mbox{Ma}=0.3$,   $L^2 =256^2$}
\end{subfigure}
\caption{Vorticity contours of doubly periodic shear layers that roll up due to an applied perturbation at $t = 1$ for different sets of lower Mach numbers of $0.05$, $0.2$ and $0.3$ (along rows) and at grid resolutions of $64^2$, $128^2$, and $256^2$ (along columns) at an extremely large bulk viscosity by setting the relaxation parameter associated with the bulk viscosity to $\omega_3 = 0.35$ and computed using the Fokker-Planck equilibria based FPC-LBM. It is seen that the FPC-LBM remains stable even for the cases with an extreme increase in bulk viscosity shown previously, where the MCM-LBM did not remain stable  (see Fig.~\ref{DPSL_MCM3}). This indicates that the FPC-LBM is numerically more stable when compared to the MCM-LBM in such cases as well.}
\label{DPSL_FPC3}
\end{figure}

Lastly, we consider cases in the lower Mach number ranges again ($\mbox{Ma}=0.05$, $0.2$ and $0.3$) and with the same sets of grid resolutions as before, but at an extremely large bulk viscosity by setting the relaxation parameter associated with the bulk viscosity to $\omega_3 = 0.35$. Figure~\ref{DPSL_MCM3} shows the vorticity contours at $t = 1$ computed using MCM-LBM. In general, increasing the bulk viscosity dampens any spurious pressure waves; however, by increasing the relaxation time $\tau_3 = 1/\omega_3$ associated with the bulk viscosity to $2.857$  from being $\sim 1.0$ or smaller, the range of beneficial limits is exceeded, and instead the simulations begin to numerically destabilize with MCM-LBM with the incipient spurious secondary vortices at the coarsest resolution even with the usage of lower range of Mach numbers (see Fig.~\ref{DPSL_MCM3}); by contrast, as shown in Fig.~\ref{DPSL_FPC3} using the FPC-LBM does not result in spurious vortices for these cases and the simulations remain stable. This again indicates that the FPC-LBM is a more robust central moment LB scheme than the MCM-LBM.

\clearpage

\subsection{Three-dimensional lid-driven cubic cavity flow: An accuracy study}
Next, let's validate the 3D FPC-LBM by performing simulations of flow inside a cubic cavity driven by the motion of one of its lids and compare the results with the benchmark numerical data available in the literature~\cite{ku1987pseudospectral,shu2003numerical}. This problem is set up with a cubic grid domain of $ L_x \times L_y \times L_z = 150 \times150 \times150$, with a lid, located at $y = L_y$, that moves with a constant velocity, $U_o$, in the $x$ direction. The characteristic plate velocity is set by specifying the Mach number to be relatively small, with $\mbox{Ma} = U_o / c_s = 0.1$, so that the flow is maintained in the weakly compressible range; the choice of the Reynolds number, defined by $\mbox{Re}=U_o L_0/\nu$ with $L_0 = L_x = L_y = L_z$, sets up the fluid viscosity $\nu$. All of the other walls are stationary, and the standard half-way bounce back boundary condition is applied to all boundaries in order to enforce the no slip condition on each wall, including a momentum correction to the moving wall.

\begin{figure}[!ht]
\centering
\begin{subfigure}{.32\textwidth}
\includegraphics[trim = 40 0 30 0, clip, width = 55mm]{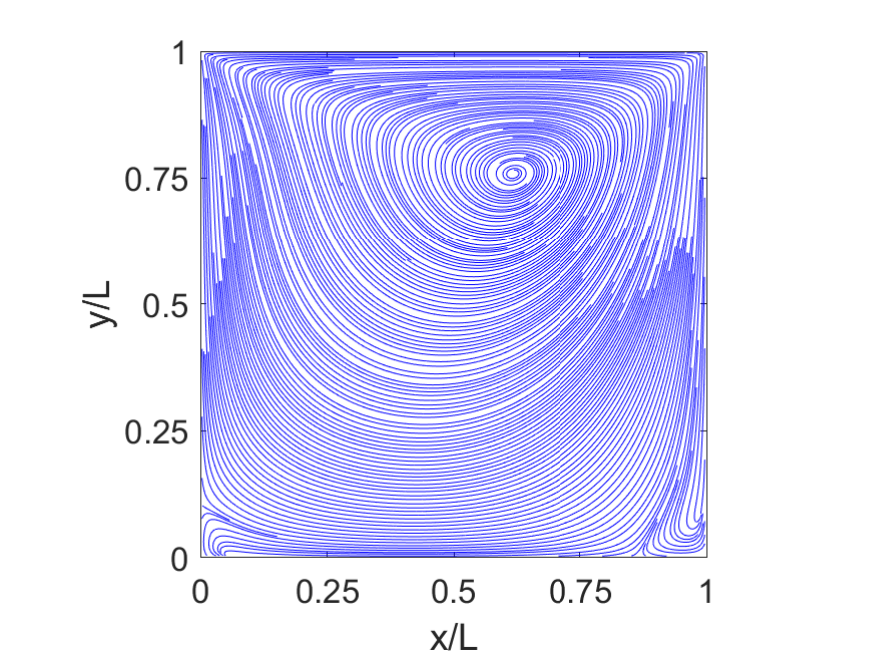}
\caption{$\mbox{Re} = 100$}
\end{subfigure}
\begin{subfigure}{.32\textwidth}
\includegraphics[trim = 40 0 30 0, clip, width =55mm]{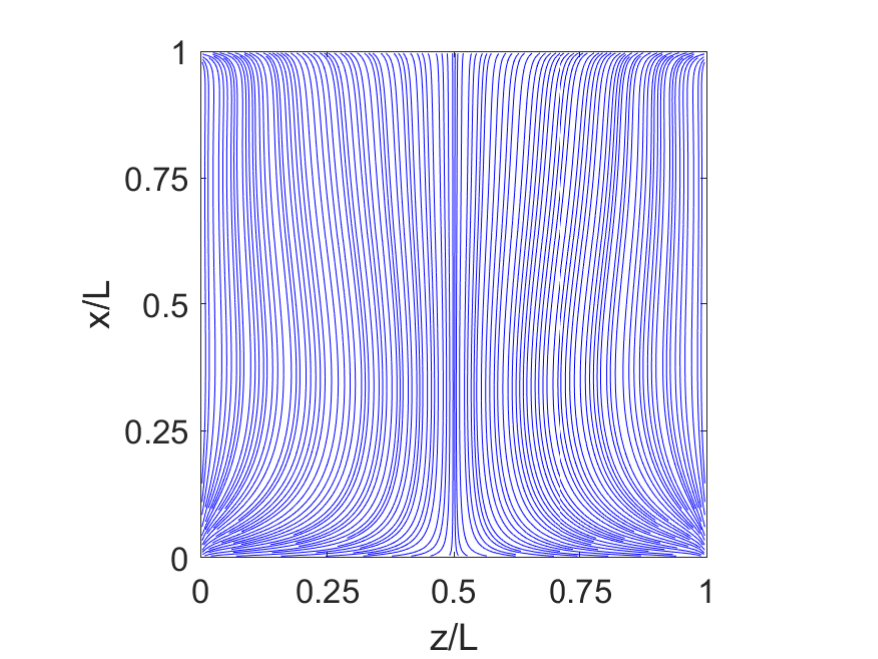}
\caption{$\mbox{Re} = 100$}
\end{subfigure}
\begin{subfigure}{.32\textwidth}
\includegraphics[trim = 40 0 30 0, clip, width = 55mm]{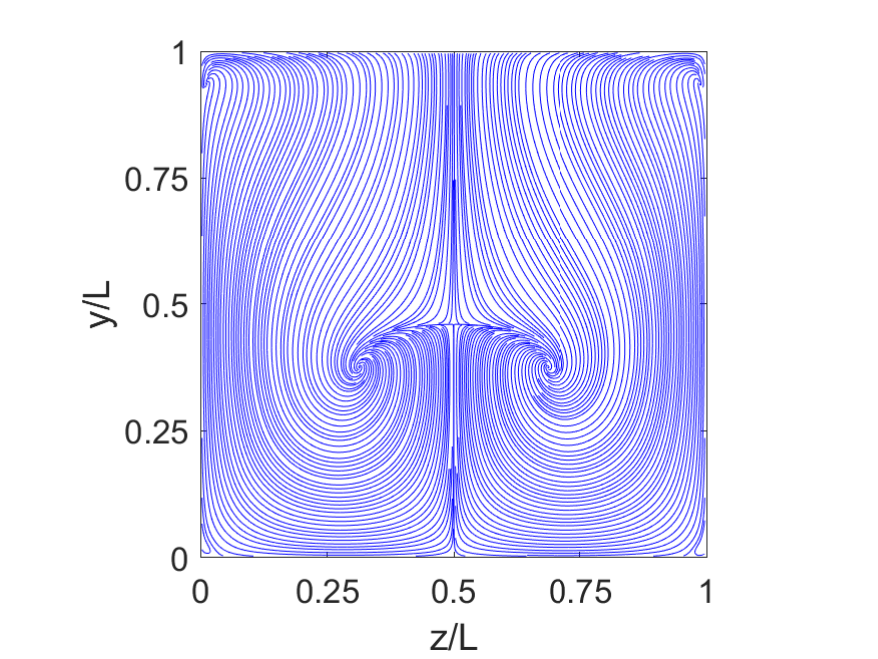}
\caption{$\mbox{Re} = 100$}
\end{subfigure}
\begin{subfigure}{.32\textwidth}
\includegraphics[trim =40 0 30 0, clip, width = 55mm]{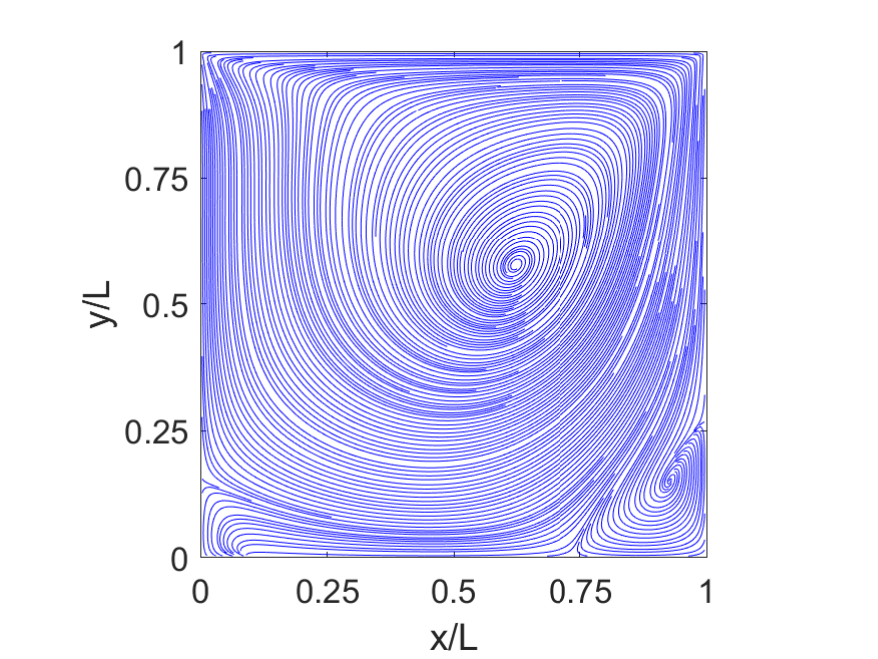}
\caption{$\mbox{Re} = 400$}
\end{subfigure}
\begin{subfigure}{.32\textwidth}
\includegraphics[trim = 40 0 30 0, clip, width = 55mm]{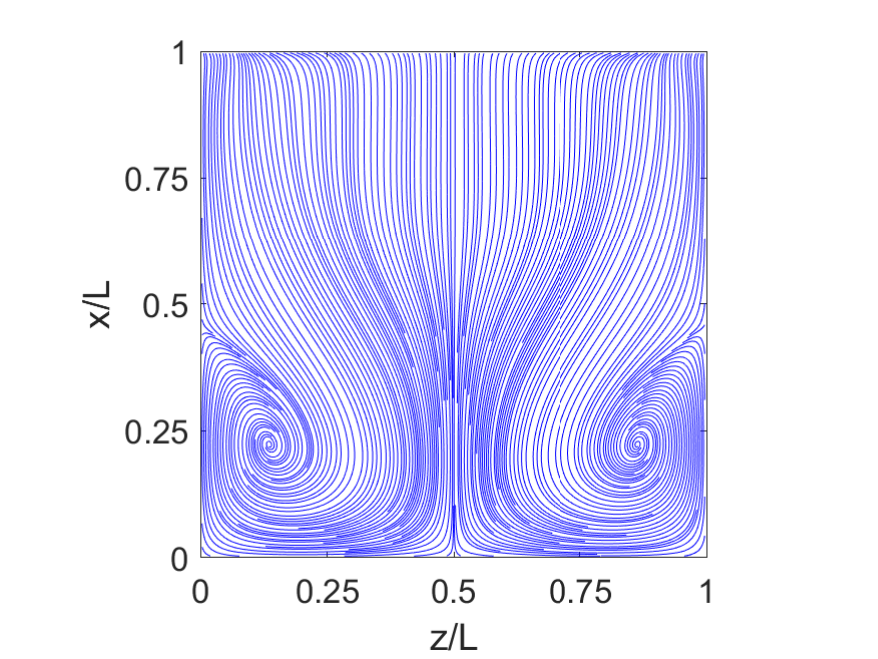}
\caption{$\mbox{Re} = 400$}
\end{subfigure}
\begin{subfigure}{.32\textwidth}
\includegraphics[trim = 40 0 30 0, clip, width = 55mm]{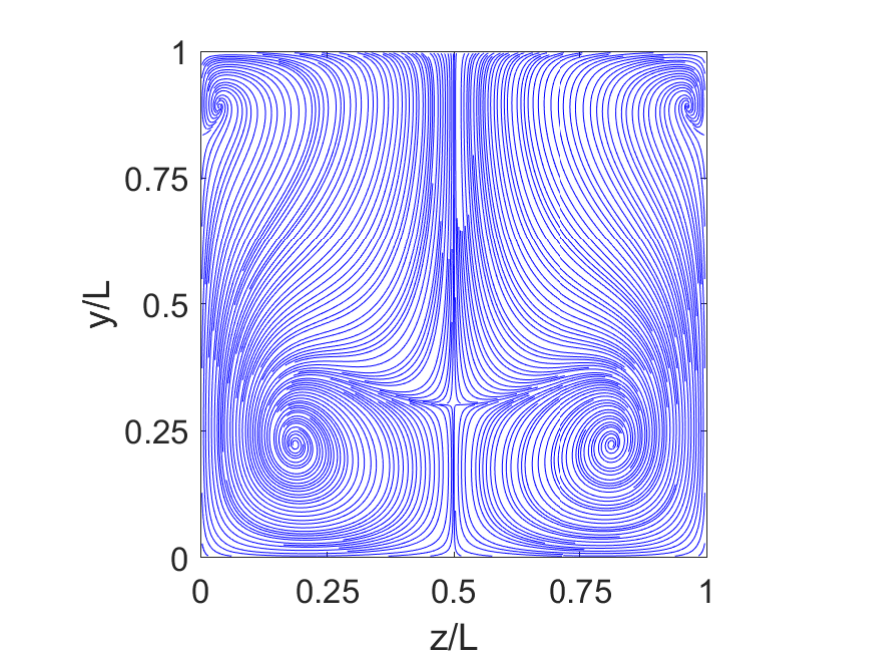}
\caption{$\mbox{Re} = 400$}
\end{subfigure}

\begin{subfigure}{.32\textwidth}
\includegraphics[trim =40 0 30 0, clip, width = 55mm]{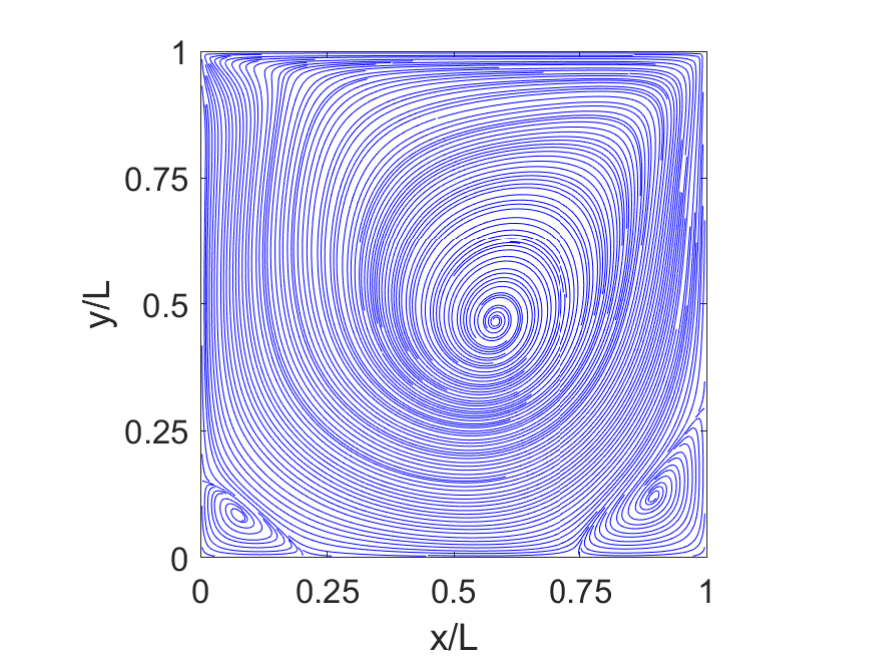}
\caption{$\mbox{Re} = 1000$}
\end{subfigure}
\begin{subfigure}{.32\textwidth}
\includegraphics[trim = 40 0 30 0, clip, width = 55mm]{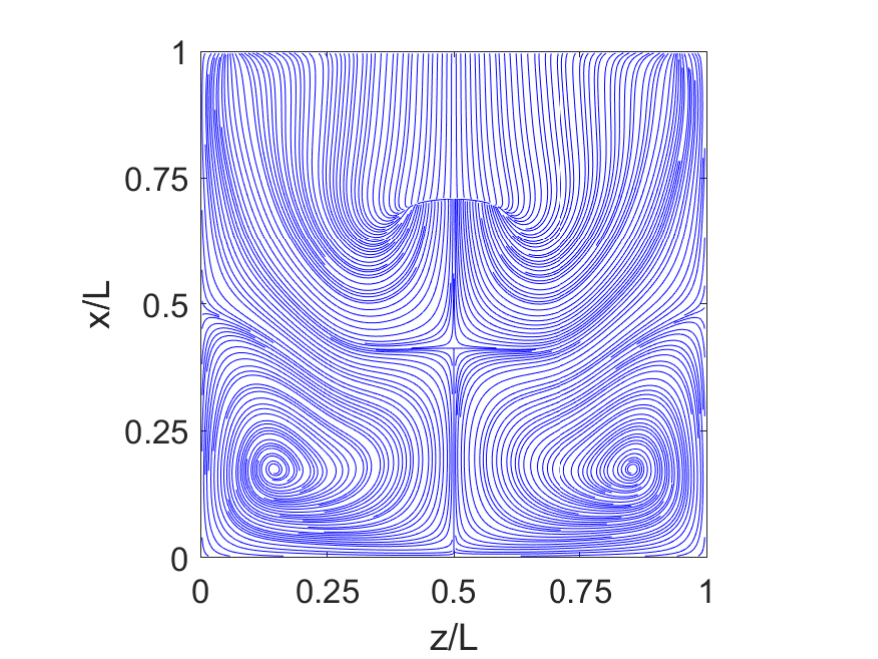}
\caption{$\mbox{Re} = 1000$}
\end{subfigure}
\begin{subfigure}{.32\textwidth}
\includegraphics[trim = 40 0 30 0, clip, width = 55mm]{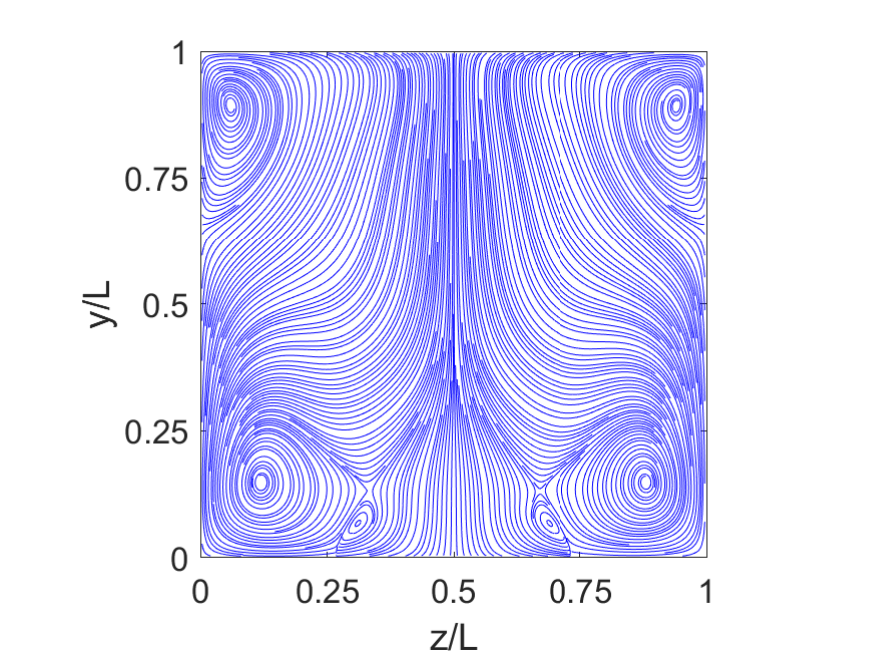}
\caption{$\mbox{Re} = 1000$}
\end{subfigure}
\caption{Streamlines for three-dimensional lid-driven cubic cavity flow at Reynolds numbers of $\mbox{Re}= 100, 400$ and $1000$ computed using the FPC-LBM along the $z=0.5L$ centerplane shown in (a), (d), and (g), the $y = 0.5L$ centerplane shown in (b), (e) and (h), and the $x = 0.5L$ centerplane shown in (c), (f) and (i).}
\label{3D_LDC}
\end{figure}

\begin{figure}[!ht]
\centering
\begin{subfigure}{.48\textwidth}
\includegraphics[trim = 0 0 0 0, clip, width = 65mm]{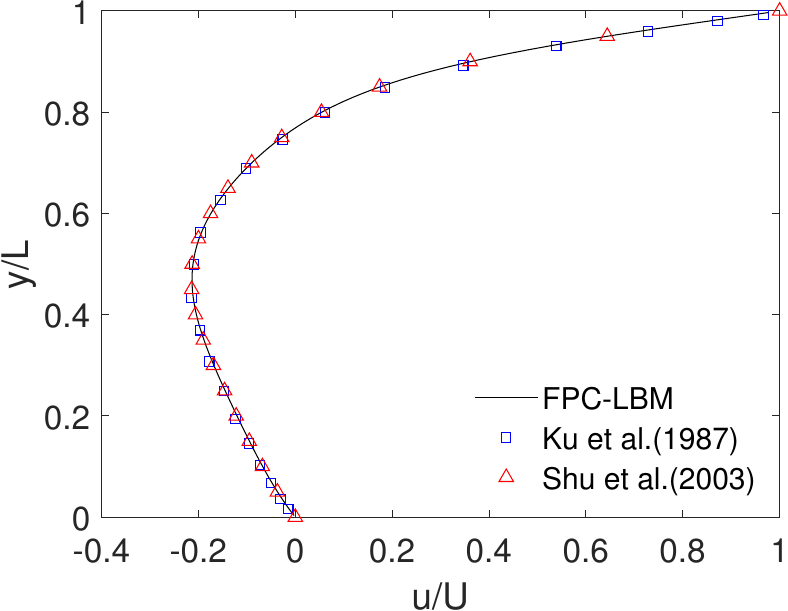}
\caption{$\mbox{Re} = 100$}
\end{subfigure}
\begin{subfigure}{.48\textwidth}
\includegraphics[trim = 0 0 0 0, clip, width =65mm]{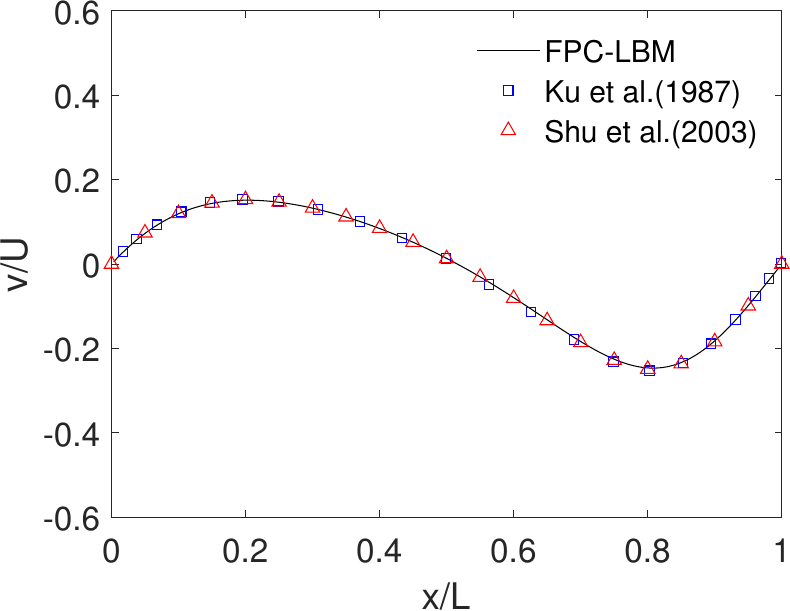}
\caption{$\mbox{Re} = 100$}
\end{subfigure}

\begin{subfigure}{.48\textwidth}
\includegraphics[trim = 0 0 0 0, clip, width = 65mm]{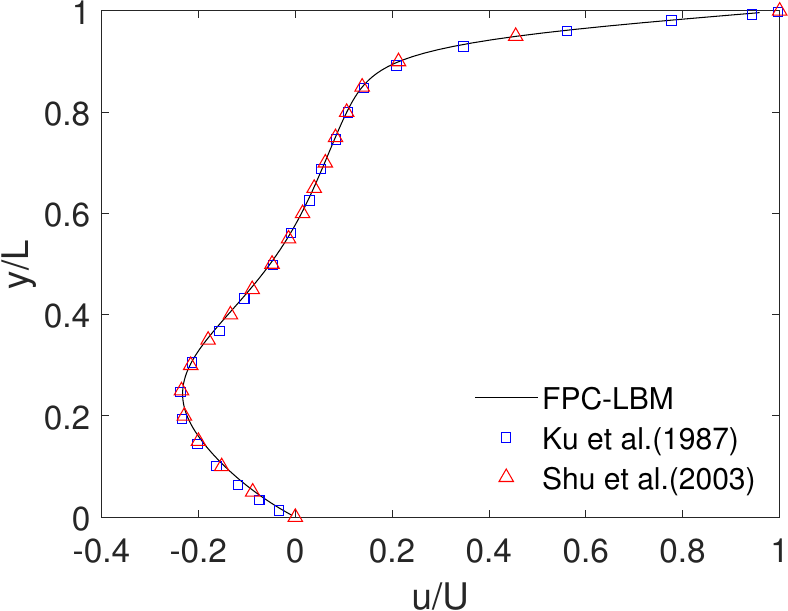}
\caption{$\mbox{Re} = 400$}
\end{subfigure}
\begin{subfigure}{.48\textwidth}
\includegraphics[trim = 0 0 0 0, clip, width =65mm]{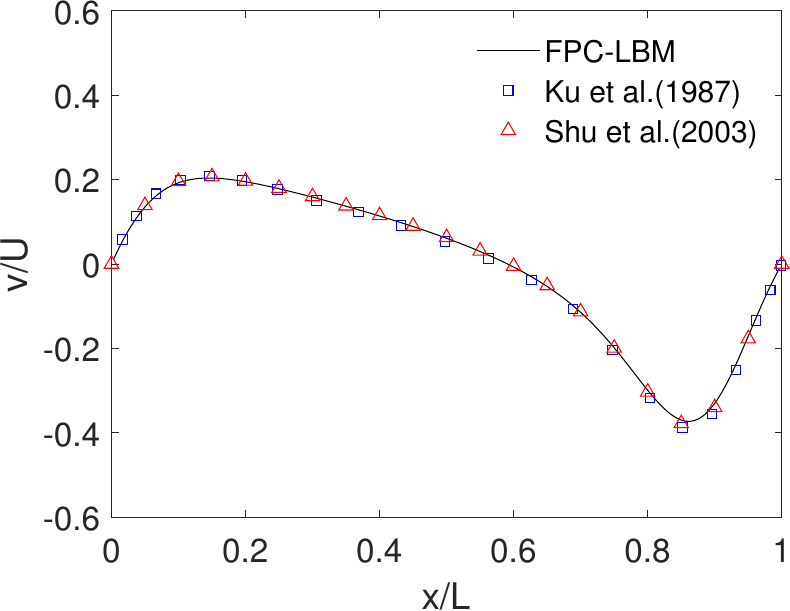}
\caption{$\mbox{Re} = 400$}
\end{subfigure}

\begin{subfigure}{.48\textwidth}
\includegraphics[trim = 0 0 0 0, clip, width = 65mm]{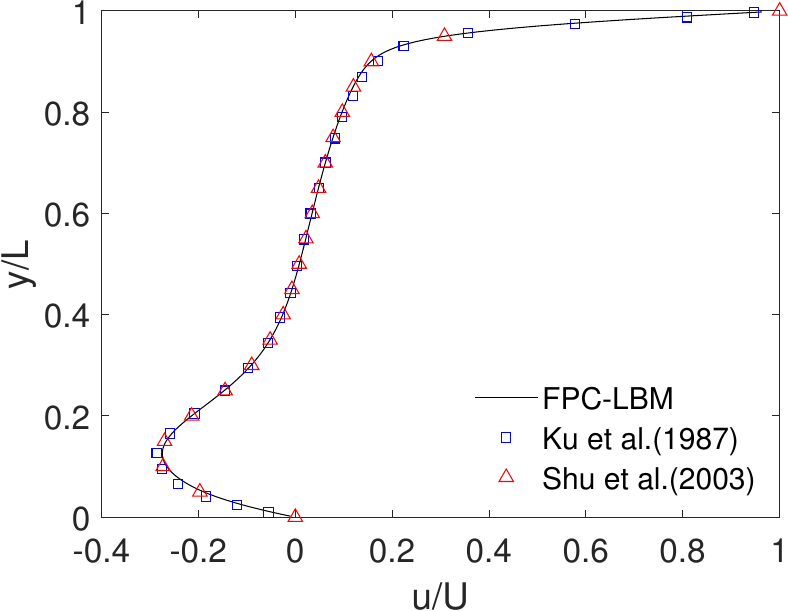}
\caption{$\mbox{Re} = 1000$}
\end{subfigure}
\begin{subfigure}{.48\textwidth}
\includegraphics[trim = 0 0 0 0, clip, width =65mm]{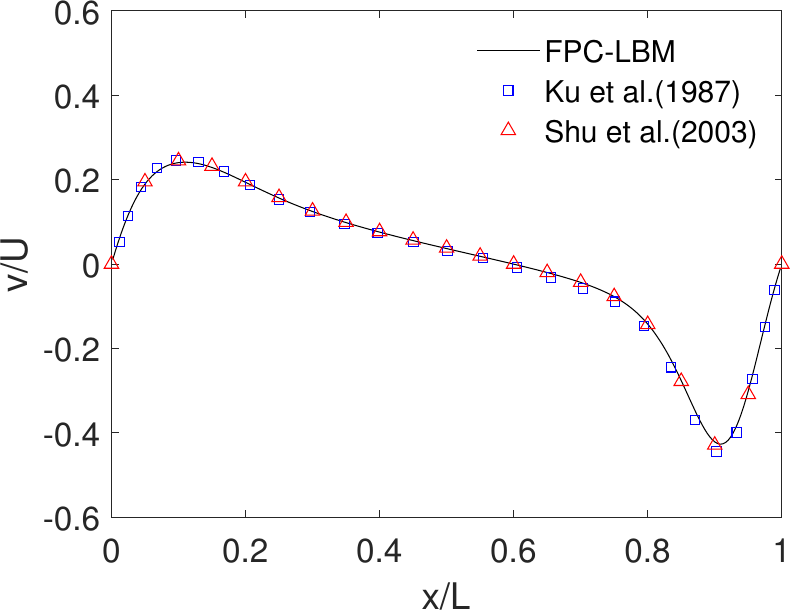}
\caption{$\mbox{Re} = 1000$}
\end{subfigure}
\caption{Comparisons of the horizontal velocity $u/U$ along the vertical coordinate $y/L$ at $x=0.5L$ and $z=0.5L$ (left) and vertical velocity $v/U$ along the horizontal coordinate $x/L$ at $y = 0.5L$ and $z=0.5L$ (right) in a three-dimensional cubic cavity flow
at different Reynolds numbers computed using the FPC-LBM with the reference results of Ku \emph{et al} (1987)~\cite{ku1987pseudospectral} and Shu \emph{et al} (2003)~\cite{shu2003numerical}. (a) and (b) $\mbox{Re}=100$,   (c) and (d) $\mbox{Re}=400$, (e) and (f) $\mbox{Re}=1000$.}
\label{3D_LDC_CL}
\end{figure}

Then, we simulate the three-dimensional lid-driven cavity flow for Reynolds numbers of $\mbox{Re} = 100, 400, 1000$. Figure~\ref{3D_LDC} shows the computed streamlines at these Reynolds numbers along three different centerplanes. As $\mbox{Re}$ increases, the flow patterns become progressively more complex with additional vortical structures near the corner edges, and the observed patterns are consistent with those reported in~\cite{shu2003numerical,ku1987pseudospectral}. In addition, comparisons of the horizontal and vertical velocity profiles computed using the FPC-LBM along different directions in the centerplanes for the above three $\mbox{Re}$ with the prior numerical results obtained from Navier-Stokes solvers~\cite{ku1987pseudospectral,shu2003numerical} are presented in Fig.~\ref{3D_LDC_CL}. Very good quantitative agreement is seen for all the cases considered. These results give an indication that the 3D FPC-LBM accurately simulates flow fields with rather complex structures.

\begin{figure}[!ht]
\centering 
\includegraphics[trim = 0 0 0 0, clip, width =75mm]{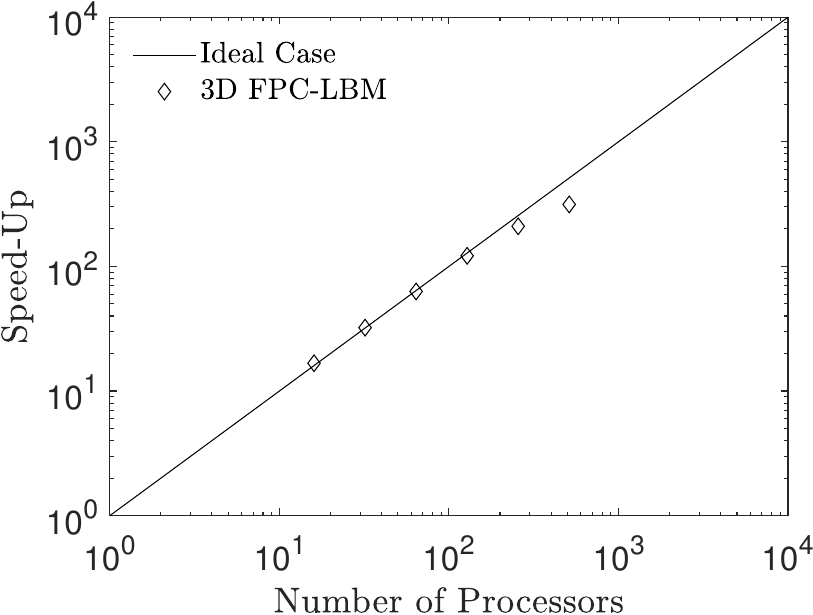}
\caption{Parallel performance of the MPI implementation of our 3D FPC-LBM for lid-driven cubic cavity flow simulations using a grid resolution of $150\times150\times150$ in our in-house computer cluster.}
\label{fig:speedup}
\end{figure}

One of the major advantages of the LBM is its natural parallelization property. We have parallelized our implementation of the 3D FPC-LBM using the Message Passing Interface (MPI) library via a domain decomposition approach. We then performed simulations of
the cubic cavity flow resolved with $150\times150\times150$ using $p= 16, 32, 64, 128, 256$, $512$ processors in our in-house computer cluster (see https://ccm-docs.readthedocs.io/en/latest/alderaan). Figure~\ref{fig:speedup} shows the parallel performance of our 3D FPC-LBM. Evidently, near-linear speed-up is seen, which is ideal for simulating large domain size problems effectively. We will exploit this feature for turbulent channel flow simulations discussed later.

\subsection{Three-dimensional lid-driven cubic cavity flow: A stability study}
In order to further illustrate the numerical properties of the 3D FPC-LBM in perspective with other existing LB collision models, next we investigate their stability characteristics in the simulation of three-dimensional, lid-driven cubic cavity flow. If $\omega_j$, where $j = 110, 101, 011, 2d1$, and $2d2$ represent the relaxation parameters of the second order moments that determine the fluid viscosity $\nu$ (see Eq.~(\ref{eqn:transportcoeff3DFPCLBM})), for brevity in discussion, we let them to be related to a relaxation time $\tau$ via $\tau = 1/\omega_j$. It is well known that as $\tau$ approaches $1/2$ or equivalently, as the viscosity becomes relatively small, the LB schemes are susceptible to numerical instabilities that can cause any errors in simulations to grow exponentially large, and different collision models can exhibit such behavior to a different degree depending on when a certain threshold viscosity or the Reynolds number is reached before they become unstable. Here, we leverage this feature to form a description for the numerical stability properties of the FPC-LBM as compared to other collision models.

\begin{figure}[ht!]
\centering
\includegraphics[trim = 20 0 50 20, clip, width =130mm]{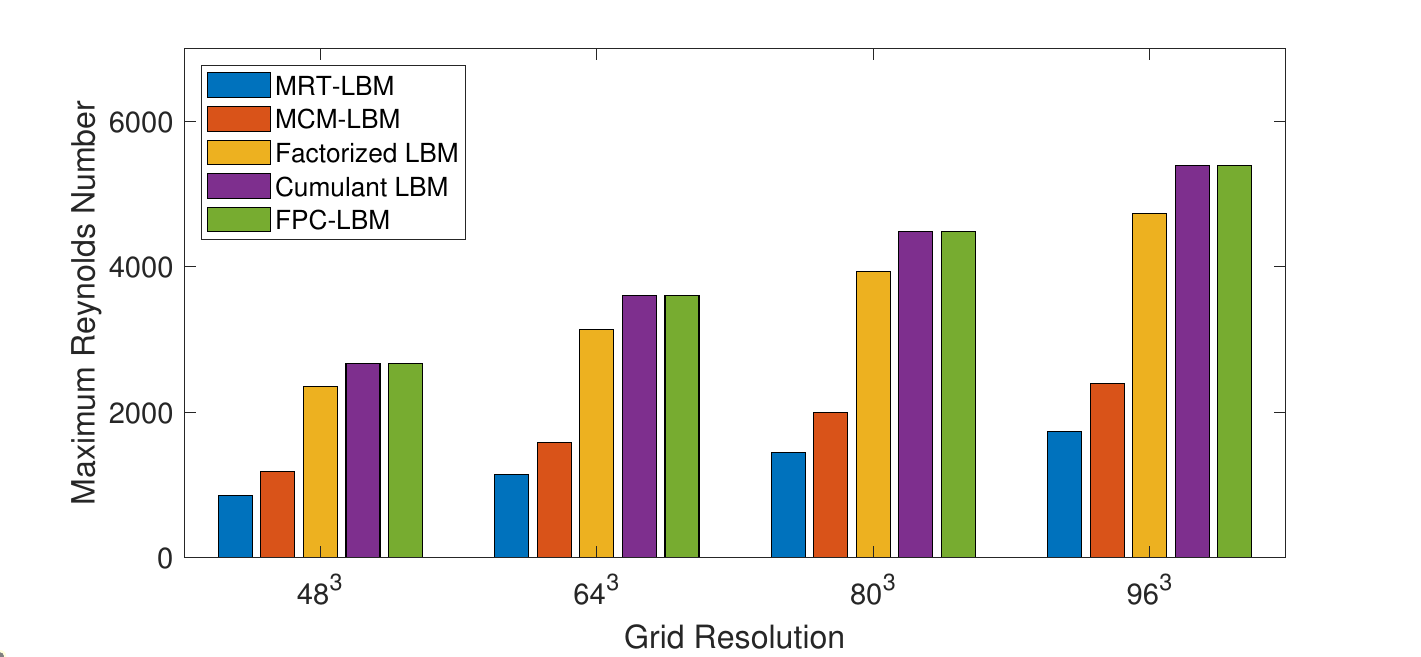}
\caption{The maximum possible Reynolds number for which the three-dimensional lid-driven cubic cavity flow simulations remain stable at different grid resolutions of $48^3$, $64^3$, $80^3$ and $96^3$ for different types of LB collision models -- raw moments-based MRT-LBM, Maxwellian central moments-based MCM-LBM, factorized central moments-based Factorized LBM, cumulant LBM, and Fokker-Planck central moments-based FPC-LBM.}
\label{LDC_Stability}
\end{figure}

We use the three-dimensional lid driven cavity flow simulation as a prototypical example in this regard (see e.g., Ref.~\cite{adam2019numerical} for a recent such simulation study using the cascaded LBM and other collision models). In this regard, we start with the case that we know is numerically unstable, namely $\tau=0.5$, which has a corresponding Reynolds number, $\mbox{Re}=\infty$, when the characteristic velocity and length scales are fixed, as the relaxation time $\tau$ is related to the fluid viscosity, $\nu$. We then systematically increase the relaxation time until the simulation can remain stable, and report the resulting threshold Reynolds number associated with that relaxation time as being the maximum possible Reynolds number for which the simulation is stable. Specifically, we increase the relaxation time as $\tau_{new} = \tau_{old} + \delta$, where $\delta$ is the amount of increase for each case after the previous simulation has become unstable. Here, we use $\delta= 5\times10^{-5}$, and we consider any simulation that reaches 500000 time steps to be stable. Furthermore, after we find the minimum value for the relaxation time that allows the simulation to remain stable for 500000 time steps, we then validate that every value above this minimum also remains stable for at least ten more increments in it. At this point, we define the simulation as having crossed the threshold from unstable to stable, and we note the Reynolds number at the edge of that threshold. We perform this study using four different grid resolutions and compare the results from five different collision models. The grid resolutions we consider for the cubic cavity of dimension $L_x \times L_y \times L_z = L^3$,  are $L = 48, 64, 80$, and $96$ and the plate velocity $U_0$ is based on the Mach number via $U_0 =\mbox{Ma}\times c_s$ by fixing $\mbox{Ma} = 0.2$ in all cases. We then compare the results of these cases across the MRT-LBM, MCM-LBM, factorized LBM, cumulant LBM, and the FPC-LBM approaches, but do not consider the SRT-LBM as it is already known to be the least stable approach in LBM in general. As a result, this study gives an indication of the numerical stability features of the FPC-LBM as compared to other collision models. Figure~\ref{LDC_Stability} shows the maximum possible Reynolds number achieved using different collision models at various resolutions. In general, the central moment-based schemes (MCM-LBM, factorized LBM, and FPC-LBM) are more stable compared to the raw moment-based MRT-LBM. Among the central moment-based schemes, MCM-LBM is the least stable and the FPC-LBM consistently outperforms it by a factor greater than 2 for the maximum threshold $\mbox{Re}$, and the latter is more stable than the ad hoc factorized LBM as well. As such, we find that the FPC-LBM has stability characteristics that are very close to those observed for the cumulant LBM (see Fig.~\ref{LDC_Stability}).

\subsection{Orthogonal Crossing Shear Waves: A numerical hyperviscosity study}

This benchmark is applied in effort to asses the presence of a numerical artifact known as hyperviscosity that has been observed in some of the LBM formulations, especially those that relax different moments at different rates during collision,~\cite{geier2009factorized, geier2015cumulant}. More specifically, when computing flows with relatively very small fluid viscosities, there can exist large disparities between the relaxation rates of the second order moments compared to the higher order moments. In turn, depending on the choice of equilibria in such cases, the contributions from such higher order moments involve terms similar to the non-equilibrium momentum fluxes (related to the strain rate tensor) which can dominate the corresponding physical contributions from the second order non-equilibrium moments, and manifest as numerical hyperviscosities~\cite{geier2015cumulant}. Such a numerical artifact can be investigated by studying the decay rate of three-dimensional crossing shear waves~\cite{geier2009factorized}.

If we consider the fluid velocities and the pressure gradients to be small, and in the absence of body forces or boundaries, for a pair of three-dimensional crossing shear waves involving orthogonal wave vectors, the Navier-Stokes equations can be simplified to obtain the following analytical solution based on the product solution of the respective shear waves:
\begin{eqnarray}\label{OCSW_solution}
u_y (t) =U_y \cos(k_x x)\cos(k_z z)\exp(-\nu k_x^2)\exp(-\nu k_z^2),
\end{eqnarray}
where $k_x = 2\pi / L_x$ and $k_z = 2\pi / L_z$ are the wavenumbers in the two orthogonal coordinate directions $x$ and $z$, respectively, $U_y$ is the initial amplitude of the wave perturbed along the $y$ direction, and $\nu$ is the fluid viscosity. Thus, starting from the initial condition in a periodic box
\begin{eqnarray}
u_x = 0, \;\;\;\; u_z=0,\;\;\;\; u_y = U_y \cos(k_x x)\cos(k_z z),
\end{eqnarray}
according to Eq.~(\ref{OCSW_solution}), the crossing shear waves decay at a rate that depends on the wavenumbers and the viscosity of the fluid. Furthermore, given that the decay rate depends on the fluid viscosity, the presence of any numerical hyperviscosities can significantly alter the amplitude decay rate from the predictions above. Here, we discretize such a periodic box with a relatively coarse grid resolution and apply the initial conditions above, so that we can study different LBM formulations and investigate their ability to predict the decay rate.

As in~\cite{geier2009factorized}, we consider an extreme case of relatively small fluid viscosity of $\nu = 1\times 10^{-7}$, with wavenumbers defined using $L_x = L_z = 30$ and an initial amplitude of $U_y = 1\times 10^{-5}$, and we study the decay rate for simulations using various LBM collision formulations relative to the analytically predicted decay rate shown above. Namely, we compare the decay rates as produced by the single relaxation time collision model or the SRT-LBM, the central moment collision model which uses the Maxwellian central moments for equilibria or the MCM-LBM, the central moment collision model with the Fokker-Planck guided collisions or the FPC-LBM, and the cumulant LBM involving relaxations based on cumulants during collision.
\begin{figure}[ht]
\centering
\begin{subfigure}{.53\textwidth}
\includegraphics[trim = 0 0 -8mm 0, clip, width = 78mm]{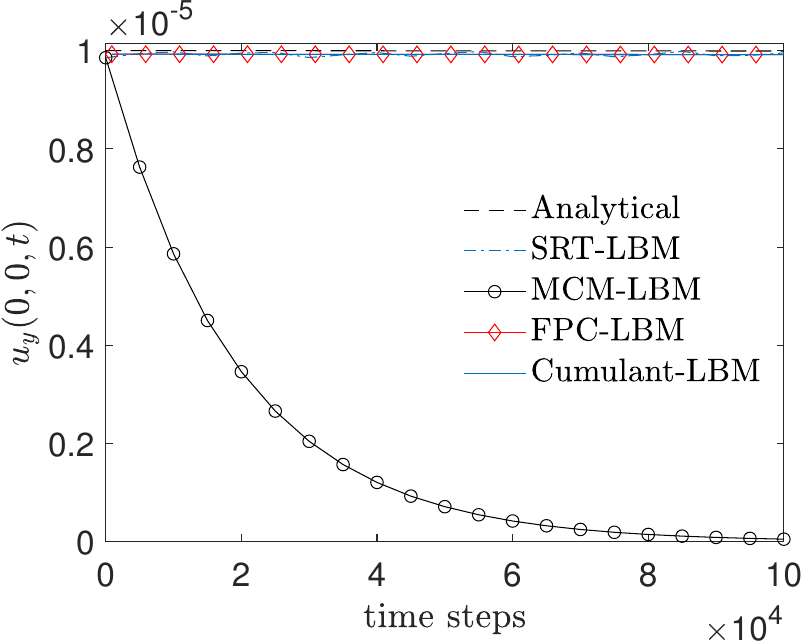}
\caption{}
\end{subfigure}
\begin{subfigure}{.45\textwidth}
\includegraphics[trim =0 0 0 0, clip, width = 75mm]{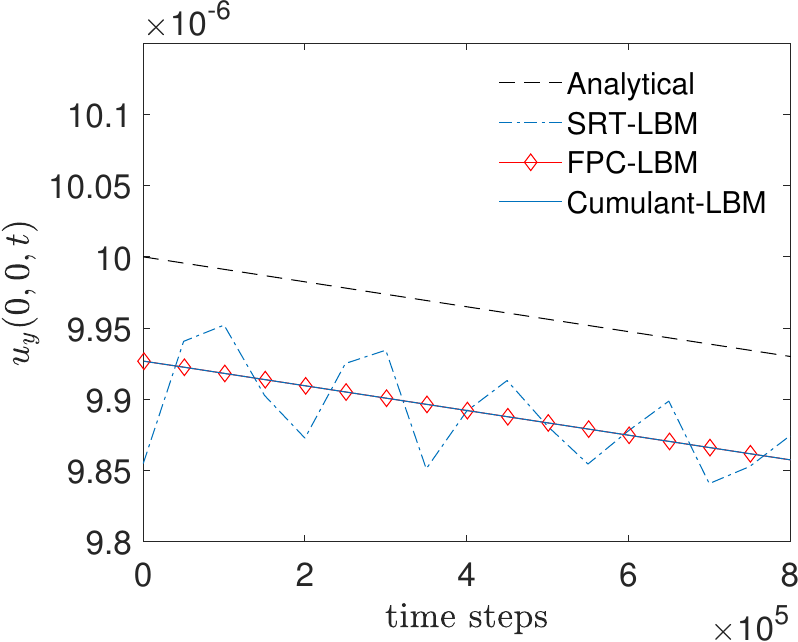}
\caption{}
\end{subfigure}
\caption{A comparison of the decay rates produced by different LB collision models -- SRT-LBM, Maxwellian central moments-based MCM-LBM, cumulant LBM, and Fokker-Planck central moments-based FPC-LBM as compared to the analytically predicted decay rate for the simulation of orthogonal crossing shear waves. Figure (a) indicates that the MCM-LBM fails to produce a decay rate similar to that of the analytical solution, and thus is not able to deal with the numerical hyperviscosity effects associated with this problem. Figure (b), which is a highly zoomed version of the left figure, indicates that the SRT-LBM can deal with the hyperviscosity effects but also contains unwanted noise. Furthermore, the cumulant LBM and the FPC-LBM are seen to have nearly identical decay rates that are consistent with the analytical solution. }
\label{OCSW}
\end{figure}
The results in Fig.~\ref{OCSW}(a) indicate that the decay rate for the MCM-LBM is much higher than the analytically derived decay rate because the velocity amplitude decays rapidly to zero, and thus is not able to effectively overcome the hyperviscosity artifact. On the other hand the SRT-LBM, FPC-LBM, and the cumulant LBM have apparently similar decay rates as compared to what is predicted analytically. Figure~\ref{OCSW}(b) shows a closer view of these results, which more clearly illuminates the key features of the other three collision models shown here. To begin with, the SRT-LBM produces a decay rate for the crossing shear waves that is consistent with what is expected; however it has other features which cause noise in the decay rate, and this is consistent with observations shown in ~\cite{geier2009factorized}. On the other hand, we observed that the FPC-LBM and cumulant LBM both produce nearly identical decay rates in this benchmark, which are in agreement with those predicted by the analytical solution. As such, it underscores the importance of equilibria in reducing the hyperviscosity effects dramatically. Specifically, with the FPC-LBM, the choice of the central moment equilibria based on the Markovian attractor given in Eq.~(\ref{eq:Mvcmattractor3D}) and its attendant tensor diffusion parameter as presented in Eq.~(\ref{eq:diffusion-tensor-parameter-3D}) (which controls the rate of diffusion of the distribution function in different directions in the velocity space) play a crucial role in this regard.

\subsection{Fully developed turbulent channel flow: Demonstration case study of FPC-LBM for turbulence simulations}
\begin{figure}[ht!]
\centering
\begin{subfigure}{.53\textwidth}
\includegraphics[trim = 0 0 0 0, clip, width = 75mm]{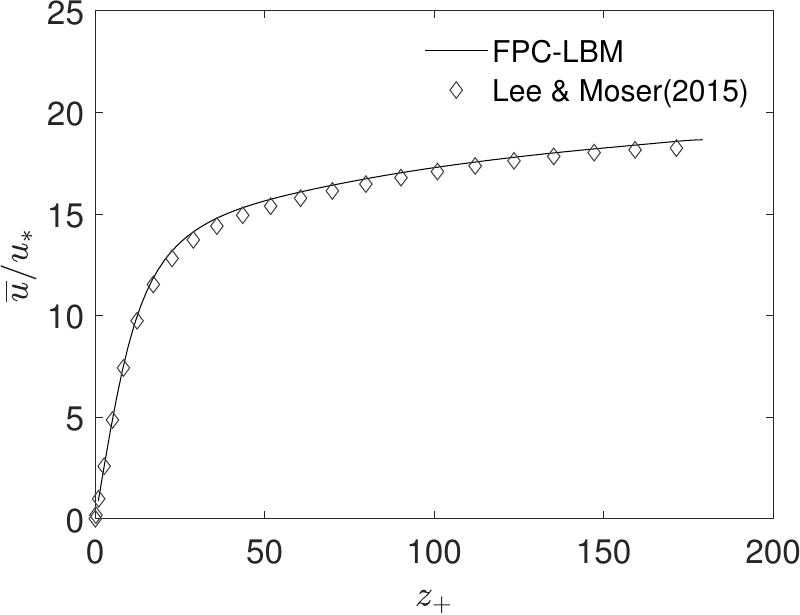}
\caption{}
\end{subfigure}
\begin{subfigure}{.45\textwidth}
\includegraphics[trim =0 0 0 0, clip, width = 75mm]{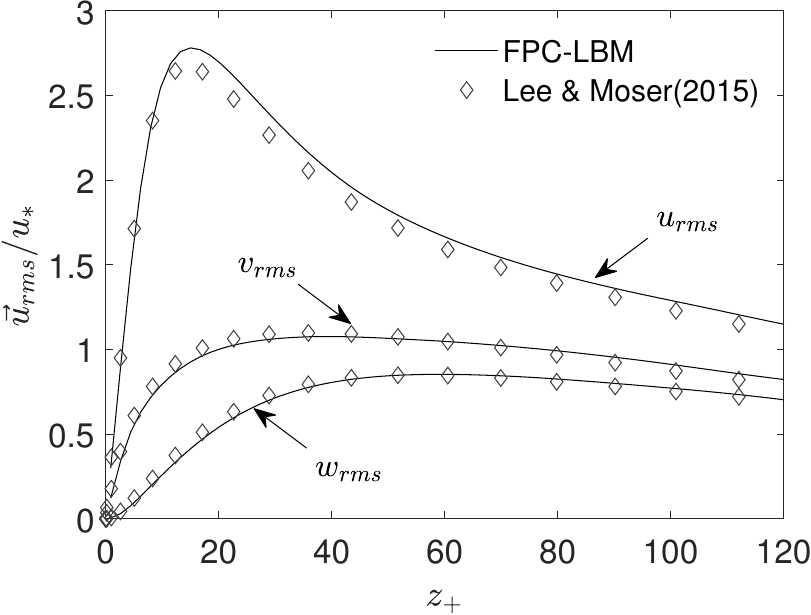}
\caption{}
\end{subfigure}
\begin{subfigure}{.45\textwidth}
\includegraphics[trim =0 0 0 0, clip, width = 75mm]{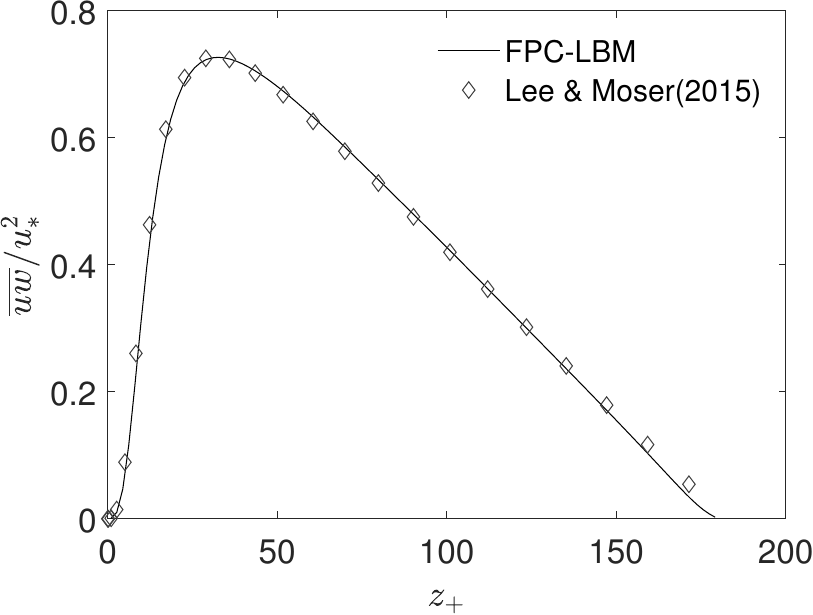}
\caption{}
\end{subfigure}
\caption{Comparisons of turbulence statistics for fully developed turbulent channel flow at a shear Reynolds number of $\mbox{Re}_*=180$ computed using the FPC-LBM and compared to the direct numerical simulations (DNS) data of Lee and Moser (2015)~\cite{lee2015direct}. (a) Mean streamwise velocity, (b) Root-mean-square (rms) velocity fluctuations, and (c) Reynolds stress along the streamwise-wall normal direction.}
\label{TCF}
\end{figure}

In our final case study, we demonstrate the capability of the FPC-LBM to perform turbulence simulations using a prototypical example involving the turbulent channel flow and compare the computed turbulence statistics against a recent DNS data of Lee and Moser (2015)~\cite{lee2015direct}. To reduce the computational effort, we perform simulations within a domain that encompasses only the half-channel height $H$ by utilizing a specular refection boundary condition on the symmetry plane~\cite{premnath2009generalized}. For resolving the turbulent flow structures adequately, the computational domain is taken to have an aspect ratio of $2 \pi H \times \pi H \times H$. The flow is driven by a body force along the periodic streamwise direction given by $F_x = -\frac{dP}{dx} = \frac{\tau_w}{H} = \frac{\rho u_*^2}{H}$, where $\tau_w$ is the wall shear stress and $u_*$ is the shear or friction velocity, which satisfies $u_* = (\tau_w/\rho)^{1/2}$. For performing turbulence simulations, we consider a shear Reynolds number $\mbox{Re}_* = 180$, where $\mbox{Re}_* = u_* H/\nu$ with $\nu$ being the molecular viscosity of the fluid, by resolving the computational domain using the above aspect ratio by taking $H = 100$ grid nodes in the wall normal direction. We choose a similar approach as that of~\cite{premnath2009generalized},  where we perform large eddy simulations (LES) by utilizing a common subgrid scale model known as the Smagorinsky model along with the van Driest wall dampening function to represent the variations in the subgrid effects near the wall. The characteristic length scale near the wall is the viscous length scale $\delta_{\nu} = \nu/u_*$, which can be used to express the characteristic time scale of eddy as $T^* = H/u_*$. The initial run for the simulations are performed using the 3D FPC-LBM for a duration of $50T^*$ until stationary turbulence statistics (such as the invariance in the Reynolds stress profiles) is achieved; then an additional run for a period of $30T^*$ is carried out to collect the turbulence statistics by averaging the flow field and turbulence fluctuations in time and in space along the homogeneous directions (i.e., the horizontal planes). Large eddy simulations performed using parallel computations with 512 processors of our in-house cluster took less than 24 hours of wall-clock time to complete and to collect the turbulence statistics. Figure~\ref{TCF} shows comparisons of the mean streamwise velocity, root-mean-square (rms) velocity fluctuations, and the Reynolds stress as a function of the distance from the wall in wall units, i.e., $z^+ = z/\delta_{\nu}$ computed using the FPC-LBM with the DNS data of Lee and Moser (2015)~\cite{lee2015direct}. Evidently, good agreement is seen, which demonstrates that the FPC-LBM is well-suited for simulating turbulent flows accurately and effectively.

\section{Summary and Conclusions\label{sec:summaryconclusions}}
Collision models play a critical role in determining the numerical accuracy and stability of simulations using LBM. Fokker-Planck (FP) kinetic equation, which represents stochastic processes such as the prototypical Brownian motion, can be used as a model for the collision integral of the Boltzmann equation and involves variations in the distribution function under collision due to its drift and diffusion in the phase space. In this paper, we have derived a new approach based on a FP-guided central moment collision operator for the LBM. It effectively involves relaxation of different central moments to their respective attractors or “equilibria” that depend on the products of lower order central moments and the components of the diffusion coefficient tensor; the latter is related to the variance of the distribution function or its second order central moments. We designate such attractors as the Markovian central moment attractors reflecting the repeated randomness nature of the collision process.

We have constructed central moment LB algorithms based on the FP collision model, which is referred to as the FPC-LBM, in 2D and 3D using D2Q9 and D3Q27 lattices, respectively, through a matching principle, viz., by matching the changes in different discrete central moments independently supported by the lattice under collision to those given by the continuous Boltzmann equation with its collision term expressed in terms of the FP model. We have shown the consistency of our approach to the Navier-Stokes equations via a Chapman-Enskog analysis based directly on expansions of central moments about their attractors. The development of the continuous central moment based FP collision model and their use in the construction of novel LB schemes based on discrete velocities highlighted the important role of the rates of diffusion of the distribution function along different directions in the velocity space, or the second order central moments, in determining the evolution of still higher order moments, which can in turn influence their overall numerical properties.

Simulations of a variety of benchmark problems in both 2D and 3D established its accuracy and demonstrated its superior numerical properties such as enhanced numerical stability and an ability to avoid numerical hyperviscosities in simulations with extremely low physical viscosities of the fluid. It is shown to have cumulant LBM-like behavior in terms of stability while being simpler, and does not involve ad hoc factorized collision formulation to reduce hyperviscosity effects in central moment LBMs as it is rooted and derived directly from a well-founded kinetic model based on the Fokker-Planck formulation. The FPC-LBM is also shown to be effective in computing wall-bounded turbulent flows as it is able to predict turbulence statistics accurately that is in very good agreement when compared to state-of-the-art direct numerical simulations (DNS) data based on a NS-based solver.

\section*{Acknowledgements}
Early stages of this research were presented virtually at the 73rd Annual Meeting of the American Physical Society (APS) Division of Fluid Dynamics (DFD) in November 2020 (https://meetings.aps.org/Meeting/DFD20/Session/F10.5) and at the 17th International Conference on Mesoscopic Methods in Engineering and Science (ICMMES) in July 2021. The authors would like to acknowledge the support of the US National Science Foundation (NSF) for research under Grant CBET-1705630. The second author would also like to thank the NSF for support of the development of a computer cluster `Alderaan' hosted at the Center for Computational Mathematics at the University of Colorado Denver under Grant OAC-2019089 (Project ``CC* Compute: Accelerating Science and Education by Campus and Grid Computing''), which was used in performing the simulations.

\newpage
\appendix

\numberwithin{equation}{section}

\setcounter{equation}{0}

\section{Chapman-Enskog analysis of 2D central moment formulation of the Boltzmann equation with Fokker-Planck collision model\label{sec:CEAnalysis2DFPC-LBM}}
We will now perform a Chapman-Enskog (C-E) analysis of the central moment system of the continuous Boltzmann equation with the FP collision model. Thus, the starting point for the consistency analysis of the continuous Boltzmann equation with the Fokker-Planck collision model to derive the macroscopic fluid dynamical equations is to convert the LHS of Eq.~(\ref{boltzmann_equation}), i.e., the streaming (Liouville) operator, in a central moment form by taking an inner product with the weight $W_{ mn}$ given in Eq.~(\ref{cm_weights}) and then evaluating the necessary definite integrals. The resulting central moment system after simplification (as formulated by Premnath~\cite{Premnath2019}), is given by
 \begin{eqnarray}\label{eqn:52}
\left<  \frac{\partial f}{\partial t} + \bm{\xi}\cdot\bm{\nabla} f , W_{ mn} \right>  \! \! \! & =& \! \! \! \frac{D\Pi_{mn}}{D t} + m \Pi_{m-1,n}\frac{Du_x}{Dt}+ n\Pi_{m,n-1}\frac{Du_y}{D t}\\[3mm]
& &+ \Pi_{mn} [(m+1)\partial_x u_x + (n+1)\partial_y u_y]+ \partial_x \Pi_{ m+1,n} \nonumber \\[3mm]
& & + \partial_y \Pi_{m,n+1}  + n\Pi_{m+1,n-1} \partial_x u_y+ m\Pi_{m-1,n+1} \partial_y u_x,\nonumber
\end{eqnarray}
where $\frac{D}{Dt}(\cdots)$ is the Lagrangian (material) derivative involving the fluid velocity and given by
\begin{equation}\label{eqn:53}
 \frac{D}{Dt} = \partial_t + \bm{u}\cdot\bm{\nabla}.
\end{equation}
Combining Eq.~(\ref{eqn:52}) with the rate of change of the central moment of order ($m+n$) due to the Fokker-Planck collision model (Eq.~(\ref{FPcollision})) and forcing (Eq.~(\ref{forcing:2D})) then provides the complete system of central moments of the continuous model Boltzmann equation.

In order to derive slow space/time variation limit of such a representation, i.e., the fluid dynamical equations, we introduce a small perturbation parameter $\varepsilon$ in the collision term $\left(\frac{\delta \Pi_{mn}}{\delta t}  \right)_{coll}^{FP}$ given in Eq.~(\ref{FPcollision}) so that the space/time derivatives are $\varepsilon$ order smaller than the relaxation time. This is the first step in the C-E analysis to obtain the so-called normal solution. Hence, we get
\begin{eqnarray} \label{eqn:54} 
\frac{D\Pi_{mn}}{D t} + m \Pi_{m-1,n}\frac{Du_x}{Dt} + n\Pi_{m,n-1}\frac{Du_y}{D t} + \Pi_{mn} [(m+1)\partial_x u_x + (n+1)\partial_y u_y]  \\[3mm]
+ \partial_x \Pi_{ m+1,n} + \partial_y \Pi_{m,n+1} + n\Pi_{m+1,n-1} \partial_x u_y + m\Pi_{m-1,n+1} \partial_y u_x \nonumber \\[3mm]
=\frac{\omega_{mn}}{\varepsilon} \left[ \Pi_{ mn}^{ Mv} - \Pi_{mn} \right] + m\frac{F_x}{\rho}\Pi_{m-1,n} + n\frac{F_y}{\rho}\Pi_{m,n-1}. \nonumber
\end{eqnarray}
The parameter $\varepsilon$ serves for book keeping by avoiding the closure problem and may be absorbed into $\omega_{mn}$ by setting $\varepsilon = 1$ at the end of the analysis. Then, expanding the distribution function about its attractor $f^{(0)} = f^{Mv}$ as well as using a multiple time expansion of $\partial_t$ in terms of the perturbation parameter as
\begin{subequations} 
\begin{eqnarray}
& &f = f^{(0)} + \varepsilon f^{(1)} + \varepsilon^{2} f^{(2)} + \dots , \label{eqn:55a}\\[3mm]
& &\partial_t = \partial_{t_{0}} + \varepsilon \partial_{t_{1}} + \varepsilon^{2} \partial_{t_{2}}+ \dots. \label{eqn:55b}
\end{eqnarray}
\end{subequations}
The first expansion can be equivalently rewritten directly in terms of central moment expansion as
\begin{equation}\label{eqn:56}
\Pi_{mn} = \Pi_{mn}^{(0)} + \varepsilon \Pi_{mn}^{(1)} + \varepsilon^{2} \Pi_{mn}^{(2)} + \dots ,
\end{equation}
where $\Pi_{mn}^{(0)} = \Pi_{mn}^{Mv}$. The solvability conditions arising via the collision invariants of mass and momentum are
\begin{eqnarray}\label{eqn:57} 
& &\Pi_{00}^{(0)} = \rho, \Pi_{10}^{ (0)} = \Pi_{01}^{(0)} = 0, \nonumber \\[3mm]
& &\Pi_{00}^{ (n)} = \Pi_{ 10}^{ (n)} = \Pi_{ 01}^{(n)} = 0, \; \; \; \; \; \; n\geq 1, \nonumber
\end{eqnarray}
which will be needed to obtain the hydrodynamics of the above central moment system shown in Eq.~(\ref{eqn:54}). Defining the leading order Lagrangian derivative
\begin{equation}\label{eqn:58}
\frac{D}{Dt_0} = \partial_{t_0} +  \bm{u}\cdot\bm{\nabla},
\end{equation}
and substituting Eq.~(\ref{eqn:55b}),  Eq.~(\ref{eqn:56}), and Eq.~(\ref{eqn:57}) in Eq.~(\ref{eqn:54}), then the leading $O(\varepsilon^{ 0})$ central moment system becomes
\begin{eqnarray}\label{eqn:59} 
\frac{D\Pi_{mn}^{ (0)}}{D t_{ 0}} + m \Pi_{m-1,n}^{ (0)}\frac{Du_x}{Dt_{ 0}} + n\Pi_{m,n-1}^{ (0)}\frac{Du_y}{D t_{ 0}}+ \Pi_{mn}^{ (0)} [(m+1)\partial_x u_x + (n+1)\partial_y u_y]\\[3mm]
 + \partial_x \Pi_{ m+1,n}^{ (0)} + \partial_y \Pi_{m,n+1}^{ (0)}+ n\Pi_{m+1,n-1}^{ (0)} \partial_x u_y + m\Pi_{m-1,n+1}^{ (0)} \partial_y u_x \nonumber \\[3mm]
= -\omega_{mn} \Pi_{ mn}^{ (1)}  + m\frac{F_x}{\rho}\Pi_{m-1,n}^{ (0)} + n\frac{F_y}{\rho}\Pi_{m,n-1}^{ (0)}.\nonumber
\end{eqnarray}
Next, the $O(\varepsilon^1)$ central moment system arising from the asymptotic expansion reads
\begin{eqnarray}\label{eqn:60} 
\frac{\partial\Pi_{mn}^{(0)}}{\partial t_{ 1}}+ \frac{D\Pi_{mn}^{ (1)}}{D t_{ 0}}+m\Pi_{m-1,n}^{ (1)}\frac{Du_x}{Dt_{ 0}}+ m\Pi_{m-1,n}^{ (0)}\frac{\partial u_x}{\partial t_{ 1}} +n\Pi_{m,n-1}^{ (1)}\frac{D u_y}{D t_{ 0}}\\[3mm]
 + n\Pi_{m,n-1}^{ (0)}\frac{\partial u_y}{\partial t_{ 1}}+ \Pi_{mn}^{ (1)} [(m+1)\partial_x u_x + (n+1)\partial_y u_y] + \partial_x \Pi_{ m+1,n}^{ (1)} \nonumber  \\[3mm]
 + \partial_y \Pi_{m,n+1}^{ (1)} +  n\Pi_{m+1,n-1}^{ (1)} \partial_x u_y+ m\Pi_{m-1,n+1}^{ (1)} \partial_y u_x \nonumber  \\[3mm]
 = -\omega_{mn} \Pi_{ mn}^{ (2)}  + m\frac{F_x}{\rho}\Pi_{m-1,n}^{ (1)} + n\frac{F_y}{\rho}\Pi_{m,n-1}^{ (1)}.\nonumber
\end{eqnarray}
It may be noted that the leading order of central moments due to the attractors, i.e., $\Pi_{ mn}^{ (0)} = \Pi_{ mn}^{ Mv}$ are given in Eq.~(\ref{eq:MvCM2D}). In order to separate out the trace of the diagonal part (isotropic part) of the momentum transfer, i.e., the bulk viscosity effects from the shear viscosity related transport, we define
\begin{eqnarray}
\Pi_{2s} = \Pi_{20}+\Pi_{02}, \quad \Pi_{2d} = \Pi_{20}-\Pi_{02}.\label{eqn:61}
\end{eqnarray}
Then, we prescribe them to relax at their own individual relaxation rates $\omega_{2s}$ and $\omega_{2d}$, respectively via modified representations of Eqs. (\ref{FPcollision}):
\begin{subequations}
\begin{eqnarray}
\left(\frac{\delta}{\delta t} \Pi_{2s} \right)_{\!\!\! coll}^{\!\!\! F\!P} = \omega_{2s} \left[ \Pi_{2s}^{Mv} - \Pi_{2s} \right],\label{eqn:62a} \\[3mm]
\left(\frac{\delta}{\delta t} \Pi_{2d} \right)_{\!\!\!  coll}^{\!\!\!  F\!P} = \omega_{2d} \left[ \Pi_{2d}^{Mv} - \Pi_{2d} \right],\label{eqn:62b}
\end{eqnarray}
\end{subequations}
where $\Pi_{ 2s}^{ Mv} = \Pi_{ 20}^{ Mv}+\Pi_{02}^{ Mv} = 2 \rho c_s^2 = \Pi_{ 2s}^{ (0)}$ and $\Pi_{ 2d}^{ Mv} = \Pi_{ 20}^{ Mv}-\Pi_{ 02}^{ Mv} = 0 = \Pi_{ 2d}^{ (0)}$ . All other higher central moments follow Eq. (\ref{FPcollision}), and are relaxed at their own relaxation rate $\omega_{mn}$, e.g., ($mn$) = ($11$), ($21$), ($12$), and ($22$).

We first list the $O(\varepsilon^{ 0})$ evolution equations for ($mn$) = (00), (10), and (01), respectively using Eq.~(\ref{eqn:59}) as
\begin{subequations}
\begin{eqnarray}
\frac{D\Pi_{ 00} }{Dt}+ \Pi_{ 00} (\partial_x u_x + \partial_y u_y) = 0, \; \; \; \; \; \; &\mbox{or}& \; \; \; \; \; \;  \frac{D\rho}{Dt_0}+ \rho (\partial_x u_x + \partial_y u_y) = 0, \label{eqn:63a} \\[3mm]
 \Pi_{ 00}\frac{Du_x}{Dt} + \partial_x \Pi_{ 20}^{ (0)} = F_x, \; \; \; \; \; \; &\mbox{or}& \; \; \; \; \; \; \rho\frac{Du_x}{Dt_{ 0}} + \partial_x (c_s^2 \rho) = F_x, \label{eqn:63b}\\[3mm]
\Pi_{ 00}\frac{Du_y}{Dt} + \partial_y \Pi_{ 02}^{ (0)} = F_y, \; \; \; \; \; \; &\mbox{or}& \; \; \; \; \; \; \rho\frac{Du_y}{Dt_{0}} + \partial_y (c_s^2 \rho) = F_y. \label{eqn:63c}
\end{eqnarray}
\end{subequations}
Then the diagonal components  ($mn$) = (20) and (02), in view of Eq.~(\ref{eqn:62a}) and Eq.~(\ref{eqn:62b}) and using $\Pi_{2s} =\Pi_{2s}^{(0)} + \varepsilon \Pi_{ 2s}^{ (1)} + \dots$ and $\Pi_{2d} =\Pi_{2d}^{ (0)} + \varepsilon \Pi_{2d}^{(1)}+ \dots$, we separately take the sums and difference of the LHS of Eq.~(\ref{eqn:59}) evaluated for ($mn$) = ($20$) and ($02$) and apply Eqs.~(\ref{eqn:62a}) and (\ref{eqn:62b}) on its RHS, respectively. As a result, and using $ \Pi_{ 20}^{ (0)} = \Pi_{ 02}^{ (0)} = \rho c_s^2$, we have the following for ($mn$) = ($2s$) and ($2d$), respectively:
\begin{subequations}
\begin{eqnarray}
\frac{D}{Dt_{ 0}} \Pi_{ 2s}^{ (0)} + 4\Pi_{20}^{ (0)}(\partial_x u_x + \partial_y u_y) + \partial_x \Pi_{30}^{ (0)}+ \partial_y \Pi_{ 03}^{ (0)} = -\omega_{ 2s} \Pi_{2s}^{ (1)}, \label{eqn:64a}\\[3mm]
\frac{D}{Dt_{ 0}} \Pi_{ 2d}^{ (0)} + 2\Pi_{20}^{ (0)}(\partial_x u_x - \partial_y u_y) + \partial_x \Pi_{30}^{ (0)}+ \partial_y \Pi_{ 03}^{ (0)} = -\omega_{2d} \Pi_{ 2d}^{ (1)}.\label{eqn:64b}
\end{eqnarray}
\end{subequations}
For the continuous case $\Pi_{30}^{(0)}=\Pi_{03}^{(0)} = 0$, which follows from $\Pi_{30}^{(0)}=\Pi_{30}^{Mv}=2D'_{20}\Pi_{10}=0$, etc (while they are $O(u^3)$ in the LB formulation using discrete velocities for the standard D2Q9 lattice due to the aliasing effects and they are usually neglected or are corrected for in some cases). Simplifying Eq.~(\ref{eqn:64a}) and Eq.~(\ref{eqn:64b}), we obtain
\begin{subequations}
\begin{eqnarray}
2 c_s^2 \frac{D\rho}{Dt_{ 0}} + 4\rho c_s^2 (\partial_x u_x + \partial_y u_y) &= - \omega_{ 2s} \Pi_{ 2s}^{ (1)},\label{eqn:65a} \\[3mm]
2\rho c_s^2(\partial_x u_x - \partial_y u_y) &=  -\omega_{ 2d} \Pi_{ 2d}^{ (1)}.\label{eqn:65b}
\end{eqnarray}
\end{subequations}
Also, evaluating Eq.~(\ref{eqn:59}) for the remaining off-diagonal central moment, i.e., $(mn)=(11)$, we get
\begin{eqnarray}
 2\Pi_{ 11}^{ (0)}(\partial_x u_x + \partial_y u_y) + \partial_x \Pi_{ 21}^{ (0)} + \partial_y \Pi_{ 12}^{ (0)} + \Pi_{ 20}^{ (0)} \partial_x u_y + \Pi_{ 02}^{ (0)} \partial_y u_x = -\omega_{ 11}\Pi_{11}^{ (1)}.\nonumber
\end{eqnarray}
Using $\Pi_{ 11}^{ (0)} = \Pi_{ 21}^{ (0)} = \Pi_{ 12}^{ (0)} = 0$ and $ \Pi_{ 20}^{ (0)}= \Pi_{ 02}^{ (0)} = c_s^2 \rho$, the above equation simplifies to
\begin{equation}\label{eqn:66}
\rho c_s^2 (\partial_x u_y + \partial_y u_x) = -\omega_{ 11}\Pi_{11}^{ (1)}.
\end{equation}

The above equations, Eqn.~(\ref{eqn:63a}) - Eqn.~(\ref{eqn:63c}), need to be respectively combined with the $O(\varepsilon)$ central moment evolution equations for the components ($mn$)= (00), (10), and (01) to obtain the desired hydrodynamic equations. Hence, the first three components of the $O(\varepsilon)$ central moment system from Eq. (\ref{eqn:60}) are given by
\begin{subequations}
\begin{eqnarray}
\frac{\partial \Pi_{00}^{ (0)}}{\partial t_{ 0}} = 0 \; \; \; \; \; \;  &\mbox{or}& \; \; \; \; \; \;   \frac{\partial \rho}{\partial t_{ 0}}= 0, \label{eqn:67a} \\[3mm]
\Pi_{00}^{ (0)}\frac{\partial u_x}{\partial t_{ 1}} + \partial_x \Pi_{ 20}^{ (1)}   + \partial_y \Pi_{ 11}^{ (1)} = 0 \; \; \; \; \; \; &\mbox{or}& \; \; \; \; \; \; \rho\frac{\partial u_x}{\partial t_{ 1}} + \partial_x \Pi_{ 20}^{ (1)}   + \partial_y \Pi_{ 11}^{ (1)} = 0, \label{eqn:67b}     \\[3mm]
\Pi_{00}^{ (0)}\frac{\partial u_y}{\partial t_{ 1}} + \partial_x \Pi_{ 11}^{ (1)}   + \partial_y \Pi_{ 02}^{ (1)} = 0 \; \; \; \; \; \; &\mbox{or}& \; \; \; \; \; \; \rho\frac{\partial u_y}{\partial t_{ 1}} + \partial_x \Pi_{ 11}^{ (1)}   + \partial_y \Pi_{ 02}^{ (1)} = 0.\label{eqn:67c}
\end{eqnarray}
\end{subequations}
where in Eq.~(\ref{eqn:67b}) and Eq.~(\ref{eqn:67c}), $\Pi_{ 20}^{ (1)}$ and $\Pi_{ 02}^{ (1)}$ can be rewritten using the definitions of the nonequilibrium combined diagonal second order moments (analogous to Eq.~(\ref{eqn:61})) $\Pi_{2s}^{ (1)} =\left( \Pi_{20}^{ (1)} + \Pi_{02}^{ (1)} \right)$ and  $\Pi_{ 2d}^{ (1)} =\left( \Pi_{ 20}^{ (1)} - \Pi_{ 02}^{ (1)} \right)$ as
\begin{equation}\label{eqn:68}
\Pi_{ 20}^{ (1)} =\frac{1}{2}\left( \Pi_{ 2s}^{ (1)} + \Pi_{ 2d}^{ (1)} \right), \; \; \; \; \; \;  \; \; \; \; \; \; \Pi_{ 02}^{ (1)} =\frac{1}{2}\left( \Pi_{ 2s}^{ (1)} -\Pi_{ 2d}^{ (1)} \right).
\end{equation}
Hence, the equations, Eq.~(\ref{eqn:67a}) - Eq.~(\ref{eqn:67c}), become
\begin{subequations}
\begin{eqnarray}
\frac{\partial \rho}{\partial t_{1}} = 0, \label{eqn:69a}\\[3mm]
\rho\frac{\partial}{\partial t_{1}}u_x + \partial_x \left[ \frac{1}{2} \Pi_{ 2s}^{(1)} +\frac{1}{2} \Pi_{2s}^{(1)}  \right]   + \partial_y \Pi_{11}^{(1)} = 0, \label{eqn:69b}      \\[3mm]
\rho\frac{\partial}{\partial t_{1}}u_y +\partial_x \Pi_{11}^{(1)}+ \partial_y \left[ \frac{1}{2} \Pi_{2s}^{(1)} -\frac{1}{2} \Pi_{ 2s}^{(1)}  \right]    = 0. \label{eqn:69c}
\end{eqnarray}
\end{subequations}
Then, we combine $O(1)$ and $O(\varepsilon)$ evolution equations for the conserved moments, i.e., $\mbox{Eqs}.~(\ref{eqn:63a}) - (\ref{eqn:63c}) + \varepsilon [\mbox{Eqs}.~(\ref{eqn:69a})-(\ref{eqn:69c})]$, by using $\partial_t = \partial_{t_{0}} +\varepsilon \partial_{t_{1}}$, and setting the book keeping parameter $\varepsilon = 1$ we get
\begin{subequations}
\begin{eqnarray}
\frac{D\rho}{Dt} + \rho(\partial_x u_x + \partial_y u_y) = 0,\label{eqn:70a} \\[3mm]
\rho \frac{Du_x}{Dt} + \partial_x (c_s^2 \rho) + \partial_x \left[ \frac{1}{2} \Pi_{ 2s}^{ (1)} + \frac{1}{2} \Pi_{ 2d}^{ (1)} \right] + \partial_y \Pi_{ 11}^{ (1)} = F_x,\label{eqn:70b} \\[3mm]
\rho \frac{Du_y}{Dt} + \partial_y (c_s^2 \rho) + \partial_x \Pi_{ 11}^{ (1)} + \partial_y \left[ \frac{1}{2} \Pi_{ 2s}^{ (1)} - \frac{1}{2} \Pi_{ 2d}^{ (1)} \right]  = F_y. \label{eqn:70c}
\end{eqnarray}
\end{subequations}
In order to get the non-equilibrium central moments $\Pi_{ 2s}^{ (1)}$, $\Pi_{ 2d}^{ (1)}$, and $\Pi_{ 11}^{ (1)}$, we consider Eqs.~(\ref{eqn:65a}), (\ref{eqn:65b}), and (\ref{eqn:66}) and use (\ref{eqn:63a}) to replace$\frac{D\rho}{Dt_0}$ in terms of $-\rho \left( \partial_x u_x + \partial_y u_y \right)$. Thus, we get
\begin{eqnarray}
&&\Pi_{ 2s}^{ (1)} = -\frac{2\rho c_s^2}{\omega_{ 2s}} \left( \partial_x u_x + \partial_y u_y \right),\quad
\Pi_{ 2d}^{ (1)} = -\frac{2\rho c_s^2}{\omega_{ 2d}} \left( \partial_x u_x - \partial_y u_y \right),\nonumber\\
&&\Pi_{ 11}^{ (1)} = -\frac{\rho c_s^2}{\omega_{ 11}} \left( \partial_x u_y + \partial_y u_x \right).\label{eqn:71}
\end{eqnarray}
Defining the bulk viscosity $\zeta$ and shear viscosity $\nu$ based on the respective relaxation parameters as
\begin{eqnarray}\label{eqn:72}
\zeta = \frac{c_s^2}{\omega_{ 2s}}, \quad \nu =\frac{c_s^2}{\omega_{ 2d}} = \frac{c_s^2}{\omega_{ 11}}. \label{eqn:72}
\end{eqnarray}
and using Eqs.~(\ref{eqn:71}) and (\ref{eqn:72}) in Eqs.~(\ref{eqn:70a})-(\ref{eqn:70c}) and simplifying we get
\begin{eqnarray}
&&\frac{D\rho}{Dt} + \rho  \bm{\nabla}\cdot\bm{u} = 0,\\
&&\rho\frac{D u_x}{Dt} = -\partial_x P + \partial_x \left[ \rho \nu (2\partial_x u_x - \bm{\nabla}\cdot\bm{u}) + \rho \zeta \bm{\nabla}\cdot\bm{u}   \right]  + \partial_y \left[  \rho \nu (\partial_x u_y + \partial_y u_x)   \right ] + F_x, \nonumber\\
&&\rho\frac{D u_y}{Dt} = -\partial_y P + \partial_x \left[  \rho \nu (\partial_x u_y + \partial_y u_x)   \right ]+ \partial_y \left[ \rho \nu (2\partial_y u_y -\bm{\nabla}\cdot\bm{u}) + \rho \zeta \bm{\nabla}\cdot\bm{u}  \right] + F_y,\nonumber
\end{eqnarray}
where $P = c_s^2 \rho$. Hence, the Navier-Stokes Equations (NSE) is a consistent long-time and coarse-grained behavior of the Boltzmann equation with the Fokker-Planck collision model specified in a central moment formulation. It may be noted that the above Chapman-Enskog analysis of the central moment system of the Boltzmann equation can also be readily extended to study the evolution of the higher order moments. In addition, we note that when we approximate the Boltzmann equation using a discrete set of particle velocities and integrate it along the particle characteristics, the resulting LBM recovers the NSE with the relation between the viscosities and the relaxation parameters modified by replacing $1/\omega_{ij}$ with $(1/\omega_{ij}-1/2)$, i.e., based on the H\'{e}non correction. Hence, for the discrete 2D FPC-LBM, Eq.~(\ref{eqn:72}) changes to
\begin{eqnarray}\label{eqn:transportcoeff2DFPCLBM}
\zeta = c_s^2\left(\frac{1}{\omega_{ 2s}}-\frac{1}{2}\right), \quad \nu =c_s^2\left(\frac{1}{\omega_{ 2d}}-\frac{1}{2}\right) = c_s^2\left(\frac{1}{\omega_{ 11}}-\frac{1}{2}\right).
\end{eqnarray}

\newpage

\section{Appendix: Algorithmic details of 2D FPC-LBM using the D2Q9 lattice\label{sec:2DFPC-LBM}}
Step 1: Convert the pre-collision distribution functions into pre-collision raw moments using $\m = \PP\mathbf{f}$ as
\begin{eqnarray}
\Kp_{00} = \f{0} + \f{1} + \f{2} + \f{3} + \f{4} + \f{5} + \f{6} + \f{7} + \f{8},
\end{eqnarray}
\vspace{-9mm}
\begin{eqnarray}
\Kp_{10} = \f{1} - \f{3} + \f{5}- \f{6} - \f{7} + \f{8}, \nonumber
\end{eqnarray}
\vspace{-9mm}
\begin{eqnarray}
\Kp_{01} = \f{2} - \f{4} + \f{5} + \f{6} - \f{7} - \f{8},\nonumber
\end{eqnarray}
\vspace{-9mm}
\begin{eqnarray}
\Kp_{20} = \f{1} + \f{3} + \f{5} + \f{6} + \f{7} + \f{8},\nonumber
\end{eqnarray}
\vspace{-9mm}
\begin{eqnarray}
\Kp_{02} = \f{2} + \f{4} + \f{5} + \f{6} + \f{7} + \f{8},\nonumber
\end{eqnarray}
\vspace{-9mm}
\begin{eqnarray}
\Kp_{11} = \f{5} - \f{6} + \f{7} - \f{8},\nonumber
\end{eqnarray}
\vspace{-9mm}
\begin{eqnarray}
\Kp_{21} = \f{5} + \f{6} - \f{7} - f{8},\nonumber
\end{eqnarray}
\vspace{-9mm}
\begin{eqnarray}
\Kp_{12} = \f{5} - \f{6} - \f{7} + \f{8},\nonumber
\end{eqnarray}
\vspace{-9mm}
\begin{eqnarray}
\Kp_{22} = \f{5} + \f{6} + \f{7} + \f{8}.\nonumber
\end{eqnarray}
Step 2: Map the pre-collision raw moments into pre-collision central moments using $\mc = \F \m$ as
\begin{eqnarray}
\Ks{00} = \Kp_{00},
\end{eqnarray}
\vspace{-9mm}
\begin{eqnarray}
\Ks{10} = \Kp_{10} - u_x \Kp_{00}, \nonumber
\end{eqnarray}
\vspace{-9mm}
\begin{eqnarray}
\Ks{01} = \Kp_{01} - u_y  \Kp_{00} \nonumber
\end{eqnarray}
\vspace{-9mm}
\begin{eqnarray}
\Ks{20} =\Kp_{20} - 2u_x  \Kp_{10} + u_x^2  \Kp_{00}, \nonumber
\end{eqnarray}
\vspace{-9mm}
\begin{eqnarray}
\Ks{02} = \Kp_{02} - 2u_y  \Kp_{01} + u_y^2  \Kp_{00}, \nonumber
\end{eqnarray}
\vspace{-9mm}
\begin{eqnarray}
\Ks{11} = \Kp_{11} - u_y  \Kp_{10} - u_x   \Kp_{01} + u_x u_y \Kp_{00}, \nonumber
\end{eqnarray}
\vspace{-9mm}
\begin{eqnarray}
\Ks{21} = \Kp_{21} - 2u_x \Kp_{11} + u_x^2 \Kp_{01} - u_y  \Kp_{20} +2 u_x u_y  \Kp_{10}- u_x^2  uy  \Kp_{00}, \nonumber
\end{eqnarray}
\vspace{-9mm}
\begin{eqnarray}
\Ks{12} = \Kp_{12} -2u_y  \Kp_{11} + u_y^2 \Kp_{10} - u_x  \Kp_{02} + 2 u_x u_y  \Kp_{01} - u_x  u_y^2  \Kp_{00}, \nonumber
\end{eqnarray}
\vspace{-9mm}
\begin{eqnarray}
\Ks{22} \! \! \! & =& \! \! \! \Kp_{22} - 2u_x \Kp_{12} + u_x^2  \Kp_{02} - 2 u_y \Kp_{21} + 4  u_x u_y  \Kp_{11} - u_x^2  2 u_y  \Kp_{01} + u_y^2  \Kp_{20} \\ \nonumber
& & -2 u_x  u_y^2  \Kp_{10} + u_x^2  u_y^2  \Kp_{00}. \nonumber
\end{eqnarray}
Step 3: Collision update - Relax different central moments to their attractors and augment them with source terms
First, to independently evolve the effects of shear and bulk viscosities, we combine the diagonal second order central moments as
\begin{eqnarray}
\Ks{2s} = \Ks{20} + \Ks{02},  \qquad \qquad
\Ks{2d} = \Ks{20} - \Ks{02}.
\end{eqnarray}
Next, we compute the lower order Markovian attractors using
\begin{gather}
\Ks{00}^{Mv} = \rho,   \qquad \qquad
\Ks{10}^{Mv} = 0,  \qquad \qquad
\Ks{01}^{Mv} = 0, \qquad \qquad
\Ks{20}^{Mv} = c_s^2 \rho, \\[2mm]
\Ks{02}^{Mv} =c_s^2 \rho, \qquad \qquad
\Ks{11}^{Mv} = 0, \qquad \qquad
\Ks{21}^{Mv} = 0, \qquad \qquad
\Ks{12}^{Mv} = 0, \nonumber
\end{gather}
and also combine the second order attractors in the same way as
\begin{gather}
\Ks{2s}^{Mv} = \Ks{20}^{Mv} + \Ks{02}^{Mv},  \qquad \qquad
\Ks{2d}^{Mv} = \Ks{20}^{Mv} - \Ks{02}^{Mv}. \nonumber
\end{gather}
If an external body force is present, combine similar second order attractors due to them through $\Ss{2s}=\Ss{20}+\Ss{02}$ and $\Ss{2d}=\Ss{20}-\Ss{02}$.
Then, we relax various central moments up to the third order (i.e., ($m+n\le 3$) to their attractors along with contributions due to the source terms and the resulting post-collision central moments read as
\begin{eqnarray}
\Kts{00} = \Ks{00} + \omega_{0} (\Ks{00}^{Mv} - \Ks{00})+ \left(1-\omega_{0}/2 \right)\Ss{00}\delta_t,
\end{eqnarray}
\vspace{-9mm}
\begin{eqnarray}
\Kts{10} = \Ks{10} + \omega_{1} (\Ks{10}^{Mv} - \Ks{10})+ \left(1-\omega_{1}/2 \right)\Ss{10}\delta_t,\nonumber
\end{eqnarray}
\vspace{-9mm}
\begin{eqnarray}
\Kts{01} = \Ks{01} + \omega_{2} (\Ks{01}^{Mv} - \Ks{01})+ \left(1-\omega_{2}/2 \right)\Ss{01}\delta_t,\nonumber
\end{eqnarray}
\vspace{-9mm}
\begin{eqnarray}
\Kts{2s} = \Ks{2s} + \omega_{3 } (\Ks{2s}^{Mv}- \Ks{2s})+ \left(1-\omega_{3}/2 \right)\Ss{2s}\delta_t,\nonumber
\end{eqnarray}
\vspace{-9mm}
\begin{eqnarray}
\Kts{2d} =\Ks{2d} + \omega_{4 } (\Ks{2d}^{Mv}- \Ks{2d})+ \left(1-\omega_{4}/2 \right)\Ss{2d}\delta_t,\nonumber
\end{eqnarray}
\vspace{-9mm}
\begin{eqnarray}
\Kts{11} = \Ks{11} + \omega_{5}  (\Ks{11}^{Mv} - \Ks{11})+ \left(1-\omega_{5}/2 \right)\Ss{11}\delta_t,\nonumber
\end{eqnarray}
\vspace{-9mm}
\begin{eqnarray}
\Kts{21} = \Ks{21} + \omega_{6}  (\Ks{21}^{Mv} - \Ks{21})+ \left(1-\omega_{6}/2 \right)\Ss{21}\delta_t,\nonumber
\end{eqnarray}
\vspace{-9mm}
\begin{eqnarray}
\Kts{12} = \Ks{12} + \omega_{7 } (\Ks{12}^{Mv} - \Ks{12})+ \left(1-\omega_{7}/2 \right)\Ss{12}\delta_t,\nonumber
\end{eqnarray}
where $\Ss{mn}$ is given in Eq.~(\ref{eq:sourceCM2D}). Then, decompose the post-collision second order combined central moments via
\begin{eqnarray}
\tilde{\K}_{20} = 0.5(\tilde{\K}_{2s} +\tilde{\K}_{2d}), \qquad \qquad
\tilde{\K}_{02} = 0.5(\tilde{\K}_{2s} -\tilde{\K}_{2d}).
\end{eqnarray}
which is then used to define the Markovian attractor for the fourth order central moment as
\begin{eqnarray}
\Ks{22}^{Mv} = \frac{1}{\rho} (\tilde{\K}_{20} \tilde{\K}_{02} + 2 \tilde{\K}_{11} \tilde{\K}_{11}),
\end{eqnarray}
Following this, the fourth order central moment is relaxed and updated as follows:
\begin{eqnarray}
\tilde{\K}_{22} = \Ks{22} + \omega_{8} (\Ks{22}^{Mv} - \Ks{22})+ \left(1-\omega_{8}/2 \right)\sigma_{22}.
\end{eqnarray}
Step 4: Transform post-collision central moments into post-collision raw moments via $\tilde{\m} = \Fi \tilde{\mc}$ as
\begin{eqnarray}
\Kpt_{00} =  \Kt_{00},
\end{eqnarray}
\vspace{-9mm}
\begin{eqnarray}
\Kpt_{10} = \Kt_{10} + u_x\Kt_{00},\nonumber
\end{eqnarray}
\vspace{-9mm}
\begin{eqnarray}
\Kpt_{01} = \Kt_{01} + u_y\Kt_{00},\nonumber
\end{eqnarray}
\vspace{-9mm}
\begin{eqnarray}
\Kpt_{20}=  \Kt_{20} + 2  u_x\Kt_{10} + u_x^2 \Kt_{00},\nonumber
\end{eqnarray}
\vspace{-9mm}
\begin{eqnarray}
\Kpt_{02} =  \Kt_{02} + 2 u_y\Kt_{01} + u_y^2\Kt_{00},\nonumber
\end{eqnarray}
\vspace{-9mm}
\begin{eqnarray}
\Kpt_{11} =  \Kt_{11} + u_x   \Kt_{01} + u_y\Kt_{10} + u_x u_y \Kt_{00},\nonumber
\end{eqnarray}
\vspace{-9mm}
\begin{eqnarray}
\Kpt_{21} =  \Kt_{21} + 2  u_x\Kt_{11} + u_y\Kt_{20} + u_x^2  \Kt_{01} + 2 u_x u_y  \Kt_{10} + u_x^2 u_y  \Kt_{00}, \nonumber
\end{eqnarray}
\vspace{-9mm}
\begin{eqnarray}
\Kpt_{12} =  \Kt_{12} + 2  u_y\Kt_{11} + u_x\Kt_{02} + 2 u_x u_y \Kt_{01} + u_y^2 \Kt_{10} + u_x u_y^2 \Kt_{00},\nonumber
\end{eqnarray}
\vspace{-9mm}
\begin{eqnarray}
\Kpt_{22}  \! \! \! & =& \! \! \!  \Kt_{22} + 2  u_x\Kt_{12} + 2 u_y\Kt_{21} + 4 u_x u_y \Kt_{11} + u_y^2 \Kt_{20} + u_x^2 \Kt_{02} + 2 u_x^2 u_y\Kt_{01}\nonumber \\
& & + 2 u_x u_y^2 \Kt_{10} + u_x^2 u_y^2 \Kt_{00}.\nonumber
\end{eqnarray}
Step 5: Transform post-collision raw moments into post-collision distribution functions through $\tilde{\mathbf{f}} = \PP^{-1}\tilde{\m}$ as
\begin{eqnarray}
\tilde{f}_0 = \Kpt_{00} - \Kpt_{20} - \Kpt_{02} + \Kpt_{22},
\end{eqnarray}
\vspace{-9mm}
\begin{eqnarray}
\tilde{f}_1 =0.5(\Kpt_{10} + \Kpt_{20} - \Kpt_{12} - \Kpt_{22}),\nonumber
\end{eqnarray}
\vspace{-9mm}
\begin{eqnarray}
\tilde{f}_2 = 0.5(\Kpt_{01} + \Kpt_{02} - \Kpt_{21} - \Kpt_{22}),\nonumber
\end{eqnarray}
\vspace{-9mm}
\begin{eqnarray}
\tilde{f}_3 =0.5(-\Kpt_{10} + \Kpt_{20} + \Kpt_{12} - \Kpt_{22}),\nonumber
\end{eqnarray}
\vspace{-9mm}
\begin{eqnarray}
\tilde{f}_4 =0.5(-\Kpt_{01}+ \Kpt_{02} + \Kpt_{21} - \Kpt_{22}),\nonumber
\end{eqnarray}
\vspace{-9mm}
\begin{eqnarray}
\tilde{f}_5 = 0.25(\Kpt_{11} + \Kpt_{21} + \Kpt_{12} + \Kpt_{22}),\nonumber
\end{eqnarray}
\vspace{-9mm}
\begin{eqnarray}
\tilde{f}_6 =  0.25(-\Kpt_{11} + \Kpt_{21} - \Kpt_{12} + \Kpt_{22}),\nonumber
\end{eqnarray}
\vspace{-9mm}
\begin{eqnarray}
\tilde{f}_7 = 0.25(\Kpt_{11} - \Kpt_{21} - \Kpt_{12} + \Kpt_{22}),\nonumber
\end{eqnarray}
\vspace{-9mm}
\begin{eqnarray}
\tilde{f}_8 = 0.25(-\Kpt_{11} - \Kpt_{21} + \Kpt_{12} + \Kpt_{22}).\nonumber
\end{eqnarray}
Step 6: Perform streaming step via lock-step advection along different discrete particle directions
\begin{eqnarray}
  f_\alpha(\bm{x},t+\delta_t) &=& \tilde{f}_\alpha(\bm{x}-\bm{e}_{\alpha}\delta_t,t), \quad \alpha = 0,1,2,\ldots, 8,
\end{eqnarray}
and apply wall boundary conditions, as appropriate. \newline
Step 7: Compute hydrodynamic fields via zeroth and first discrete velocity moments as
\begin{eqnarray}
\rho = \sum_{\alpha=0}^{8} f_{\alpha}, \qquad \rho \bm{u} = \sum_{\alpha=0}^{8} f_{\alpha}\bm{e}_{\alpha} + \frac{1}{2}\bm{F} \delta_t,
\end{eqnarray}
and $P = \rho c_s^2$. The Steps 1 through 7 represent the computations involved in the FPC-LBM during one time step $\delta_t$, which are repeated as many times as required depending on the nature of the flow simulation. A note regarding the relaxation parameters $\omega_j$, which satisfy $0 < \omega_j < 2$, is in order here. For the relaxation parameters for the second order moments $\omega_3 = \omega_{2s}$ and $\omega_4=\omega_5 =\omega_{2d}$, they are based on the choice of the bulk and shear viscosities, $\zeta$ and $\nu$, respectively, according to Eq.~(\ref{eqn:transportcoeff2DFPCLBM}) to recover the Navier-Stokes equations. The rest of $\omega_j$, where $j = 0, 1, 2, 6, 7, 8$ are free parameters and are selected based on numerical stability considerations of simulations.

\newpage
\section{Chapman-Enskog analysis of 3D central moment formulation of the Boltzmann equation with Fokker-Planck collision model\label{sec:CEAnalysis3DFPC-LBM}}

From Eq.~(\ref{boltzmann_equation}), taking $\left<  \frac{\partial f}{\partial t} + \bm{\xi}\cdot\bm{\nabla}f,W_{mnp} \right>$ and setting it equal to the sum of  $\left(\frac{\delta \Pi_{mnp}}{\delta t}  \right)_{coll}^{FP}$ and $\left(\frac{\delta \Pi_{mnp}}{\delta t}  \right)_{forcing}$, and, as in the 2D case, a small perturbation parameter $\varepsilon$ is introduced into the collision term to obtain the normal solutions in the low-frequency hydrodynamical limit for the C-E expansion, we get~\cite{Premnath2019}
\begin{eqnarray}\label{eqn:97}
 \frac{D \Pi_{mnp}}{D t} + m \Pi_{m-1,n,p}\frac{D  u_x}{Dt} + n\Pi_{m,n-1,p}\frac{D u_y}{D t} + p\Pi_{m,n,p-1}\frac{D u_z}{D t}\\[3mm]
+  \Pi_{mnp} [(m+1)\partial_x u_x + (n+1)\partial_y u_y+ (p+1)\partial_z u_z]+ \partial_x \Pi_{ m+1,n,p}  \nonumber \\[3mm]
+  \partial_y \Pi_{m,n+1,p}+\partial_z \Pi_{m,n,p+1}+  n\Pi_{m+1,n-1,p} \partial_x u_y  + p\Pi_{m+1,n,p-1} \partial_x u_z \nonumber \\[3mm]
+m\Pi_{m-1,n+1,p} \partial_y u_x+  p\Pi_{m,n+1,p-1} \partial_y u_z + m\Pi_{m-1,n,p+1} \partial_z u_x \nonumber \\[3mm]
 +  n\Pi_{m,n-1,p+1} \partial_z u_y = \frac{\omega_{mnp}}{\varepsilon} \left[ \Pi_{mnp}^{ Mv} -  \Pi_{ mnp} \right] + \Gamma_{mnp},\nonumber
\end{eqnarray}
where $\Gamma_{mnp}$ is given in Eq.~(\ref{forcing:3D}) and the material derivative $\frac{D}{Dt}$ is defined in Eq.~(\ref{eqn:53}). As in the 2D case, we expand $f$ and $\partial_t$ given in Eq.~(\ref{eqn:55a}) and Eq.~(\ref{eqn:55b}), respectively, and the corresponding central moment expansion is
\begin{equation}\label{eqn:100}
\Pi_{ mnp}= \Pi_{ mnp}^{ (0)} +  \varepsilon\Pi_{ mnp}^{ (1)} + \varepsilon^2 \Pi_{ mnp}^{ (2)} +\cdots,
\end{equation}
where $\Pi_{ mnp}^{ (0)} = \Pi_{ mnp}^{ Mv}$, with the solvability conditions arising via the collision invariants of mass and momentum given by
\begin{gather}\label{eqn:101}
\Pi_{ 000}^{ (0)} =\rho, \Pi_{001}^{ (0)} =\Pi_{ 010}^{ (0)} = \Pi_{ 001}^{ (0)} = 0, \\[3mm]
\Pi_{ 000}^{ (n)} = \Pi_{001}^{ (n)} =\Pi_{ 010}^{ (n)} = \Pi_{ 001}^{ (n)} = 0.  \; \; \; n\geq 1 \nonumber
\end{gather}
Using the definition of the leading order Lagrangian derivative $\frac{D}{Dt_0}$ given in Eq.~(\ref{eqn:58}), and substituting Eq.~(\ref{eqn:100}) in Eq.~(\ref{eqn:97}) by invoking Eq.~(\ref{eqn:101}), we get the evolution equations of the central moment of ($m+n+p$) at various $O(\varepsilon^n)$, where $n=0,1,\dots$.

At $O(\varepsilon^0)$, we obtain
\begin{eqnarray}\label{eqn:102} 
 \frac{D \Pi_{mnp}^{ (0)}}{D t_{ 0}} + m \Pi_{m-1,n,p}^{ (0)}\frac{D u_x}{Dt_{0}} + n\Pi_{m,n-1,p}^{ (0)}\frac{D u_y}{D t_{ 0}} + p\Pi_{m,n,p-1}^{ (0)}\frac{D u_z}{D t_{0}}   \\[3mm]
+ \Pi_{mnp}^{ (0)} [(m+1)\partial_x u_x + (n+1)\partial_y u_y+ (p+1)\partial_z u_z] + \partial_x \Pi_{ m+1,n,p}^{ (0)} \nonumber \\[3mm]
 + \partial_y \Pi_{m,n+1,p}^{ (0)} + \partial_z \Pi_{ m,n,p+1}^{ (0)}+ n\Pi_{m+1,n-1,p}^{ (0)} \partial_x u_y + p\Pi_{m+1,n,p-1}^{ (0)} \partial_x u_z  \nonumber \\[3mm]
+  m\Pi_{m-1,n+1,p}^{ (0)} \partial_y u_x +  p\Pi_{m,n+1,p-1}^{ (0)} \partial_y u_z+  m\Pi_{m-1,n,p+1}^{ (0)} \partial_z u_x  \nonumber \\[3mm]
+  n\Pi_{m,n-1,p+1}^{ (0)} \partial_z u_y = -\omega_{mnp} \Pi_{ mnp}^{ (1)}  + \Gamma_{mnp}^{ (0)}, \nonumber
\end{eqnarray}
where
\begin{equation}\label{eqn:103}
 \Gamma_{mnp}^{ (0)} = m\frac{F_x}{\rho}\Pi_{m-1,n,p}^{ (0)} + n\frac{F_y}{\rho}\Pi_{m,n-1,p}^{ (0)}+ p\frac{F_z}{\rho}\Pi_{m,n,p-1}^{ (0)}.
\end{equation}
Then, the $O(\varepsilon^{(1)})$ system reads as
\begin{eqnarray}\label{eqn:firstorderCMsystem3D}
\frac{\partial \Pi_{mnp}^{ (0)}}{\partial t_{ 1}} + \frac{D \Pi_{mnp}^{ (1)}}{D t_{ 0}}+ m \Pi_{m-1,n,p}^{ (1)}\frac{D u_x}{Dt_{0}}+ m \Pi_{m-1,n,p}^{ (0)}\frac{\partial u_x}{\partial t_{1}} \\[3mm]
+ n\Pi_{m,n-1,p}^{ (1)}\frac{D u_y}{D t_{ 0}}+ n\Pi_{m,n-1,p}^{ (0)}\frac{\partial u_y}{\partial t_{ 1}} + p\Pi_{m,n,p-1}^{ (1)}\frac{D u_z}{D t_{0}}+ p\Pi_{m,n,p-1}^{ (0)}\frac{\partial u_z}{\partial t_{1}} \nonumber \\[3mm]
+ \Pi_{mnp}^{ (1)} [(m+1)\partial_x u_x+ (n+1)\partial_y u_y+ (p+1)\partial_z u_z] + \partial_x \Pi_{ m+1,n,p}^{ (1)} \nonumber \\[3mm]
 + \partial_y \Pi_{m,n+1,p}^{ (1)}+ \partial_z \Pi_{ m,n,p+1}^{ (1)}+ n\Pi_{m+1,n-1,p}^{ (1)} \partial_x u_y+ p\Pi_{m+1,n,p-1}^{ (1)} \partial_x u_z  \nonumber \\[3mm]
 +  m\Pi_{m-1,n+1,p}^{ (1)} \partial_y u_x+  p\Pi_{m,n+1,p-1}^{ (1)} \partial_y u_z+  m\Pi_{m-1,n,p+1}^{ (1)} \partial_z u_x \nonumber \\[3mm]
 +  n\Pi_{m,n-1,p+1}^{ (1)} \partial_z u_y= -\omega_{mnp} \Pi_{ mnp}^{ (2)}  + \Gamma_{mnp}^{ (1)},\nonumber
\end{eqnarray}
where
\[ \Gamma_{mnp}^{ (1)} = m\frac{F_x}{\rho}\Pi_{m-1,n,p}^{ (1)} + n\frac{F_y}{\rho}\Pi_{m,n-1,p}^{ (1)}+ p\frac{F_z}{\rho}\Pi_{m,n,p-1}^{ (1)}.\]

Next, in order to separate out the isotropic or the trace of the diagonal part of the viscous momentum transfer, i.e., the bulk viscosity effects from those due to the shear viscosity, we define the following combinations of second order moments:
\begin{eqnarray}
\Pi_{2s1} = \Pi_{ 200}+\Pi_{ 020}+\Pi_{ 002}, \quad \Pi_{2d1} = \Pi_{ 200}-\Pi_{ 020}, \quad \Pi_{2d2} = \Pi_{ 200}-\Pi_{ 002}.\label{eqn:104}
\end{eqnarray}
Conversely, from the combined moments $\Pi_{ 2s1}$, $\Pi_{ 2d1}$, and $\Pi_{ 2d2}$, the separate components $\Pi_{200}$, $\Pi_{020}$, and $\Pi_{002}$ may be retrieved via
\begin{subequations}
\begin{eqnarray}
& &\Pi_{200} = \frac{1}{3}\left(\Pi_{ 2s1}+\Pi_{2d1}+\Pi_{ 2d2}\right),\label{eqn:105a}  \\[3mm]
& &\Pi_{020} =  \frac{1}{3}\left(\Pi_{ 2s1}-2\Pi_{2d1}+\Pi_{ 2d2}\right),\label{eqn:105b} \\[3mm]
& &\Pi_{002} =  \frac{1}{3}\left(\Pi_{ 2s1}+\Pi_{2d1}-2\Pi_{ 2d2}\right).\label{eqn:105c}
\end{eqnarray}
\end{subequations}
Then, we prescribe those combined moments to relax at their own individual relaxation rates $\omega_{2s1}$, $\omega_{2d1}$, and $\omega_{2d2}$ under collision rather than undergoing changes separately as
\begin{subequations}
\begin{eqnarray}
& &\left(\frac{\delta}{\delta t} \Pi_{ 2s} \right)_{\!\!\!  coll}^{\!\!\! F\!P} = \omega_{ 2s1} \left[ \Pi_{ 2s1}^{ Mv} - \Pi_{ 2s} \right],\label{eqn:106a} \\[3mm]
& &\left(\frac{\delta}{\delta t} \Pi_{ 2d1} \right)_{\!\!\!  coll}^{\!\!\!  F\!P} = \omega_{ 2d1} \left[ \Pi_{ 2d1}^{ Mv} - \Pi_{ 2d1} \right],\label{eqn:106b} \\[3mm]
& &\left(\frac{\delta}{\delta t} \Pi_{ 2d2} \right)_{\!\!\!  coll}^{\!\!\!  F\!P} = \omega_{ 2d2} \left[ \Pi_{ 2d2}^{ Mv} - \Pi_{ 2d2} \right],\label{eqn:106c}
\end{eqnarray}
\end{subequations}
where, from Eq.~(\ref{eqn:104}), the attractors of the combined components of the diagonal parts of the second order moments are given by $\Pi_{ 2s1}^{ Mv} = \Pi_{ 200}^{ Mv}+\Pi_{ 020}^{ Mv}+\Pi_{ 002}^{ Mv} = 3 c_s^2 \rho$, $\Pi_{ 2d1}^{ Mv} = \Pi_{ 200}^{ Mv}-\Pi_{ 020}^{ Mv}=0$, and $\Pi_{ 2d2}^{ Mv} = \Pi_{ 200}^{ Mv}-\Pi_{ 002}^{ Mv} = 0$. All other second order moments and higher, such as ($mnp$) = (110), (101), (011), (120), (102), (210), (012), (201), (021), etc relax according to Eq.~(\ref{FPcoll3d}).

We first list $O(\varepsilon^0)$ evolution equations for the conserved moments, i.e., ($mnp$) = (000), (100), (010), and (001) respectively, using Eq.~(\ref{eqn:102}) as
\begin{subequations}
\begin{eqnarray}
\frac{D\Pi_{ 000}^{ (0)}}{Dt_0} + \Pi_{ 000}^{ (0)}\bm{\nabla \cdot u} = 0 \; \; \; \; \; \; &\mbox{or}& \; \; \; \; \; \; \frac{D\rho}{Dt_0} + \rho\bm{\nabla \cdot u}=0,\label{eqn:107a} \\[3mm]
\Pi_{ 000}^{ (0)}\frac{D u_x}{Dt_0}+ \partial_x\Pi_{ 200}^{ (0)}  = F_x \; \; \; \; \; \; &\mbox{or}& \; \; \; \; \; \; \rho\frac{D u_x}{Dt_0}+ \partial_x(c_s^2 \rho)  = F_x, \label{eqn:107b}\\[3mm]
\Pi_{ 000}^{ (0)}\frac{Du_y}{Dt_0} + \partial_y\Pi_{ 020}^{ (0)}  = F_y \; \; \; \; \; \; &\mbox{or}& \; \; \; \; \; \; \rho\frac{D u_y}{Dt_0}+ \partial_y(c_s^2 \rho)  = F_y, \label{eqn:107c}\\[3mm]
\Pi_{ 000}^{ (0)}\frac{Du_z}{Dt_0} + \partial_z\Pi_{ 002}^{ (0)}  = F_z \; \; \; \; \; \; &\mbox{or}& \; \; \; \; \; \; \rho\frac{D u_z}{Dt_0}+ \partial_z(c_s^2 \rho)  = F_z .\label{eqn:107d}
\end{eqnarray}
\end{subequations}
Similarly, for the second order off-diagonal moments ($mnp$) = ($110$), ($101$), and  ($011$) respectively , we have
\begin{subequations}
\begin{eqnarray}
\Pi_{ 200}^{ (0)} \partial_x u_y + \Pi_{ 020}^{ (0)}\partial_y u_x = -\omega_{ 110}\Pi_{ 110}^{ (1)}\; \; \; \; \; \; \mbox{or} \; \; \; \; \; \; \rho c_s^2 (\partial_x u_y + \partial_y u_x) = -\omega_{ 110}\Pi_{ 110}^{ (1)},\label{eqn:eqnoffdiaga} \\[3mm]
\Pi_{ 200}^{ (0)} \partial_x u_z + \Pi_{ 002}^{ (0)}\partial_z u_x = -\omega_{ 101}\Pi_{ 101}^{ (1)}\; \; \; \; \; \; \mbox{or} \; \; \; \; \; \; \rho c_s^2 (\partial_x u_z + \partial_z u_x) = -\omega_{ 101}\Pi_{ 101}^{ (1)},\label{eqn:eqnoffdiagb} \\[3mm]
\Pi_{ 020}^{ (0)} \partial_y u_z + \Pi_{ 002}^{ (0)}\partial_z u_y = -\omega_{ 011}\Pi_{ 011}^{ (1)}\; \; \; \; \; \; \mbox{or} \; \; \; \; \; \; \rho c_s^2 (\partial_y u_z + \partial_z u_y) = -\omega_{ 011}\Pi_{ 011}^{ (1)}.\label{eqn:eqnoffdiagc}
\end{eqnarray}
\end{subequations}
As in the 2D case, we consider the combinations of the second order diagonal moments (200), (020), and (002) in the LHS of Eq.~(\ref{eqn:102}) for $\Pi_{ 2s1}$, $\Pi_{ 2d1}$, and $\Pi_{ 2d2}$, where $\Pi_{ 2s1} = \Pi_{ 2s}^{ (0)} + \varepsilon \Pi_{ 2s1}^{ (0)} + \dots$, $\Pi_{ 2d1} = \Pi_{ 2d1}^{ (0)} + \varepsilon \Pi_{ 2d1}^{ (0)} + \cdots$, and $\Pi_{ 2d2} = \Pi_{ 2d2}^{ (0)} + \varepsilon \Pi_{ 2d2}^{ (0)} + \cdots$, and then use them in equations, Eq.~(\ref{eqn:106a})-Eq.~(\ref{eqn:106c}), for the collision terms on their RHS. In view of this, and using  $\Pi_{ 200}^{ (0)} = \Pi_{ 020}^{ (0)} = \Pi_{ 002}^{ (0)} = \rho c_s^2$, and by setting $\Pi_{ 300}^{ (0)} = \Pi_{ 030}^{ (0)} = \Pi_{ 003}^{ (0)} = 0$ in the continuous case (see the Chapman-Enksog analysis for the 2D case given earlier for related discussion), we get the following evolution equations for the moments with ($mnp$) = ($2s1$), ($2d1$), and ($2d2$), respectively, as
\begin{subequations}
\begin{eqnarray}
\frac{D}{Dt_0}\Pi_{ 2s1}^{ (0)} + 5 \Pi_{200}^{ (0)}\left( \partial_x u_x + \partial_y u_y + \partial_z u_z \right)  \! \! \! & =& \! \! \! - \omega_{ 2s}\Pi_{ 2s}^{ (1)},\label{eqn:108a} \\[3mm]
\frac{D}{Dt_0}\Pi_{ 2d1}^{ (0)} + 2 \Pi_{200}^{ (0)}\left( \partial_x u_x - \partial_y u_y  \right) \! \! \! & =& \! \! \! - \omega_{ 2d1}\Pi_{ 2d1}^{ (1)},\label{eqn:108b}\\[3mm]
\frac{D}{Dt_0}\Pi_{ 2d2}^{ (0)} + 2 \Pi_{200}^{ (0)}\left( \partial_x u_x -\partial_z u_z \right)  \! \! \! & =& \! \! \! - \omega_{ 2d2}\Pi_{ 2d2}^{ (1)}.\label{eqn:108c}
\end{eqnarray}
\end{subequations}
Simplifying the above three equations, we get
\begin{subequations}
\begin{eqnarray}
3 c_s^2\frac{D\rho}{Dt_0} + 5 \rho c_s^2 \left( \partial_x u_x + \partial_y u_y + \partial_z u_z \right)  \! \! \! & =& \! \! \! -\omega_{ 2s} \Pi_{ 2s}^{ (1)},\label{eqn:109a} \\[3mm]
2\rho c_s^2 (\partial_x u_x - \partial_y u_y)  \! \! \! & =& \! \! \!  -\omega_{ 2d1} \Pi_{ 2d1}^{ (1)},\label{eqn:109b} \\[3mm]
2\rho c_s^2 (\partial_x u_x - \partial_z u_z)  \! \! \! & =& \! \! \!  -\omega_{ 2d2} \Pi_{ 2d2}^{ (1)}.\label{eqn:109c}
\end{eqnarray}
\end{subequations}

From equations, Eq.~(\ref{eqn:eqnoffdiaga})-Eq.~(\ref{eqn:eqnoffdiagc}) and Eq.~(\ref{eqn:109a})-Eq.~(\ref{eqn:109c}), and after using Eq.~(\ref{eqn:107a}) to replace $\frac{D\rho}{Dt_0}$ in terms of $-\rho \bm{\nabla}\cdot\bm{u}$ in Eq.~(\ref{eqn:109a}), we can then obtain the non-equilibrium second order central moment components, which read as follows:
\begin{eqnarray}
&&\Pi_{ 110}^{ (1)}  = -\frac{\rho c_s^2}{\omega_{ 110}} (\partial_x u_y + \partial_y u_x), \quad
\Pi_{ 101}^{ (1)} = -\frac{\rho c_s^2}{\omega_{ 101}} (\partial_x u_z + \partial_z u_x), \nonumber \\
&&\Pi_{ 011}^{ (1)}  = -\frac{\rho c_s^2}{\omega_{ 011}} (\partial_y u_z + \partial_z u_y),\nonumber \\[3mm]
&&\Pi_{ 2s1}^{ (1)} = -2\frac{\rho c_s^2}{\omega_{ 2s}}\bm{\nabla}\cdot\bm{u}, \quad
\Pi_{ 2d1}^{ (1)}  = -2\frac{\rho c_s^2}{\omega_{ 2d1}}  (\partial_x u_x - \partial_y u_y ), \nonumber\\
&&\Pi_{ 2d2}^{ (1)}  = -2\frac{\rho c_s^2}{\omega_{ 2d2}}  (\partial_x u_x - \partial_z u_z).\label{eqn:noneqmmoments3D}
\end{eqnarray}
Analogous to Eqs.~(\ref{eqn:107a})-(\ref{eqn:107d}), we obtain the evolution equations for the conserved central moments with components (000), (100), (010), and (001) at $O(\varepsilon)$ level from Eq.~(\ref{eqn:firstorderCMsystem3D}), respectively, as
\begin{subequations}
\begin{eqnarray}\label{eqn:111}
\frac{\partial \rho}{\partial t_1}  \! \! \! & =& \! \! \!0,\label{eqn:111a}\\[3mm]
\rho \frac{\partial}{\partial t_1}u_x + \partial_x \Pi_{200}^{ (1)}+\partial_y \Pi_{ 110}^{ (1)}+\partial_z \Pi_{ 101}^{ (1)} \! \! \! & =& \! \! \!0,\label{eqn:111b}\\[3mm]
\rho \frac{\partial u_y}{\partial t_1} + \partial_x \Pi_{110}^{ (1)}+\partial_y \Pi_{ 020}^{ (1)}+\partial_z \Pi_{ 011}^{ (1)} \! \! \! & =& \! \! \!0,\label{eqn:111c}\\[3mm]
\rho \frac{\partial u_z }{\partial t_1}+ \partial_x \Pi_{101}^{ (1)}+\partial_y \Pi_{ 011}^{ (1)}+\partial_z \Pi_{ 002}^{ (1)} \! \! \! & =& \! \! \!0.\label{eqn:111d}
\end{eqnarray}
\end{subequations}
The individual diagonal components $\Pi_{ 200}^{ (1)}$, $\Pi_{ 020}^{ (1)}$, and $\Pi_{ 002}^{ (1)}$ in Eqs.~(\ref{eqn:111a})-(\ref{eqn:111d}) will then be rewritten in terms of the known combined non equilibrium central moments $\Pi_{ 2s1}^{ (1)}$, $\Pi_{ 2d1}^{ (1)}$, and $\Pi_{ 2d2}^{ (1)}$ via using the definitions Eqs.~(\ref{eqn:105a})-(\ref{eqn:105c}) and rewriting for the non-equilibrium parts as
\begin{subequations}
\begin{eqnarray}\label{eqn:112}
\frac{\partial \rho}{\partial t_1}  \! \! \! & =& \! \! \! 0,\label{eqn:112a} \\[3mm]
\rho \frac{\partial u_x}{\partial t_1} + \partial_x\left[\frac{1}{3}(\Pi_{ 2s1}^{ (1)}+ \Pi_{ 2d1}^{ (1)}+\Pi_{ 2d2}^{ (1)})\right]+\partial_y \Pi_{ 110}^{ (1)}+\partial_z \Pi_{ 101}^{ (1)}  \! \! \! & =& \! \! \! 0,\label{eqn:112b}\\[3mm]
\rho \frac{\partial u_y }{\partial t_1}+ \partial_x \Pi_{110}^{ (1)}+\partial_y \left[\frac{1}{3}(\Pi_{ 2s1}^{ (1)}-2 \Pi_{ 2d1}^{ (1)}+\Pi_{ 2d2}^{ (1)})\right]+\partial_z \Pi_{ 011}^{ (1)}  \! \! \! & =& \! \! \! 0,\label{eqn:112c}\\[3mm]
\rho \frac{\partial u_z}{\partial t_1} + \partial_x \Pi_{101}^{ (1)}+\partial_y \Pi_{ 011}^{ (1)}+\partial_z \left[\frac{1}{3}(\Pi_{ 2s1}^{ (1)}+ \Pi_{ 2d1}^{ (1)}-2\Pi_{ 2d2}^{ (1)})\right] \! \! \! & =& \! \! \!0.\label{eqn:112d}
\end{eqnarray}
\end{subequations}

We then combine $O(1)$ and $O(\varepsilon)$ evolution equations for the conserved moments, i.e., Eq.~(\ref{eqn:107a}) + $\varepsilon \times$  Eqs.~(\ref{eqn:112a}),  Eq.~(\ref{eqn:107b}) + $\varepsilon \times$ Eqs.~(\ref{eqn:112b}),  Eq.~(\ref{eqn:107c}) + $\varepsilon \times$ Eq.~(\ref{eqn:112c}), and  Eq.~(\ref{eqn:107d}) + $\varepsilon \times$  Eq.~(\ref{eqn:112d}) and then using $\partial_t = \partial_{t_0} + \varepsilon \partial_{t_1}$ and $D_t = D_{t_0} + \varepsilon \partial_{t_1}$, and setting the book keeping parameter $\varepsilon = 1$. These steps then lead to the following equations for the conserved fields:
\begin{subequations}
\begin{eqnarray}
\frac{D \rho}{D t} + \rho\bm{\nabla}\cdot\bm{u} \! \! \! & =& \! \! \! 0\label{eqn:113a} \\[3mm]
\rho \frac{D}{D t}u_x + \partial_x(c_s^2 \rho)+ \partial_x\left[\frac{1}{3}(\Pi_{ 2s1}^{ (1)}+ \Pi_{ 2d1}^{ (1)}+\Pi_{ 2d2}^{ (1)})\right]+\partial_y \Pi_{ 110}^{ (1)}+\partial_z \Pi_{ 101}^{ (1)}  \! \! \! & =& \! \! \! F_x,\label{eqn:113b}\\[3mm]
\rho  \frac{D}{D t}u_y + \partial_y(c_s^2 \rho)+ \partial_x \Pi_{110}^{ (1)}+\partial_y \left[\frac{1}{3}(\Pi_{ 2s1}^{ (1)}-2 \Pi_{ 2d1}^{ (1)}+\Pi_{ 2d2}^{ (1)})\right]+\partial_z \Pi_{ 011}^{ (1)}  \! \! \! & =& \! \! \! F_y,\label{eqn:113c}\\[3mm]
\rho\frac{D}{D t}u_z+ \partial_z(c_s^2 \rho) + \partial_x \Pi_{101}^{ (1)}+\partial_y \Pi_{ 011}^{ (1)}+\partial_z \left[\frac{1}{3}(\Pi_{ 2s1}^{ (1)}+ \Pi_{ 2d1}^{ (1)}-2\Pi_{ 2d2}^{ (1)})\right] \! \! \! & =& \! \! \!F_z.\label{eqn:113d}
\end{eqnarray}
\end{subequations}
In view of Eq.~(\ref{eqn:noneqmmoments3D}) for the non-equilibrium moments, and in anticipation of arriving at the Navier-Stokes equations (NSE) from Eqs.~(\ref{eqn:113a})-(\ref{eqn:113d}), we define the bulk viscosity $\zeta$ and shear viscosity $\nu$ as
\begin{subequations}
\begin{eqnarray}
\zeta  \! \! \! & =& \! \! \! \frac{2}{3}\frac{c_s^2}{\omega_{2s}}, \label{eqn:114a}\\[3mm]
\nu = \frac{c_s^2}{\omega_{110}} =  \frac{c_s^2}{\omega_{101}} =  \frac{c_s^2}{\omega_{011}}=  \frac{c_s^2}{\omega_{2d1}} \! \! \! & =& \! \! \!  \frac{c_s^2}{\omega_{2d2}}.\label{eqn:114b}
\end{eqnarray}
\end{subequations}
Then, using Eq.~(\ref{eqn:noneqmmoments3D}) in Eqs.~(\ref{eqn:113b})-(\ref{eqn:113d}) and simplifying by utilizing Eqs.~(\ref{eqn:114a}) and (\ref{eqn:114b}), we finally get the emergent equations for the conserved fields, which read as
\begin{eqnarray}
\frac{D \rho}{D t} +\rho \bm{\nabla}\cdot\bm{u} = 0 ,\label{eqn:115a}\nonumber
\end{eqnarray}
\begin{eqnarray}
\rho\frac{D u_x}{Dt}  \! \! \! & =& \! \! \! -\partial_x P + \partial_x \left[ \rho \zeta \bm{\nabla}\cdot\bm{u}+ \rho \nu \left(2\partial_x u_x -\frac{2}{3}\bm{\nabla}\cdot\bm{u}\right) \right] \nonumber \\[3mm]
& &+ \partial_y \left[  \rho \nu (\partial_x u_y + \partial_y u_x)   \right ]+  \partial_z \left[  \rho \nu (\partial_x u_z + \partial_z u_x)   \right ]+ F_x,\nonumber
\end{eqnarray}
\begin{eqnarray}
\rho\frac{D u_y}{Dt}  \! \! \! & =& \! \! \! -\partial_y P + \partial_x \left[  \rho \nu (\partial_x u_y + \partial_y u_x)   \right ]  + \partial_y \left[ \rho \zeta \bm{\nabla \cdot u}+\rho \nu \left(2\partial_y u_y -\frac{2}{3} \bm{\nabla \cdot u}\right)    \right] \nonumber \\[3mm]
& &+ \partial_z \left[  \rho \nu (\partial_y u_z + \partial_z u_y)   \right] + F_y, \nonumber
\end{eqnarray}
\begin{eqnarray}
 \rho\frac{D u_z}{Dt} \! \! \! & =& \! \! \! -\partial_z P + \partial_x \left[  \rho \nu (\partial_x u_z + \partial_z u_x)   \right ]  + \partial_y \left[ \rho \nu (\partial_y u_z+ \partial_z u_y )   \right] \nonumber \\[3mm]
 & &+ \partial_z \left[ \rho \zeta \bm{\nabla \cdot u}+ \rho \nu \left(2\partial_z u_z -\frac{2}{3} \bm{\nabla \cdot u}\right)   \right ] + F_z, \nonumber
\end{eqnarray}
where $P=\rho c_s^2$. Hence, the NSE in 3D can be derived consistently from the central moment formulation of the continuous Boltzmann equation with the Fokker-Planck collision model. The above analysis may also be readily extended to study the evolution of the higher order moments in 3D. Also, as noted at the end of Appendix~\ref{sec:CEAnalysis2DFPC-LBM}, when this continuous analysis is extended for a discrete formulation involving the LBM, the relationships between the viscosities and the relaxation parameters require the so-called  H\'{e}non corrections, i.e., by
replacing $1/\omega_{ij}$ with $(1/\omega_{ij}-1/2)$. Hence, for the discrete 3D FPC-LBM, Eqs.~(\ref{eqn:114a}) and (\ref{eqn:114b}) is modified to
\begin{eqnarray}\label{eqn:transportcoeff3DFPCLBM}
\zeta = \frac{2}{3}c_s^2\left(\frac{1}{\omega_{ 2s}}-\frac{1}{2}\right), \; \nu =c_s^2\left(\frac{1}{\omega_{j}}-\frac{1}{2}\right) \; \mbox{for} \; j = (110), (101), (011), (2d1), (2d2).
\end{eqnarray}

\newpage
\section{Appendix: Algorithmic details of 3D FPC-LBM using the D3Q27 lattice\label{sec:3DFPC-LBM}}
Step 1: Convert the pre-collision distribution functions into pre-collision raw moments using $\m = \PP\mathbf{f}$ as
\begin{eqnarray}
\Kps{000} \! \! \! & =& \! \! \!  \f{0} + \f{1} + \f{2} + \f{3} + \f{4} + \f{5} + \f{6} + \f{7} +  \f{8} + \f{9} + \f{10} + \f{11} + \f{12}\\
& & + \f{13} + \f{14} +\f{15} + \f{16} + \f{17} + \f{18} + \f{19} + \f{20} + \f{21} + \f{22} + \f{23}\nonumber\\
& &  +\f{24} + \f{25} +\f{26},\nonumber
\end{eqnarray}
\vspace{-9mm}
\begin{eqnarray}
\Kps{100}  \! \! \! & =& \! \! \! \f{1} -\f{2} + \f{7} - \f{8} + \f{9} - \f{10} + \f{11}-\f{12} + \f{13} - \f{14}+\f{19} - \f{20}\nonumber\\
  & & + \f{21} - \f{22} + \f{23} - \f{24} + \f{25} -\f{26}, \nonumber
\end{eqnarray}
\vspace{-9mm}
\begin{eqnarray}
\Kps{010}  \! \! \! & =& \! \! \! \f{3} - \f{4} + \f{7} + \f{8} - \f{9} - \f{10} + \f{15}-\f{16} + \f{17} - \f{18} + \f{19} + \f{20}\nonumber\\
 & & - \f{21}- \f{22} + \f{23}+\f{24} - \f{25} -\f{26}, \nonumber
\end{eqnarray}
\vspace{-9mm}
\begin{eqnarray}
\Kps{001} \! \! \! & =& \! \! \! \f{5} - \f{6} + \f{11} + \f{12} - \f{13} - \f{14} + \f{15}+ \f{16} - \f{17} - \f{18} + \f{19} + \f{20}\nonumber\\
 & & + \f{21}+ \f{22} - \f{23} -\f{24}-\f{25} - \f{26}, \nonumber
\end{eqnarray}
\vspace{-9mm}
\begin{eqnarray}
\Kps{110}  \! \! \! & =& \! \! \! \f{7} - \f{8} - \f{9} + \f{10} + \f{19} - \f{20} - \f{21}+ \f{22} + \f{23} - \f{24}- \f{25}+ \f{26},\nonumber
\end{eqnarray}
\vspace{-9mm}
\begin{eqnarray}
\Kps{101}  \! \! \! & =& \! \! \! \f{11} - \f{12} - \f{13} + \f{14} + \f{19} - \f{20} + \f{21}- \f{22} - \f{23} + \f{24}- \f{25}+ \f{26},\nonumber
\end{eqnarray}
\vspace{-9mm}
\begin{eqnarray}
\Kps{011} \! \! \! & =& \! \! \! \f{15} - \f{16} - \f{17} + \f{18} + \f{19} + \f{20} - \f{21}- \f{22} - \f{23} - \f{24}+ \f{25}+ \f{26},\nonumber
\end{eqnarray}
\vspace{-9mm}
\begin{eqnarray}
\Kps{200} \! \! \! & =& \! \! \! \f{1} + \f{2} +\f{7} +  \f{8} + \f{9} + \f{10} +\f{11} + \f{12} + \f{13} + \f{14}+\f{19} + \f{20} + \f{21}\nonumber \\
  & & + \f{22} + \f{23} +\f{24} + \f{25} +\f{26},\nonumber
\end{eqnarray}
\vspace{-9mm}
\begin{eqnarray}
\Kps{020}  \! \! \! & =& \! \! \! \f{3} + \f{4} +\f{7} +  \f{8} + \f{9} + \f{10} +\f{15} + \f{16} + \f{17} + \f{18}+\f{19} + \f{20} + \f{21} \nonumber \\
  & &+ \f{22}+ \f{23} +\f{24} + \f{25} +\f{26},\nonumber
\end{eqnarray}
\vspace{-9mm}
\begin{eqnarray}
\Kps{002} \! \! \! & =& \! \! \! \f{5} + \f{6} +\f{11} + \f{12} + \f{13} + \f{14} +\f{15} + \f{16} + \f{17} + \f{18}+\f{19} + \f{20}\nonumber \\
   & &+ \f{21}+ \f{22}+ \f{23} +\f{24} + \f{25} +\f{26},\nonumber
\end{eqnarray}
\vspace{-9mm}
\begin{eqnarray}
\Kps{120} \! \! \! & =& \! \! \! \f{7} - \f{8} +\f{9} - \f{10}+\f{19} - \f{20} + \f{21} - \f{22} + \f{23} - \f{24} + \f{25} -\f{26},\nonumber
\end{eqnarray}
\vspace{-9mm}
\begin{eqnarray}
\Kps{102}  \! \! \! & =& \! \! \! \f{11} - \f{12} +\f{13} - \f{14}+\f{19} - \f{20} + \f{21} - \f{22} + \f{23} - \f{24} + \f{25} -\f{26},\nonumber
\end{eqnarray}
\vspace{-9mm}
\begin{eqnarray}
\Kps{210}  \! \! \! & =& \! \! \! \f{7} + \f{8} -\f{9} - \f{10}+ \f{19} + \f{20}- \f{21} - \f{22} + \f{23} +\f{24} - \f{25} -\f{26},\nonumber
\end{eqnarray}
\vspace{-9mm}
\begin{eqnarray}
\Kps{012}  \! \! \! & =& \! \! \! \f{15} - \f{16} +\f{17} - \f{18}+ \f{19} + \f{20}- \f{21} - \f{22} + \f{23} +\f{24} - \f{25} -\f{26},\nonumber
\end{eqnarray}
\vspace{-9mm}
\begin{eqnarray}
\Kps{201} \! \! \! & =& \! \! \! \f{11} + \f{12} -\f{13} - \f{14}+\f{19} + \f{20} + \f{21} + \f{22} - \f{23}-\f{24}-\f{25} - \f{26},\nonumber
\end{eqnarray}
\vspace{-9mm}
\begin{eqnarray}
\Kps{021} \! \! \! & =& \! \! \! \f{15} + \f{16} -\f{17} - \f{18}+\f{19} + \f{20} + \f{21} + \f{22} - \f{23}-\f{24}-\f{25} - \f{26},\nonumber
\end{eqnarray}
\vspace{-9mm}
\begin{eqnarray}
\Kps{111}  \! \! \! & =& \! \! \! \f{19} - \f{20} -\f{21} + \f{22}-\f{23} + \f{24}+\f{25} - \f{26},\nonumber
\end{eqnarray}
\vspace{-9mm}
\begin{eqnarray}
\Kps{220} = \f{7} +  \f{8} + \f{9} + \f{10} +\f{19} + \f{20} + \f{21} + \f{22} + \f{23} +\f{24} + \f{25} +\f{26},\nonumber
\end{eqnarray}
\vspace{-9mm}
\begin{eqnarray}
\Kps{202} = \f{11} + \f{12} + \f{13} + \f{14} +\f{19} + \f{20} + \f{21} + \f{22} + \f{23} +\f{24} + \f{25} +\f{26}, \nonumber
\end{eqnarray}
\vspace{-9mm}
\begin{eqnarray}
\Kps{022} = \f{15} + \f{16} + \f{17} + \f{18}+\f{19} + \f{20} + \f{21} + \f{22} + \f{23} +\f{24} + \f{25} +\f{26},\nonumber
\end{eqnarray}
\vspace{-9mm}
\begin{eqnarray}
\Kps{211} = \f{19} + \f{20} -\f{21} - \f{22}-\f{23} - \f{24}+\f{25} + \f{26},\nonumber
\end{eqnarray}
\vspace{-9mm}
\begin{eqnarray}
\Kps{121} = \f{19} - \f{20} +\f{21} - \f{22}-\f{23} + \f{24}-\f{25} + \f{26},\nonumber
\end{eqnarray}
\vspace{-9mm}
\begin{eqnarray}
\Kps{112} = \f{19} - \f{20} -\f{21} + \f{22}+\f{23} - \f{24}-\f{25} + \f{26},\nonumber
\end{eqnarray}
\vspace{-9mm}
\begin{eqnarray}
\Kps{122} = \f{19} - \f{20} + \f{21} - \f{22} + \f{23} - \f{24} + \f{25} -\f{26},\nonumber
\end{eqnarray}
\vspace{-9mm}
\begin{eqnarray}
\Kps{212} = \f{19} + \f{20}- \f{21} - \f{22} + \f{23} +\f{24} - \f{25} -\f{26},\nonumber
\end{eqnarray}
\vspace{-9mm}
\begin{eqnarray}
\Kps{221} = \f{19} + \f{20} + \f{21} + \f{22} - \f{23}-\f{24}-\f{25} - \f{26},\nonumber
\end{eqnarray}
\vspace{-9mm}
\begin{eqnarray}
\Kps{222} = \f{19} + \f{20} + \f{21} + \f{22} + \f{23} +\f{24} + \f{25} +\f{26}.\nonumber
\end{eqnarray}

\noindent
Step 2: Map the pre-collision raw moments into pre-collision central moments using $\mc = \F \m$ as
\begin{eqnarray}
\Ks{000} = \Kps{000},
\end{eqnarray}
\vspace{-9mm}
\begin{eqnarray}
\Ks{100} = \Kps{100} - \ux\Kps{000},\nonumber
\end{eqnarray}
\vspace{-9mm}
\begin{eqnarray}
\Ks{010} = \Kps{010} - \uy\Kps{000},\nonumber
\end{eqnarray}
\vspace{-9mm}
\begin{eqnarray}
\Ks{001} = \Kps{001} -\uz\Kps{000},\nonumber
\end{eqnarray}
\vspace{-9mm}
\begin{eqnarray}
\Ks{110} = \Kps{110} - \ux\Kps{010}- \uy\Kps{100} + \ux\uy\Kps{000},\nonumber
\end{eqnarray}
\vspace{-9mm}
\begin{eqnarray}
\Ks{101} = \Kps{101} - \ux\Kps{001} - \uz\Kps{100} + \ux\uz\Kps{000},\nonumber
\end{eqnarray}
\vspace{-9mm}
\begin{eqnarray}
\Ks{011} = \Kps{011} -  \uy\Kps{001} - \uz\Kps{010} +  \uy\uz\Kps{000},\nonumber
\end{eqnarray}
\vspace{-9mm}
\begin{eqnarray}
\Ks{200} = \Kps{200}  - 2\ux\Kps{100} + \uxx\Kps{000 },\nonumber
\end{eqnarray}
\vspace{-9mm}
\begin{eqnarray}
\Ks{020} = \Kps{020}  - 2\uy\Kps{010} + \uyy\Kps{000} ,\nonumber
\end{eqnarray}
\vspace{-9mm}
\begin{eqnarray}
\Ks{002} = \Kps{002}  - 2\uz \Kps{001} +\uzz\Kps{000} ,\nonumber
\end{eqnarray}
\vspace{-9mm}
\begin{eqnarray}
\Ks{120} = \Kps{120} -\ux\Kps{020} - 2\uy \Kps{110}+ 2\ux\uy\Kps{010}+\uyy\Kps{100}- \ux\uyy\Kps{000},\nonumber
\end{eqnarray}
\vspace{-9mm}
\begin{eqnarray}
\Ks{102} = \Kps{102} - \ux\Kps{002} - 2\uz \Kps{101} + 2\ux\uz \Kps{001}+ \uzz\Kps{100}  - \ux\uzz\Kps{000},\nonumber
\end{eqnarray}
\vspace{-9mm}
\begin{eqnarray}
\Ks{210} = \Kps{210} -\uy\Kps{200}- 2\ux\Kps{110}  + \uxx\Kps{010}+ 2\ux\uy\Kps{100} - \uxx\uy\Kps{000},\nonumber
\end{eqnarray}
\vspace{-9mm}
\begin{eqnarray}
\Ks{012} = \Kps{012} -\uy\Kps{002} - 2\uz\Kps{011}+ \uzz\Kps{010} + 2\uy\uz\Kps{001} - \uy\uzz\Kps{000},\nonumber
\end{eqnarray}
\vspace{-9mm}
\begin{eqnarray}
\Ks{201} = \Kps{201}- \uz\Kps{200}  - 2\ux\Kps{101} + \uxx\Kps{001} + 2\ux\uz\Kps{100}- \uxx\uz\Kps{000},\nonumber
\end{eqnarray}
\vspace{-9mm}
\begin{eqnarray}
\Ks{021} = \Kps{021} - \uz\Kps{020}- 2\uy\Kps{011} + \uyy\Kps{001}+ 2\uy\uz\Kps{010} - \uyy\uz\Kps{000},\nonumber
\end{eqnarray}
\vspace{-9mm}
\begin{eqnarray}
\Ks{111} \! \! \! & =& \! \! \! \Kps{111} - \ux\Kps{011} -\uy\Kps{101} - \uz\Kps{110} + \ux\uy\Kps{001} + \ux\uz\Kps{010} + \uy\uz\Kps{100}\nonumber \\
& & - \ux\uy\uz\Kps{000},\nonumber
\end{eqnarray}
\vspace{-9mm}
\begin{eqnarray}
\Ks{220}  \! \! \! & =& \! \! \! \Kps{220} - 2\uy \Kps{210}- 2\ux \Kps{120}+ \uxx\Kps{020}+ \uyy\Kps{200}+ 4\ux\uy \Kps{110} - 2\uxx\uy \Kps{010}\nonumber \\
& &  - 2\ux\uyy \Kps{100} +\uxx\uyy\Kps{000},\nonumber
\end{eqnarray}
\vspace{-9mm}
\begin{eqnarray}
\Ks{202} \! \! \! & =& \! \! \!  \Kps{202}-2\uz\Kps{201} - 2\ux \Kps{102}+ \uxx\Kps{002} + \uzz\Kps{200}+ 4\ux\uz\Kps{101}- 2\uxx\uz\Kps{001}\nonumber \\
& &- 2\ux\uzz\Kps{100}+ \uxx\uzz\Kps{000} ,\nonumber
\end{eqnarray}
\vspace{-9mm}
\begin{eqnarray}
\Ks{022} \! \! \! & =& \! \! \! \Kps{022}- 2\uz \Kps{021}- 2\uy \Kps{012}+\uzz \Kps{020}+\uyy\Kps{002} + 4\uy\uz\Kps{011}- 2\uyy\uz \Kps{001} \nonumber \\
& &- 2\uy\uzz\Kps{010}+ \uyy\uzz\Kps{000},\nonumber
\end{eqnarray}
\vspace{-9mm}
\begin{eqnarray}
\Ks{211} \! \! \! & =& \! \! \! \Kps{211} -2\ux \Kps{111}- \uy\Kps{201} -\uz\Kps{210} + \uxx\Ks{011} + 2\ux\uy \Kps{101}+ \uy\uz\Kps{200}\nonumber \\
& & + 2\ux\uz \Kps{110}  -\uxx\uy\Kps{001} -\uxx\uz\Kps{010}-2\ux\uy\uz\Kps{100} + \uxx\uy\uz\Kps{000},\nonumber
\end{eqnarray}
\vspace{-9mm}
\begin{eqnarray}
\Ks{121} \! \! \! & =& \! \! \! \Kps{121} - 2\uy\Kps{111}  -\ux\Kps{021}- \uz\Kps{120}+ \ux\uz\Kps{020} + 2\ux\uy\Kps{011} + \uyy\Kps{101}\nonumber \\
& &  + 2\uy\uz\Kps{110}- \ux\uyy\Kps{001} -2\ux\uy\uz\Kps{010}- \uyy\uz\Kps{100}  +\ux\uyy\uz\Kps{000},\nonumber
\end{eqnarray}
\vspace{-9mm}
\begin{eqnarray}
\Ks{112} \! \! \! & =& \! \! \! \Kps{112}-2\uz \Kps{111}- \ux\Kps{012} -\uy\Kps{102}+ \ux\uy\Kps{002}+ 2\ux\uz\Kps{011} + 2\uy\uz \Kps{101}\nonumber \\
& &  + \uzz\Kps{110}-2\ux\uy\uz\Kps{001}- \ux\uzz\Kps{010} - \uy\uzz\Kps{100} +\ux\uy\uzz\Kps{000},\nonumber
\end{eqnarray}
\vspace{-9mm}
\begin{eqnarray}
\Ks{122} \! \! \! & =& \! \! \! \Kps{122}- 2\uy\Kps{112} - 2\uz \Kps{121}-\ux\Kps{022} + 4\uy\uz \Kps{111}  + 2\ux\uz\Kps{021}\nonumber\\
& & + 2\ux\uy\Kps{012} + \uyy\Kps{102}+ \uzz\Kps{120}- \ux\uyy\Kps{002} -\ux\uzz\Kps{020}- 4\ux\uy\uz \Kps{011} \nonumber\\
& & - 2\uyy\uz\Kps{101}- 2\uy\uzz\Kps{110}+ 2\ux\uyy\uz\Kps{001}  + 2\ux\uy\uzz \Kps{010}+ \uyy\uzz\Kps{100}\nonumber \\
& & - \ux\uyy\uzz\Kps{000}, \nonumber
\end{eqnarray}
\vspace{-9mm}
\begin{eqnarray}
\Ks{212} \! \! \! & =& \! \! \! \Kps{212}- 2\ux\Kps{112}- 2\uz \Kps{211} -\uy\Kps{202} + 4\ux\uz \Kps{111} + 2\uy\uz \Kps{201}\nonumber \\
& &  + \uxx\Kps{012} + \uzz\Kps{210} + 2\ux\uy \Kps{102}- \uxx\uy\Kps{002}-\uy\uzz\Kps{200}- 2\uxx\uz\Kps{011} \nonumber \\
& & - 4\ux\uy\uz \Kps{101} -2\ux\uzz\Kps{110}+ 2\uxx\uy\uz \Kps{001}+ \uxx\uzz\Kps{010} + 2\ux\uy\uzz \Kps{100}\nonumber \\
& & - \uxx\uy\uzz\Kps{000},\nonumber
\end{eqnarray}
\vspace{-9mm}
\begin{eqnarray}
\Ks{221} \! \! \! & =& \! \! \!  \Kps{221}- 2\ux \Kps{121} - 2\uy\Kps{211}- \uz\Kps{220}+ 4\ux\uy \Kps{111} + \uxx\Kps{021}\nonumber \\
& & + \uyy\Kps{201}+ 2\uy\uz \Kps{210}+ 2\ux\uz \Kps{120} - \uxx\uz\Kps{020} - \uyy\uz\Kps{200}  - 2\uxx\uy\Kps{011}\nonumber \\
& &-2\ux\uyy \Kps{101} - 4\ux\uy\uz \Kps{110}+\uxx\uyy\Kps{001}+ 2\uxx\uy\uz \Kps{010}+ 2 \ux\uyy\uz\Kps{100}\nonumber \\
& & - \uxx\uyy\uz\Kps{000},\nonumber
\end{eqnarray}
\vspace{-9mm}
\begin{eqnarray}
\Ks{222} \! \! \! & =& \! \! \!\Kps{222} - 2\uz\Kps{221}- 2\uy\Kps{212}- 2 \ux\Kps{122}+ 4\ux\uy\Kps{112}+ 4\ux\uz \Kps{121}\nonumber \\
& &+ 4\uy\uz\Kps{211}+\uxx\Kps{022}+ \uyy\Kps{202}+ \uzz\Kps{220}- 8\ux\uy\uz\Kps{111}- 2\uxx\uz\Kps{021}\nonumber \\
& &- 2\uyy\uz\Kps{201}- 2\uxx\uy \Kps{012}- 2\uy\uzz\Kps{210} - 2\ux\uyy\Kps{102} - 2\ux\uzz\Kps{120}\nonumber \\
& & +\uxx\uyy\Kps{002}+ \uxx\uzz\Kps{020}+\uyy\uzz\Kps{200}+ 4\uxx\uy\uz\Kps{011} + 4\ux\uyy\uz\Kps{101}\nonumber \\
& &+ 4\ux\uy\uzz\Kps{110}- 2\uxx\uyy\uz\Kps{001} - 2\uxx\uy\uzz\Kps{010} - 2\ux\uyy\uzz\Kps{100}\nonumber \\
& & +\uxx\uyy\uzz\Kps{000}.
\end{eqnarray}
\noindent
Step 3: Collision update - Relax different central moments to their attractors and augment them with source terms

Then, based on various combinations that are usually considered in the moment basis itself, we perform those combinations here on the pre-collision central moments to relax those groups together later during collision, as
\begin{eqnarray}
\Ks{2s1} =\Ks{200}+ \Ks{020} + \Ks{002},\qquad
\Ks{2d1} = \Ks{200}- \Ks{020},\qquad
\Ks{2d2} = \Ks{200}- \Ks{002},
\end{eqnarray}
\vspace{-9mm}
\begin{eqnarray}
\Ks{3s1} =  \Ks{120} + \Ks{102}, \qquad
\Ks{3m1} = \Ks{120} - \Ks{102}, \qquad
\Ks{3s2} = \Ks{210} + \Ks{012}, \qquad \nonumber
\end{eqnarray}
\vspace{-9mm}
\begin{eqnarray}
\Ks{3m2} = \Ks{210} - \Ks{012},\qquad
\Ks{3s3} =  \Ks{201} + \Ks{021}, \qquad
\Ks{3m3} =\Ks{201} -\Ks{021},\nonumber
\end{eqnarray}
\vspace{-9mm}
\begin{eqnarray}
\Ks{4s1}  = \Ks{220} + \Ks{202} + \Ks{022}, \qquad
\Ks{4d1}=\Ks{220} +\Ks{202}-\Ks{022}, \qquad
\Ks{4d2} =\Ks{220}-\Ks{202}. \nonumber
\end{eqnarray}
We then define all of the lower order, up to the 3rd, Markovian central moment attractors as
\begin{eqnarray}
\KMv{000} =\rho, \qquad
\KMv{100} =0, \qquad
\KMv{010} =0, \qquad
\KMv{001} =0, \nonumber
\end{eqnarray}
\vspace{-9mm}
\begin{eqnarray}
\KMv{110} = 0 , \qquad
\KMv{101} = 0, \qquad
\KMv{011} = 0, \qquad
\KMv{200} = c_s^2 \rho, \nonumber
\end{eqnarray}
\vspace{-9mm}
\begin{eqnarray}
\KMv{020} = c_s^2 \rho, \qquad
\KMv{002} = c_s^2 \rho, \qquad
\KMv{120} = 0, \qquad
\KMv{102} = 0 , \nonumber
\end{eqnarray}
\vspace{-9mm}
\begin{eqnarray}
\KMv{210} = 0, \qquad
\KMv{012} = 0, \qquad
\KMv{201} = 0, \qquad
\KMv{021} = 0, \qquad
\KMv{111} = 0, \nonumber
\end{eqnarray}
and then perform the same combinations as above for the Markovian attractors as
\begin{eqnarray}
\Ks{2s1}^{Mv} =\Ks{200}^{Mv}+ \Ks{020}^{Mv} + \Ks{002}^{Mv}, \qquad
\Ks{2d1}^{Mv} = \Ks{200}^{Mv}- \Ks{020}^{Mv},\qquad
\Ks{2d2}^{Mv} = \Ks{200}^{Mv}- \Ks{002}^{Mv}
\end{eqnarray}
\vspace{-9mm}
\begin{eqnarray}
\Ks{3s1}^{Mv} =  \Ks{120}^{Mv} + \Ks{102}^{Mv}, \qquad
\Ks{3m1}^{Mv} = \Ks{120}^{Mv} - \Ks{102}^{Mv},  \qquad
\Ks{3s2}^{Mv} = \Ks{210}^{Mv} + \Ks{012}^{Mv},  \nonumber
\end{eqnarray}
\vspace{-9mm}
\begin{eqnarray}
\Ks{3m2}^{Mv} = \Ks{210}^{Mv} - \Ks{012}^{Mv},  \qquad
\Ks{3s3}^{Mv} =  \Ks{201}^{Mv} + \Ks{021}^{Mv}, \qquad
\Ks{3m3}^{Mv} =\Ks{201}^{Mv} -\Ks{021}^{Mv}.\nonumber
\end{eqnarray}
Then, we relax various central moments to their attractors along with contributions due to the source terms via $\sigma_{mnp}$ (from Eq.~(\ref{eq:cmupdatesourceD3Q27})) to obtain the respective post-collision central moments as follows.
\begin{eqnarray}
\Kts{000} = \Ks{000} + \omega_{0} (\Ks{000}^{Mv} - \Ks{000})+ \left(1-\omega_{0}/2 \right) \Ss{000}\delta_t,
\end{eqnarray}
\vspace{-9mm}
\begin{eqnarray}
\Kts{100} = \Ks{100} + \omega_{1} (\Ks{100}^{Mv} - \Ks{100})+ \left(1-\omega_{1}/2 \right) \Ss{100}\delta_t, \nonumber
\end{eqnarray}
\vspace{-9mm}
\begin{eqnarray}
\Kts{010} = \Ks{010} + \omega_{2} (\Ks{010}^{Mv} - \Ks{010})+ \left(1-\omega_{2}/2 \right) \Ss{010}\delta_t, \nonumber
\end{eqnarray}
\vspace{-9mm}
\begin{eqnarray}
\Kts{001} = \Ks{001} + \omega_{3} (\Ks{001}^{Mv} - \Ks{001})+ \left(1-\omega_{3}/2 \right) \Ss{001}\delta_t, \nonumber
\end{eqnarray}
\vspace{-9mm}
\begin{eqnarray}
\Kts{110} = \Ks{110} + \omega_{4} (\Ks{110}^{Mv} - \Ks{110})+ \left(1-\omega_{4}/2 \right) \Ss{110}\delta_t, \nonumber
\end{eqnarray}
\vspace{-9mm}
\begin{eqnarray}
\Kts{101} = \Ks{101} + \omega_{5} (\Ks{101}^{Mv} - \Ks{101})+ \left(1-\omega_{5}/2 \right) \Ss{101}\delta_t, \nonumber
\end{eqnarray}
\vspace{-9mm}
\begin{eqnarray}
\Kts{011} = \Ks{011} + \omega_{6} (\Ks{011}^{Mv} - \Ks{011})+ \left(1-\omega_{6}/2 \right) \Ss{011}\delta_t, \nonumber
\end{eqnarray}
\vspace{-9mm}
\begin{eqnarray}
\Kts{2s1} = \Ks{2s} + \omega_{7} (\Ks{2s1}^{Mv} - \Ks{2s1})+ \left(1-\omega_{7}/2 \right) \Ss{2s1}\delta_t, \nonumber
\end{eqnarray}
\vspace{-9mm}
\begin{eqnarray}
\Kts{2d1} = \Ks{2d1} + \omega_{8} (\Ks{2d1}^{Mv} - \Ks{2d1})+ \left(1-\omega_{8}/2 \right) \Ss{2d1}\delta_t, \nonumber
\end{eqnarray}
\vspace{-9mm}
\begin{eqnarray}
\Kts{2d2} = \Ks{2d2} + \omega_{9} (\Ks{2d2}^{Mv} - \Ks{2d2})+ \left(1-\omega_{9}/2 \right) \Ss{2d2}\delta_t. \nonumber
\end{eqnarray}
We then decompose the respective combined second order post-collision central moment combinations as
\begin{eqnarray}
\Kts{200} = (\Kts{2s1} + \Kts{2d1} + \Kts{2d2})/3, \qquad
\Kts{020} = (\Kts{2s1} -2\Kts{2d1} + \Kts{2d2})/3, \nonumber
\end{eqnarray}
\vspace{-9mm}
\begin{eqnarray}
\Kts{002} = (\Kts{2s1} + \Kts{2d1} -2\Kts{2d2})/3, \nonumber
\end{eqnarray}
and then relax the third order central moments towards their respective Markovian attractors as
\begin{eqnarray}
\Kts{3s1} = \Ks{3s1} + \omega_{10} (\Ks{3s1}^{Mv} - \Ks{3s1})+ \left(1-\omega_{10}/2 \right) \Ss{3s1}\delta_t,
\end{eqnarray}
\vspace{-9mm}
\begin{eqnarray}
\Kts{3m1} = \Ks{3m1} + \omega_{11} (\Ks{3m1}^{Mv} - \Ks{3m1})+ \left(1-\omega_{11}/2 \right) \Ss{3m1}\delta_t, \nonumber
\end{eqnarray}
\vspace{-9mm}
\begin{eqnarray}
\Kts{3s2} = \Ks{3s2} + \omega_{12} (\Ks{3s2}^{Mv} - \Ks{3s2})+ \left(1-\omega_{12}/2 \right) \Ss{3s2}\delta_t, \nonumber
\end{eqnarray}
\vspace{-9mm}
\begin{eqnarray}
\Kts{3m2} = \Ks{3m2} + \omega_{13} (\Ks{3m2}^{Mv} - \Ks{3m2})+ \left(1-\omega_{13}/2 \right) \Ss{3m2}\delta_t, \nonumber
\end{eqnarray}
\vspace{-9mm}
\begin{eqnarray}
\Kts{3s3} = \Ks{3s3} + \omega_{14} (\Ks{3s3}^{Mv} - \Ks{3s3})+ \left(1-\omega_{14}/2 \right) \Ss{3s3}\delta_t, \nonumber
\end{eqnarray}
\vspace{-9mm}
\begin{eqnarray}
\Kts{3m3} = \Ks{3m3} + \omega_{15} (\Ks{3m3}^{Mv} - \Ks{3m3})+ \left(1-\omega_{15}/2 \right) \Ss{3m3}\delta_t, \nonumber
\end{eqnarray}
\vspace{-9mm}
\begin{eqnarray}
\Kts{111} = \Ks{111} + \omega_{16} (\Ks{111}^{Mv} - \Ks{111})+ \left(1-\omega_{16}/2 \right) \Ss{111}\delta_t. \nonumber
\end{eqnarray}
The resulting combined third order post-collision central moments are then decomposed via
\begin{eqnarray}
\Kts{120} = (\Kts{3s1} + \Kts{3m1})/2, \qquad
\Kts{102} = (\Kts{3s1} - \Kts{3m1})/2,   \qquad
\Kts{210} = (\Kts{3s2} + \Kts{3m2})/2, \nonumber
\end{eqnarray}
\vspace{-9mm}
\begin{eqnarray}
\Kts{012} = (\Kts{3s2} - \Kts{3m2})/2,  \qquad
\Kts{201} =(\Kts{3s3} + \Kts{3m3})/2, \qquad
\Kts{112} = (\Kts{3s3} - \Kts{3m3})/2.\nonumber
\end{eqnarray}
We can then define the fourth order Markovian Attractors using the post collision values of second order central moments as
\begin{eqnarray}
\KMv{220} = \frac{1}{\rho} (\Kts{200}\Kts{020}+2\Kts{110}\Kts{110}), \qquad
\KMv{202} = \frac{1}{\rho} (\Kts{200}\Kts{002}+2\Kts{101}\Ks{101}),
\end{eqnarray}
\vspace{-3mm}
\begin{eqnarray}
\KMv{022} = \frac{1}{\rho}(\Kts{020}\Kts{002}+2\Kts{011}\Kts{011}), \qquad
\KMv{211} = \frac{1}{\rho}(\Kts{200}\Kts{011}+2\Kts{110}\Kts{101}), \nonumber
\end{eqnarray}
\vspace{-3mm}
\begin{eqnarray}
\KMv{121} = \frac{1}{\rho}(\Kts{020}\Kts{101}+2\Kts{110}\Kts{011}), \qquad
\KMv{112} = \frac{1}{\rho}(\Kts{002}\Kts{110}+2\Kts{011}\Kts{101}),  \nonumber
\end{eqnarray}
and the 4th order combinations as
\begin{gather}
\Ks{4s1}^{Mv}  = \Ks{220}^{Mv} + \Ks{202}^{Mv} + \Ks{022}^{Mv}, \qquad
\Ks{4d1}^{Mv}=\Ks{220}^{Mv} +\Ks{202}^{Mv}-\Ks{022}^{Mv}, \qquad
\Ks{4d2}^{Mv} =\Ks{220}^{Mv}-\Ks{202}^{Mv}. \nonumber
\end{gather}
Subsequently, we relax those fourth order central moments towards their respective Markovian attractors as
\begin{eqnarray}
\Kts{4s1} = \Ks{4s1} + \omega_{17} (\Ks{4s1}^{Mv} - \Ks{4s1})+ \left(1-\omega_{17}/2 \right) \Ss{4s1}\delta_t,
\end{eqnarray}
\vspace{-9mm}
\begin{eqnarray}
\Kts{4d1} = \Ks{4d1} + \omega_{18} (\Ks{4d1}^{Mv} - \Ks{4d1})+ \left(1-\omega_{18}/2 \right) \Ss{4d1}\delta_t, \nonumber
\end{eqnarray}
\vspace{-9mm}
\begin{eqnarray}
\Kts{4d2} = \Ks{4d2} + \omega_{19} (\Ks{4d2}^{Mv} - \Ks{4d2})+ \left(1-\omega_{19}/2 \right) \Ss{4d2}\delta_t, \nonumber
\end{eqnarray}
\vspace{-9mm}
\begin{eqnarray}
\Kts{211} = \Ks{211} + \omega_{20} (\Ks{211}^{Mv} - \Ks{211})+ \left(1-\omega_{20}/2 \right) \Ss{211}\delta_t, \nonumber
\end{eqnarray}
\vspace{-9mm}
\begin{eqnarray}
\Kts{121} = \Ks{121} + \omega_{21} (\Ks{121}^{Mv} - \Ks{121})+ \left(1-\omega_{21/2} \right) \Ss{121}\delta_t, \nonumber
\end{eqnarray}
\vspace{-9mm}
\begin{eqnarray}
\Kts{112} = \Ks{112} + \omega_{22} (\Ks{112}^{Mv} - \Ks{112})+ \left(1-\omega_{22}/2 \right) \Ss{112}\delta_t. \nonumber
\end{eqnarray}
Following these, we decompose those fourth order post-collision combined moments through
\begin{eqnarray}
\Kts{220} = (\Kts{4s1} + \Kts{4d1} + 2\Kts{4d2})/4, \qquad
\Kts{202} =  (\Kts{4s1} +\Kts{4d1} - 2 \Kts{4d2})/4, \nonumber
\end{eqnarray}
\vspace{-9mm}
\begin{eqnarray}
\Kts{022} =  (\Kts{4s1} - \Kts{4d1})/2.\nonumber
\end{eqnarray}
Next, we define the fifth order central moment Markovian attractors, which are based on various combinations of products of lower order, i.e., second and third, post-collision central moments as
\begin{eqnarray}
\KMv{122} = \frac{2}{5\rho}(\Kts{020}\Kts{102}+\Kts{002}\Kts{120} + 4\Kts{011}\Kts{111}+2(\Kts{101}\Kts{021}+\Kts{011}\Kts{012})),
\end{eqnarray}
\vspace{-3mm}
\begin{eqnarray}
\KMv{212} = \frac{2}{5\rho}(\Kts{200}\Kts{012}+\Kts{002}\Kts{210} + 4\Kts{101}\Kts{111}+2(\Kts{110}\Kts{102}+\Kts{011}\Kts{201})), \nonumber
\end{eqnarray}
\vspace{-3mm}
\begin{eqnarray}
\KMv{221} = \frac{2}{5\rho}(\Kts{200}\Kts{021}+\Kts{020}\Kts{201} + 4\Kts{110}\Kts{111}+2(\Kts{011}\Kts{210}+\Kts{101}\Kts{120})), \nonumber
\end{eqnarray}
and then relax the fifth order central moments towards their respective Markovian attractors as
\begin{eqnarray}
\Kts{122} = \Ks{122} + \omega_{23} (\Ks{122}^{Mv} - \Ks{122})+ \left(1-\omega_{23}/2 \right) \Ss{122}\delta_t,
\end{eqnarray}
\vspace{-9mm}
\begin{eqnarray}
\Kts{212} = \Ks{212} + \omega_{24} (\Ks{212}^{Mv} - \Ks{212})+ \left(1-\omega_{24}/2 \right) \Ss{212}\delta_t, \nonumber
\end{eqnarray}
\vspace{-9mm}
\begin{eqnarray}
\Kts{221} = \Ks{221} + \omega_{25} (\Ks{221}^{Mv} - \Ks{221})+ \left(1-\omega_{25}/2 \right) \Ss{221}\delta_t. \nonumber
\end{eqnarray}
Finally, we define the sixth order central moment Markovian attractor as products of second and fourth order moments as
\begin{eqnarray}
\KMv{222} = \frac{1}{3\rho}(\Kts{200}\Kts{022}+\Kts{020}\Kts{202}+\Kts{002}\Kts{220}+4(\Kts{110}\Kts{112}+\Kts{101}\Kts{121}+\Kts{011}\Kts{211})), \nonumber
\end{eqnarray}
and then relax the sixth order central moments as
\begin{eqnarray}
\Kts{222} = \Ks{222} + \omega_{26} (\Ks{222}^{Mv} - \Ks{222})+ \left(1-\omega_{26}/2 \right) \Ss{222}\delta_t. \nonumber
\end{eqnarray}
Step 4: Transform post-collision central moments into post-collision raw moments via $\tilde{\m} = \Fi \tilde{\mc}$ as
\begin{eqnarray}
\Ktps{000} = \Kts{000},
\end{eqnarray}
\vspace{-9mm}
\begin{eqnarray}
\Ktps{100} = \Kts{100} + \ux\Kts{000},\nonumber
\end{eqnarray}
\vspace{-9mm}
\begin{eqnarray}
\Ktps{010} = \Kts{010} + \uy\Kts{000},\nonumber
\end{eqnarray}
\vspace{-9mm}
\begin{eqnarray}
\Ktps{001} = \Kts{001} +\uz\Kts{000},\nonumber
\end{eqnarray}
\vspace{-9mm}
\begin{eqnarray}
\Ktps{110} = \Kts{110} + \ux\Kts{010}+ \uy\Kts{100} + \ux\uy\Kts{000},\nonumber
\end{eqnarray}
\vspace{-9mm}
\begin{eqnarray}
\Ktps{101} = \Kts{101} + \ux\Kts{001} + \uz\Kts{100} + \ux\uz\Kts{000},\nonumber
\end{eqnarray}
\vspace{-9mm}
\begin{eqnarray}
\Ktps{011} = \Kts{011} +  \uy\Kts{001} +  \uz\Kts{010} +  \uy\uz\Kts{000},\nonumber
\end{eqnarray}
\vspace{-9mm}
\begin{eqnarray}
\Ktps{200} = \Kts{200}  + 2\ux\Kts{100} + \uxx\Kts{000 },\nonumber
\end{eqnarray}
\vspace{-9mm}
\begin{eqnarray}
\Ktps{020} = \Kts{020}  + 2\uy\Kts{010} + \uyy\Kts{000} ,\nonumber
\end{eqnarray}
\vspace{-9mm}
\begin{eqnarray}
\Ktps{002} = \Kts{002}  + 2\uz \Kts{001} +\uzz\Kts{000} ,\nonumber
\end{eqnarray}
\vspace{-9mm}
\begin{eqnarray}
\Ktps{120} = \Kts{120} +\ux\Kts{020} + 2\uy \Kts{110}+ 2\ux\uy\Kts{010}+\uyy\Kts{100}+ \ux\uyy\Kts{000}\nonumber
\end{eqnarray}
\vspace{-9mm}
\begin{eqnarray}
\Ktps{102} = \Kts{102} + \ux\Kts{002} + 2\uz \Kts{101} + 2\ux\uz \Kts{001}+ \uzz\Kts{100}  + \ux\uzz\Kts{000},\nonumber
\end{eqnarray}
\vspace{-9mm}
\begin{eqnarray}
\Ktps{210} = \Kts{210} +\uy\Kts{200}+ 2\ux\Kts{110}  + \uxx\Kts{010}+ 2\ux\uy\Kts{100} + \uxx\uy\Kts{000},\nonumber
\end{eqnarray}
\vspace{-9mm}
\begin{eqnarray}
\Ktps{012} = \Kts{012} +\uy\Kts{002} + 2\uz\Kts{011}+ \uzz\Kts{010} + 2\uy\uz\Kts{001} + \uy\uzz\Kts{000},\nonumber
\end{eqnarray}
\vspace{-9mm}
\begin{eqnarray}
\Ktps{201} = \Kts{201}+ \uz\Kts{200}  + 2\ux\Kts{101} + \uxx\Kts{001} + 2\ux\uz\Kts{100}+ \uxx\uz\Kts{000},\nonumber
\end{eqnarray}
\vspace{-9mm}
\begin{eqnarray}
\Ktps{021} = \Kts{021} + \uz\Kts{020} + 2\uy\Kts{011} + \uyy\Kts{001}+ 2\uy\uz\Kts{010} + \uyy\uz\Kts{000},\nonumber
\end{eqnarray}
\vspace{-9mm}
\begin{eqnarray}
\Ktps{111} \! \! \! & =& \! \! \! \Kts{111} + \ux\Kts{011} +\uy\Kts{101} + \uz\Kts{110} + \ux\uy\Kts{001} + \ux\uz\Kts{010}\nonumber \\
& & + \uy\uz\Kts{100} + \ux\uy\uz\Kts{000},\nonumber
\end{eqnarray}
\vspace{-9mm}
\begin{eqnarray}
\Ktps{220}  \! \! \! & =& \! \! \! \Kts{220} + 2\uy \Kts{210}+ 2\ux \Kts{120}+ \uxx\Kts{020}+ \uyy\Kts{200}+ 4\ux\uy \Kts{110}\nonumber \\
& & + 2\uxx\uy \Kts{010}  + 2\ux\uyy \Kts{100} +\uxx\uyy\Kts{000},\nonumber
\end{eqnarray}
\vspace{-9mm}
\begin{eqnarray}
\Ktps{202} \! \! \! & =& \! \! \!  \Kts{202}+2\uz\Kts{201} + 2\ux \Kts{102}+ \uxx\Kts{002} + \uzz\Kts{200}+ 4\ux\uz\Kts{101}\nonumber \\
& &+ 2\uxx\uz\Kts{001}+ 2\ux\uzz\Kts{100}+ \uxx\uzz\Kts{000} ,\nonumber
\end{eqnarray}
\vspace{-9mm}
\begin{eqnarray}
\Ktps{022}  \! \! \! & =& \! \! \! \Kts{022}+ 2\uz \Kts{021}+ 2\uy \Kts{012}+\uzz \Kts{020}+\uyy\Kts{002} + 4\uy\uz\Kts{011}\nonumber \\
& &+ 2\uyy\uz \Kts{001} + 2\uy\uzz\Kts{010}+ \uyy\uzz\Kts{000},\nonumber
\end{eqnarray}
\vspace{-9mm}
\begin{eqnarray}
\Ktps{211}  \! \! \! & =& \! \! \! \Kts{211} +2\ux \Kts{111}+ \uy\Kts{201} +\uz\Kts{210} + \uxx\Ks{011} + 2\ux\uy \Kts{101}\nonumber \\
& &+ \uy\uz\Kts{200} + 2\ux\uz \Kts{110}+\uxx\uy\Kts{001} +\uxx\uz\Kts{010}+2\ux\uy\uz\Kts{100}\nonumber \\
& & + \uxx\uy\uz\Kts{000},\nonumber
\end{eqnarray}
\vspace{-9mm}
\begin{eqnarray}
\Ktps{121}  \! \! \! & =& \! \! \! \Kts{121} + 2\uy\Kts{111}  +\ux\Kts{021}+ \uz\Kts{120}+ \ux\uz\Kts{020} + 2\ux\uy\Kts{011}\nonumber \\
& & + \uyy\Kts{101}  + 2\uy\uz\Kts{110} - \ux\uyy\Kts{001} +2\ux\uy\uz\Kts{010}+ \uyy\uz\Kts{100}\nonumber \\
& &  +\ux\uyy\uz\Kts{000},\nonumber
\end{eqnarray}
\vspace{-9mm}
\begin{eqnarray}
\Ktps{112}  \! \! \! & =& \! \! \! \Kts{112}+2\uz \Kts{111}+ \ux\Kts{012} +\uy\Kts{102}+ \ux\uy\Kts{002}+ 2\ux\uz\Kts{011}\nonumber \\
& & + 2\uy\uz \Kts{101}  + \uzz\Kts{110}+2\ux\uy\uz\Kts{001}+ \ux\uzz\Kts{010} + \uy\uzz\Kts{100}\nonumber \\
& & +\ux\uy\uzz\Kts{000},\nonumber
\end{eqnarray}
\vspace{-9mm}
\begin{eqnarray}
\Ktps{122} \! \! \! & =& \! \! \! \Kts{122}+ 2\uy\Kts{112} + 2\uz \Kts{121}+ \ux\Kts{022} + 4\uy\uz \Kts{111}  + 2\ux\uz\Kts{021}\nonumber \\
& &+ 2\ux\uy\Kts{012}+ \uyy\Kts{102} + \uzz\Kts{120}+ \ux\uyy\Kts{002} +\ux\uzz\Kts{020}+ 4\ux\uy\uz \Kts{011}\nonumber \\
& & + 2\uyy\uz\Kts{101} + 2\uy\uzz\Kts{110} + 2\ux\uyy\uz\Kts{001}  + 2\ux\uy\uzz \Kts{010}+ \uyy\uzz\Kts{100}\nonumber \\
& & + \ux\uyy\uzz\Kts{000},\nonumber
\end{eqnarray}
\vspace{-9mm}
\begin{eqnarray}
\Ktps{212}  \! \! \! & =& \! \! \! \Kts{212}+ 2\ux\Kts{112}+ 2\uz \Kts{211} +\uy\Kts{202} + 4\ux\uz \Kts{111} + 2\uy\uz \Kts{201}\nonumber \\
& &  + \uxx\Kts{012} + \uzz\Kts{210} + 2\ux\uy \Kts{102}+ \uxx\uy\Kts{002}+\uy\uzz\Kts{200}+ 2\uxx\uz\Kts{011}\nonumber \\
& &+ 4\ux\uy\uz \Kts{101}  +2\ux\uzz\Kts{110}+ 2\uxx\uy\uz \Kts{001}+ \uxx\uzz\Kts{010} + 2\ux\uy\uzz \Kts{100}\nonumber \\
& & + \uxx\uy\uzz\Kts{000},\nonumber
\end{eqnarray}
\vspace{-9mm}
\begin{eqnarray}
\Ktps{221}  \! \! \! & =& \! \! \!  \Kts{221}+ 2\ux \Kts{121} + 2\uy\Kts{211}+ \uz\Kts{220}+ 4\ux\uy \Kts{111} + \uxx\Kts{021}\nonumber \\
& & + \uyy\Kts{201}+ 2\uy\uz \Kts{210}+ 2\ux\uz \Kts{120} + \uxx\uz\Kts{020} + \uyy\uz\Kts{200}  + 2\uxx\uy\Kts{011}\nonumber \\
& &+2\ux\uyy \Kts{101}+ 4\ux\uy\uz \Kts{110}+\uxx\uyy\Kts{001}+ 2\uxx\uy\uz \Kts{010}+ 2 \ux\uyy\uz\Kts{100}\nonumber \\
& &+ \uxx\uyy\uz\Kts{000}, \nonumber
\end{eqnarray}
\vspace{-9mm}
\begin{eqnarray}
\Ktps{222}  \! \! \! & =& \! \! \!\Kts{222} + 2\uz\Kts{221}+ 2\uy\Kts{212}+ 2 \ux\Kts{122}+ 4\ux\uy\Kts{112}+ 4\ux\uz \Kts{121}\nonumber \\
& &+ 4\uy\uz\Kts{211}+\uxx\Kts{022}+ \uyy\Kts{202}+ \uzz\Kts{220}+ 8\ux\uy\uz\Kts{111}+ 2\uxx\uz\Kts{021}\nonumber \\
& &+ 2\uyy\uz\Kts{201}+ 2\uxx\uy \Kts{012}  + 2\uy\uzz\Kts{210} + 2\ux\uyy\Kts{102} + 2\ux\uzz\Kts{120}\nonumber \\
& &+\uxx\uyy\Kts{002}+ \uxx\uzz\Kts{020}+\uyy\uzz\Kts{200} + 4\uxx\uy\uz\Kts{011} + 4\ux\uyy\uz\Kts{101}\nonumber \\
& &+ 4\ux\uy\uzz\Kts{110}+ 2\uxx\uyy\uz\Kts{001} + 2\uxx\uy\uzz\Kts{010}+ 2\ux\uyy\uzz\Kts{100}\nonumber \\
& & +\uxx\uyy\uzz\Kts{000}.\nonumber
\end{eqnarray}
Step 5: Transform post-collision raw moments into post-collision distribution functions through $\tilde{\mathbf{f}} = \PP^{-1}\tilde{\m}$ as
\begin{eqnarray}
\ft{0} = \Ktps{000} - \Ktps{200} - \Ktps{020}-\Ktps{002} + \Ktps{220}+\Ktps{202} + \Ktps{022}- \Ktps{222},
\end{eqnarray}
\vspace{-9mm}
\begin{eqnarray}
\ft{1} = \left(\Ktps{100} + \Ktps{200} -\Ktps{120}-\Ktps{102} - \Ktps{220} - \Ktps{202} +\Ktps{122}+ \Ktps{222}\right)/2,\nonumber
\end{eqnarray}
\vspace{-9mm}
\begin{eqnarray}
\ft{2} = \left(-\Ktps{100} + \Ktps{200} +\Ktps{120}+\Ktps{102} - \Ktps{220} - \Ktps{202} -\Ktps{122}+ \Ktps{222}\right)/2,\nonumber
\end{eqnarray}
\vspace{-9mm}
\begin{eqnarray}
\ft{3} =  \left( \Ktps{010} + \Ktps{020}- \Ktps{210}- \Ktps{012} - \Ktps{220}- \Ktps{022} +\Ktps{212} + \Ktps{222}\right)/2,\nonumber
\end{eqnarray}
\vspace{-9mm}
\begin{eqnarray}
\ft{4} = \left(\Ktps{010} + \Ktps{020}+ \Ktps{210}+ \Ktps{012} - \Ktps{220}- \Ktps{022} -\Ktps{212} +\Ktps{222}\right)/2,\nonumber
\end{eqnarray}
\vspace{-9mm}
\begin{eqnarray}
\ft{5} =\left(\Ktps{001} + \Ktps{002}- \Ktps{201}- \Ktps{021} - \Ktps{202} - \Ktps{022} + \Ktps{221}+ \Ktps{222}\right)/2,\nonumber
\end{eqnarray}
\vspace{-9mm}
\begin{eqnarray}
\ft{6} = \left(-\Ktps{001}+\Ktps{002} + \Ktps{201}+ \Ktps{021} - \Ktps{202}- \Ktps{022} - \Ktps{221}+ \Ktps{222}\right)/2,\nonumber
\end{eqnarray}
\vspace{-9mm}
\begin{eqnarray}
\ft{7} = \left(\Ktps{110} + \Ktps{120} +\Ktps{210}+ \Ktps{220}- \Ktps{112} - \Ktps{122} - \Ktps{212} - \Ktps{222}\right)/4,\nonumber
\end{eqnarray}
\vspace{-9mm}
\begin{eqnarray}
\ft{8} =  \left(-\Ktps{110} - \Ktps{120}+\Ktps{210}+ \Ktps{220}+ \Ktps{112} + \Ktps{122} - \Ktps{212} - \Ktps{222}\right)/4,\nonumber
\end{eqnarray}
\vspace{-9mm}
\begin{eqnarray}
\ft{9} = \left(-\Ktps{110} + \Ktps{120} -\Ktps{210}+ \Ktps{220}+\Ktps{112}- \Ktps{122} + \Ktps{212} - \Ktps{222}\right)/4,\nonumber
\end{eqnarray}
\vspace{-9mm}
\begin{eqnarray}
\ft{10} = \left(\Ktps{110} - \Ktps{120} -\Ktps{210}+ \Ktps{220}-\Ktps{112}+ \Ktps{122} + \Ktps{212} - \Ktps{222}\right)/4,\nonumber
\end{eqnarray}
\vspace{-9mm}
\begin{eqnarray}
\ft{11} =  \left(\Ktps{101} + \Ktps{102} +\Ktps{201}+ \Ktps{202}- \Ktps{121} - \Ktps{122} - \Ktps{221} - \Ktps{222}\right)/4,\nonumber
\end{eqnarray}
\vspace{-9mm}
\begin{eqnarray}
& \ft{12} = \left(-\Ktps{101} - \Ktps{102} +\Ktps{201}+ \Ktps{202}+ \Ktps{121} + \Ktps{122} - \Ktps{221} - \Ktps{222}\right)/4,\nonumber
\end{eqnarray}
\vspace{-9mm}
\begin{eqnarray}
\ft{13} =  \left( -\Ktps{101} +\Ktps{102}-\Ktps{201} + \Ktps{202}+ \Ktps{121} - \Ktps{122} + \Ktps{221} - \Ktps{222}\right)/4,\nonumber
\end{eqnarray}
\vspace{-9mm}
\begin{eqnarray}
\ft{14} =  \left(\Ktps{101} -\Ktps{102}-\Ktps{201} + \Ktps{202}- \Ktps{121} + \Ktps{122} + \Ktps{221} - \Ktps{222}\right)/4,\nonumber
\end{eqnarray}
\vspace{-9mm}
\begin{eqnarray}
\ft{15} = \left(\Ktps{011} +\Ktps{012} + \Ktps{021} +\Ktps{022}- \Ktps{211} - \Ktps{212} - \Ktps{221} - \Ktps{222}\right)/4,\nonumber
\end{eqnarray}
\vspace{-9mm}
\begin{eqnarray}
\ft{16} = \left(-\Ktps{011} -\Ktps{012} + \Ktps{021} +\Ktps{022}+ \Ktps{211} + \Ktps{212} - \Ktps{221} - \Ktps{222}\right)/4,\nonumber
\end{eqnarray}
\vspace{-9mm}
\begin{eqnarray}
\ft{17} = \left(-\Ktps{011}+ \Ktps{012} -\Ktps{021} + \Ktps{022}+ \Ktps{211} - \Ktps{212} + \Ktps{221} - \Ktps{222}\right)/4,\nonumber
\end{eqnarray}
\vspace{-9mm}
\begin{eqnarray}
\ft{18} =  \left(\Ktps{011}- \Ktps{012} -\Ktps{021} + \Ktps{022}- \Ktps{211} + \Ktps{212} + \Ktps{221} - \Ktps{222}\right)/4,\nonumber
\end{eqnarray}
\vspace{-9mm}
\begin{eqnarray}
\ft{19} =\left(\Ktps{111} + \Ktps{211}+ \Ktps{121} + \Ktps{112}+ \Ktps{122} + \Ktps{212} + \Ktps{221} + \Ktps{222}\right)/8,\nonumber
\end{eqnarray}
\vspace{-9mm}
\begin{eqnarray}
\ft{20}=\left(-\Ktps{111} + \Ktps{211}- \Ktps{121} - \Ktps{112}- \Ktps{122} + \Ktps{212} + \Ktps{221} + \Ktps{222}\right)/8,\nonumber
\end{eqnarray}
\vspace{-9mm}
\begin{eqnarray}
\ft{21} = \left(- \Ktps{111}-\Ktps{211}+\Ktps{121} - \Ktps{112} + \Ktps{122} - \Ktps{212} + \Ktps{221} + \Ktps{222}\right)/8,\nonumber
\end{eqnarray}
\vspace{-9mm}
\begin{eqnarray}
\ft{22} =  \left(\Ktps{111}-\Ktps{211}-\Ktps{121} + \Ktps{112} - \Ktps{122} - \Ktps{212} + \Ktps{221} + \Ktps{222}\right)/8,\nonumber
\end{eqnarray}
\vspace{-9mm}
\begin{eqnarray}
\ft{23} =\left(-\Ktps{111}-\Ktps{211}- \Ktps{121}+ \Ktps{112} + \Ktps{122} + \Ktps{212} - \Ktps{221} + \Ktps{222}\right)/8,\nonumber
\end{eqnarray}
\vspace{-9mm}
\begin{eqnarray}
\ft{24}= \left(\Ktps{111}-\Ktps{211}+ \Ktps{121}- \Ktps{112} - \Ktps{122} + \Ktps{212} - \Ktps{221} + \Ktps{222}\right)/8,\nonumber
\end{eqnarray}
\vspace{-9mm}
\begin{eqnarray}
\ft{25} = \left(\Ktps{111} + \Ktps{211}- \Ktps{121}- \Ktps{112} + \Ktps{122} - \Ktps{212} - \Ktps{221} + \Ktps{222}\right)/8,\nonumber
\end{eqnarray}
\vspace{-9mm}
\begin{eqnarray}
\ft{26} = \left(-\Ktps{111} + \Ktps{211}+ \Ktps{121}+ \Ktps{112} - \Ktps{122} - \Ktps{212} - \Ktps{221} + \Ktps{222}\right)/8.\nonumber
\end{eqnarray}
Step 6: Perform streaming step via lock-step advection along different discrete particle directions
\begin{eqnarray}
  f_\alpha(\bm{x},t+\delta_t) &=& \tilde{f}_\alpha(\bm{x}-\bm{e}_{\alpha}\delta_t,t), \quad \alpha = 0,1,2,\ldots, 26,
\end{eqnarray}
and apply wall boundary conditions, as appropriate. \newline
Step 7: Compute hydrodynamic fields via zeroth and first discrete velocity moments as
\begin{eqnarray}
\rho = \sum_{\alpha=0}^{26} f_{\alpha}, \qquad \rho \bm{u} = \sum_{\alpha=0}^{26} f_{\alpha}\bm{e}_{\alpha} + \frac{1}{2}\bm{F} \delta_t,
\end{eqnarray}
and $P = \rho c_s^2$. We remark the following regarding the selection of the relaxation parameters $\omega_j$, where  $0 < \omega_j < 2$. For the relaxation parameters for the second order moments $\omega_4 = \omega_{110}$, $\omega_5 = \omega_{101}$, $\omega_6 = \omega_{011}$, $\omega_7 = \omega_{2s}$, $\omega_8 = \omega_{2d1}$ and $\omega_9 =\omega_{2d2}$, they are based on the choice of the bulk and shear viscosities, $\zeta$ and $\nu$, respectively, according to Eq.~(\ref{eqn:transportcoeff3DFPCLBM}) to recover the Navier-Stokes equations. The rest are free parameters and are selected based on numerical stability considerations of simulations.

\bibliographystyle{unsrt}

\end{document}